\documentclass[namedreferences]{solarphysics}
\usepackage[optionalrh]{spr-sola-addons} 
\usepackage{epsfig}          
\usepackage{graphicx}        
\usepackage{natbib}         
\usepackage{url}             
\usepackage[pdfborder={0 0 0 },urlcolor=blue,breaklinks]{hyperref}
\usepackage[usenames,dvipsnames]{color}		
\usepackage{xcolor}
\usepackage[normalem]{ulem}
\def\arcsec{$^{\prime\prime}$}

\DeclareRobustCommand{\ion}[2]{%
\relax\ifmmode
\ifx\testbx\f@series
{\mathbf{#1\,\mathsc{#2}}}\else
{\mathrm{#1\,\mathsc{#2}}}\fi
\else\textup{#1\,{\mdseries\textsc{#2}}}%
\fi}

\newcommand{\aap}{    {\it Astron. Astrophys.}}

\newcommand{\araa}{   {\it Ann. Review of Astron. and Astrophys.}}

\newcommand{\aj}{     {\it Astron. J.}} 
\newcommand{\apj}{    {\it Astrophys. J.}}
\newcommand{\apjl}{   {\it Astrophys. J. Lett.}}
\newcommand{\apjs}{    {\it Astrophys. J. Suppl.}}
\newcommand{\apss}{   {\it Astrophys. Space Sci.}}

\newcommand{\jgr}{    {\it J. Geophys. Res.}}
\newcommand{\mnras}{  {\it Mon. Not. Roy. Astron. Soc.}}
\newcommand{\nat}{    {\it Nature}}
\newcommand{\pasp}{   {\it Pub. Astron. Soc. Pac.}}
\newcommand{\pasj}{   {\it Pub. Astron. Soc. Japan}}

\newcommand{\solphys}{{\it Solar Phys.}}
 
\newcommand{\ssr}{    {\it Space Sci. Rev.}}

\newcommand{\alb}{{\em IRIS$^2$}}
\newcommand{\alma}{{\em ALMA}}
\newcommand{\rhessi}{{\em RHESSI}}
\newcommand{\nustar}{{\em NuSTAR}}

\newcommand{\hinode}{{\em Hinode}}
\newcommand{\hinodes}{{\em Hinode/SOT}}
\newcommand{\hinodesp}{{\em Hinode/SP}}
\newcommand{\hinodee}{{\em Hinode/EIS}}
\newcommand{\hinodex}{{\em Hinode/XRT}}
\newcommand{\sot}{{\em SOT}}
\newcommand{\eis}{{\em EIS}}
\newcommand{\xrt}{{\em XRT}}
\newcommand{\dst}{{\em DST}}
\newcommand{\iris}{{\em IRIS}}
\newcommand{\sdo}{{\em SDO}}
\newcommand{\aia}{{\em AIA}}
\newcommand{\hmi}{{\em HMI}}
\newcommand{\stereo}{{\em STEREO}}
\newcommand{\saia}{{\em SDO/AIA}}
\newcommand{\shmi}{{\em SDO/HMI}}
\newcommand{\sst}{{\em SST}}
\newcommand{\bbso}{{\em BBSO}}
\newcommand{\gregor}{{\em GREGOR}}
\newcommand{\dkist}{{\em DKIST}}
\newcommand{\psp}{{\em PSP}}
\newcommand{\solo}{{\em SOLO}}
\newcommand{\soloe}{{\em SOLO/EUI}}
\newcommand{\solos}{{\em SOLO/SPICE}}
\newcommand{\solop}{{\em SOLO/PHI}}

\newcommand{\alfvenic}{{Alfv\'{e}nic}~}
\newcommand{\solstice}{{SOLSTICE}}

\newcommand{\hic}{{\em Hi-C}}
\DeclareRobustCommand{\ion}[2]{%
\relax\ifmmode
\ifx\testbx\f@series
{\mathbf{#1\,\mathsc{#2}}}\else
{\mathrm{#1\,\mathsc{#2}}}\fi
\else\textup{#1\,{\mdseries\textsc{#2}}}%
\fi}

\definecolor{dblue}{rgb}{.098 , .40, 0.69 }

\definecolor{org}{rgb}{.910 , .376, 0.11 }

\begin{document}

\begin{article}

\begin{opening}

\title{A new view of the solar interface region from the {\it Interface Region Imaging Spectrograph} (\iris)}

\author{B.~\surname{De Pontieu}$^{1,2,3}$\sep
V.~\surname{Polito}$^{4,1}$\sep
V.~\surname{Hansteen}$^{4,1,2,3}$\sep
P.~\surname{Testa}$^{5}$\sep
K.K.~\surname{Reeves}$^{5}$\sep
P.~\surname{Antolin}$^6$\sep
D.~\surname{N\'obrega-Siverio}$^{2,3,13,14}$\sep
A.~\surname{Kowalski}$^{7,8,9}$\sep
J.~\surname{Martinez-Sykora}$^{4,1,2,3}$\sep
M.~\surname{Carlsson}$^{2,3}$\sep
S.W.~\surname{McIntosh}$^{10}$\sep
W.~\surname{Liu}$^{4,1}$\sep
A.~\surname{Daw}$^{11}$\sep
C.C.~\surname{Kankelborg}$^{12}$\sep}
\runningauthor{B. De Pontieu {\it et al.}}
\runningtitle{{\it New views from Interface Region Imaging Spectrograph}}

   \institute{$^{1}$ Lockheed Martin Solar \& Astrophysics Laboratory,
      Org. A021S, Bldg. 252, 3251 Hanover St., Palo Alto, CA 94304, USA 
                     www:
\href{http://iris.lmsal.com/}{\textsf{iris.lmsal.com/}}
email: \href{mailto:bdp@lmsal.com}{\textsf{bdp@lmsal.com}}\\ 
 $^{2}$ Institute of Theoretical Astrophysics, University
              of Oslo, P.O. Box 1029 Blindern, 0315 Oslo, Norway
                     \\
                     $^{3}$ Rosseland Centre for Solar Physics, University of Oslo, P.O. Box 1029 Blindern, 0315 Oslo, Norway
                     \\
              $^{4}$ Bay Area Environmental Research Institute,
NASA Research Park, Moffett Field, CA 94035, USA
                   \\
            $^{5}$ Harvard-Smithsonian Center for Astrophysics, 60 Garden
              Street, Cambridge, MA 02138, USA
                      \\
            $^{6}$ Department of Mathematics, Physics and Electrical Engineering, Northumbria University, Newcastle Upon Tyne, NE1 8ST, UK
            \\
          $^{7}$ National Solar Observatory, University of Colorado Boulder, 3665 Discovery Drive, Boulder, CO 80303, USA
          \\
          $^{8}$ Department of Astrophysical and Planetary Sciences, University of Colorado, Boulder, 2000 Colorado Ave, CO 80305, USA
          \\
          $^{9}$ Laboratory for Atmospheric and Space Physics, University of Colorado Boulder, 3665 Discovery Drive, Boulder, CO 80303, USA
          \\
                        $^{10}$ High Altitude Observatory, National Center for
             Atmospheric Research, P.O. Box 3000, Boulder, CO 80307, USA
\\
$^{11}$NASA Goddard Space Flight Center, Greenbelt, MD 20771, USA
\\
              $^{12}$ Department of Physics, Montana State University,
              Bozeman, P.O. Box 173840, Bozeman, MT 59717, USA
                     \\
                     $^{13}$ Instituto de Astrof\'isica de Canarias, E-38200 La Laguna, Tenerife, Spain\\
$^{14}$ Departamento de Astrof\'isica, Universidad de La Laguna, E-38206 La Laguna, Tenerife, Spain\\
     }

\begin{abstract}
The Interface Region Imaging Spectrograph (\iris) has been obtaining near- and far-ultraviolet images and
spectra of the solar atmosphere since July 2013. \iris\ is the highest
resolution observatory to provide seamless coverage of spectra and
images from the photosphere into the low corona. The unique combination of
near and far-ultraviolet spectra and images at subarcsecond resolution
and
high cadence allows the tracing of mass and energy through the 
critical interface between the solar surface and the corona or solar wind.  \iris\ has enabled research into the fundamental physical processes thought to play a role in the low solar atmosphere such as ion-neutral interactions, magnetic reconnection (e.g., resulting from braiding or driving flares and smaller scale events), the generation, propagation, and dissipation of various types of waves, the acceleration of non-thermal particles, and various small-scale instabilities. These new findings have helped provide novel insights into a wide range of phenomena including the discovery of non-thermal particles in coronal nanoflares, the formation and impact of spicules and other jets, resonant absorption and dissipation of Alfv\'enic waves, energy release and jet-like dynamics associated with braiding of magnetic field lines, the importance of thermal instability in the chromosphere-corona mass and energy cycle, the role of turbulence and the tearing mode instability in magnetic reconnection, the contribution
 of waves, turbulence, and non-thermal particles in the energy deposition during flares and smaller-scale events such as Ellerman bombs or UV bursts, and the role of flux ropes and various other mechanisms in triggering and driving CMEs. \iris\ observations have also been used to elucidate the physical mechanisms driving the solar irradiance that impacts Earth's upper atmosphere, and have advanced studies of the connections between solar and stellar physics. Advances in numerical modeling, inversion codes, and machine learning techniques have played a key role in driving these new insights in how the Sun's atmosphere is energized.
With the advent of exciting new instrumentation both on the ground (e.g., \dkist, \alma) and space-based (e.g., Parker Solar Probe, Solar Orbiter), we aim to review new insights based on \iris\ observations or \iris\ related modeling, and highlight some of the outstanding challenges that have been brought to the fore.
\end{abstract}

\end{opening}

\section{Introduction}
     \label{S-Introduction} 

The Interface Region Imaging Spectrograph (\iris) is a NASA small
explorer consisting of a 20 cm telescope that feeds far (FUV) and
near ultraviolet (NUV) light into a high-resolution spectrograph. \iris\  obtains spectra and images at unprecedented cadence (down to 1.3 s) and
resolution both spatially (0.33\arcsec\ in FUV,
0.4\arcsec\ in NUV) and spectrally (2.7 km/s pixels), allowing flexible
rastering of fields of view that cover a large fraction of an active region \citep{DePontieu2014}. \iris\ was launched in June 2013 and after a
two-year prime mission, is now in the extended mission phase. \iris\
observations continue to provide high-resolution images and spectra of
the photosphere, chromosphere, transition region and corona.

In the almost eight years since launch, \iris\ has opened a new window in the
complex physics of the solar interface region, a region sandwiched
between the solar photosphere and the
corona. It is through this region that all of the non-thermal energy
that powers the outer solar atmosphere and solar wind is processed and
propagated. The interface region includes the chromosphere, transition region and
low corona and is characterized by several physical transitions that render
analysis, interpretation and understanding challenging. The plasma
undergoes a transition from 
partially ionized in the chromosphere and low transition region to fully ionized in the corona. Starting from magnetic field strengths similar to those observed, numerical models conclude that, apart from sunspots, the plasma $\beta$ is largely $<<1$ at roughly $1000$~km above the photosphere and $>1$ below this height \citep[ e.g.][]{Hansteen2007,Hansteen2015}. Thus, outside of sunspots, the plasma dominates the
dynamics of the magnetic field in the photosphere; forcing the
field into flux concentrations 
in the downflows of the convective flow
pattern, while the magnetic field becomes more dominant with increasing height and is volume
filling in the corona. 
This transition from high to low
plasma $\beta$ leads to a variety of complex physical processes,
including wave-mode coupling. When trying to diagnose conditions in
the low solar atmosphere, we are limited to remote sensing of the
radiation emanating from this region. However, most
chromospheric diagnostics are optically thick and require non-LTE
radiative transfer modeling to be properly understood. This can complicate
the interpretation of some of the \iris\ diagnostics significantly.
All of these complex issues imply that, in order to obtain a deeper understanding of the
dominant physical processes that drive the dynamics and energetics in
the interface region, numerical modeling needs to go
hand-in-hand with observations. The \iris\ science investigation has included a
numerical modeling component from the beginning, and we will thus
provide an overview of the most exciting developments in observations as well as modeling of the interface region in what follows below.

The \iris\ bandpasses have previously been studied at lower resolution using rockets
\citep{Bates69,Fredga69,Kohl76,Allen78,Morrill08,West11,Dere84,Brekke1993}, balloons
\citep{Lemaire69,Lemaire73,Samain85,Staath95}, or satellites
\citep{Doschek77,Bonnet78,Woodgate80,Roussel82,Billings77,Cohen1981,Poland83,Kingston82,Hayes1987,Curdt2001}. 
However, the advances of \iris\ in spatial, spectral, and temporal
resolution compared to previous missions are large enough that a
comparison with results from such missions is less useful. Instead, we
will focus  on the physical insights that have resulted from
\iris\ observations, and frame those within the context of what was
previously known about the physical mechanisms in this region of the
solar atmosphere. Many of the \iris\ results were obtained through coordinated observations with a variety of other instruments, including the Atmospheric Imaging Assembly \citep[\aia, ][]{Lemen2012} and the Helioseismic Magnetic Imager \citep[\hmi,][]{Scherrer2012} onboard the Solar Dynamics Observatory \citep[\sdo,][]{Pesnell2012}; the Solar Optical Telescope \citep[\sot,][]{Tsuneta2008}, the Extreme-ultraviolet Imaging Spectrometer \citep[\eis,][]{Culhane2007}, and the X-Ray Telescope \citep[\xrt,][]{Golub2007} onboard the \hinode\ spacecraft \footnote{Coordination between \iris\ and \hinode\ is organized through the \iris--\hinode\ Operations Plans or IHOPs. A list can be found here: \url{http://www.isas.jaxa.jp/home/solar/hinode_op/hop_list.php}} \citep{Kosugi2007}; the \stereo/SECCHI instrument \citep[\stereo,][]{Howard2008}; the Reuven Ramaty High Energy Solar Spectroscopic Imager mission \citep[\rhessi,][]{Lin2002}; the Nuclear Spectroscopic Telescope Array mission \citep[\nustar,][]{Harrison2013}; the High Resolution Coronal Imager sounding rocket \citep[\hic,][]{Kobayashi2014}; as well as the ground-based observatories the Swedish 1m Solar Telescope \citep[\sst,][]{Scharmer2003}, the Goode Solar Telescope at Big Bear Solar Observatory \citep[\bbso,][]{Goode2003}, the GREGOR telescope \citep[\gregor,][]{Schmidt2012}, and the Atacama Large Millimeter/submillimeter Array (\alma) radio telescope. In addition, we will describe outstanding challenges, also in light of the advent of novel instrumentation such as NSF's 4m Daniel K. Inouye Solar Telescope \citep[\dkist,][]{Rimmele2020}, the Solar Orbiter \citep[\solo,][]{SOLO2020} and Parker Solar Probe \citep[\psp,][]{Fox2016} missions, as well as advances in numerical modeling.

The data analysis and interpretation of the \iris\ observations, as with any spectroscopic data, depend on a careful and accurate calibration. A detailed description of the \iris\ calibration is provided by \citet{Wuelser2018}.

The paper is divided into several sections that are aligned with the prioritized science goals of the \iris\ extended mission. 
We first discuss the progress made in diagnosing physical conditions from the \iris\ observables (\S~\ref{diagnostics}) and briefly describe the numerical models that have been made publicly available as part of the \iris\ mission (\S~\ref{models}). Section \S~\ref{physical} focuses on the advances made in understanding the fundamental physical processes that
play a role in the interface region. We further discuss the large-scale instability of the solar atmosphere (\S~\ref{instability}) through flares and coronal mass ejections (CMEs). We also cover contributions from \iris\ to studies of solar irradiance (\S~\ref{irradiance}) and the solar-stellar connection (\S~\ref{stellar}). We finish with a brief summary  (\S~\ref{conclusions}). 

\section{Diagnostics}
\label{diagnostics}
\subsection{Photospheric and Chromospheric Lines}
\label{diagnostics_thick}
The chromosphere is a highly complex, finely structured, and dynamic region in which non-LTE radiative transfer and time dependent ionization play a major role. Determining the physical parameters in this region has traditionally been a major challenge. Two different approaches have been used to significantly increase the diagnostics power of spectral lines such as \ion{Mg}{ii} h (2803\AA), k (2796\AA), and UV triplet (2798.9\AA, 2791.6\AA), or \ion{C}{ii} 1334 and 1335\AA. We outline these two different approaches below.

\begin{figure}[tp]
   \includegraphics[width=0.32\textwidth]{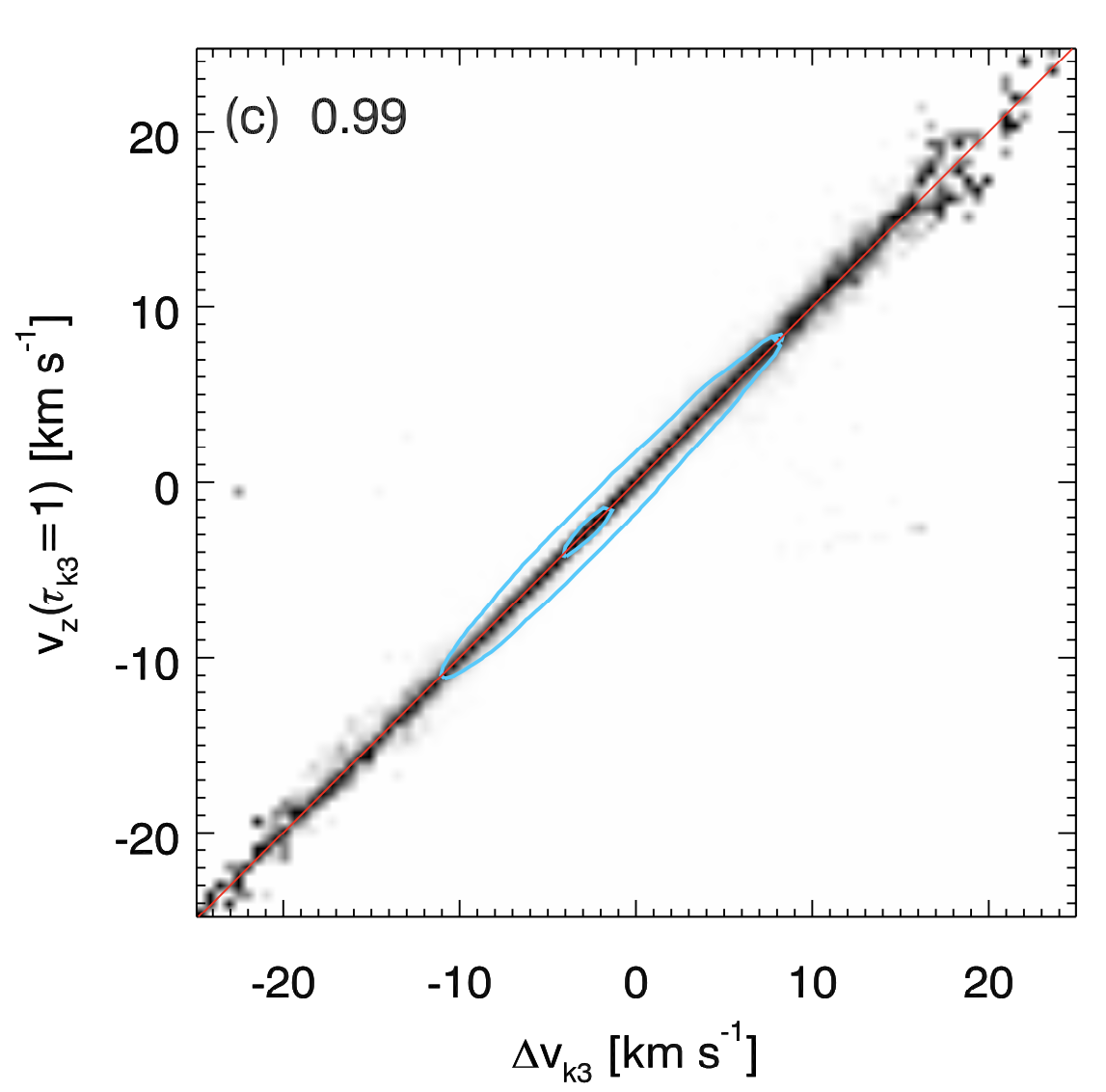}
   \includegraphics[width=0.32\textwidth,height=0.25\textwidth]{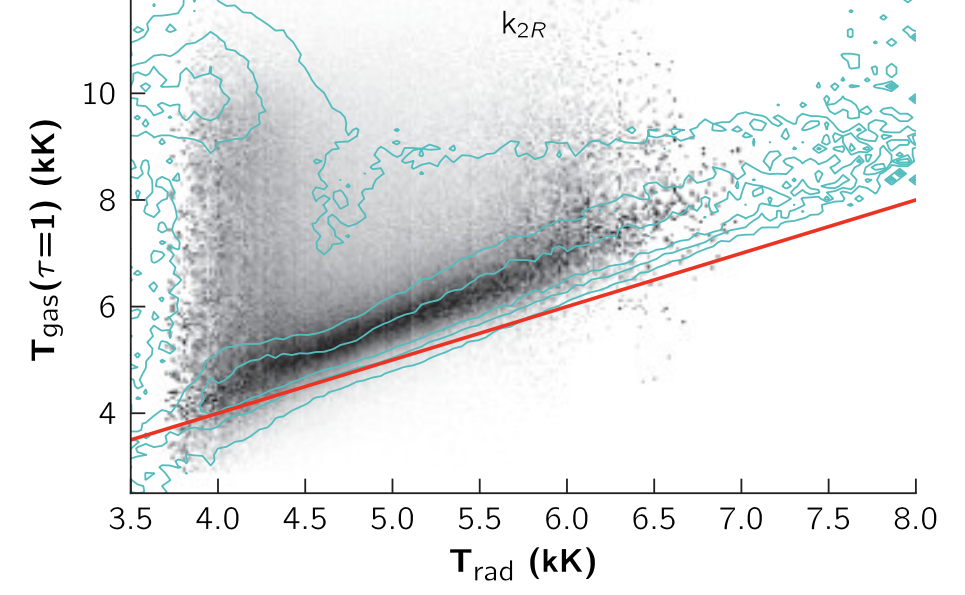}
   \includegraphics[width=0.31\textwidth]{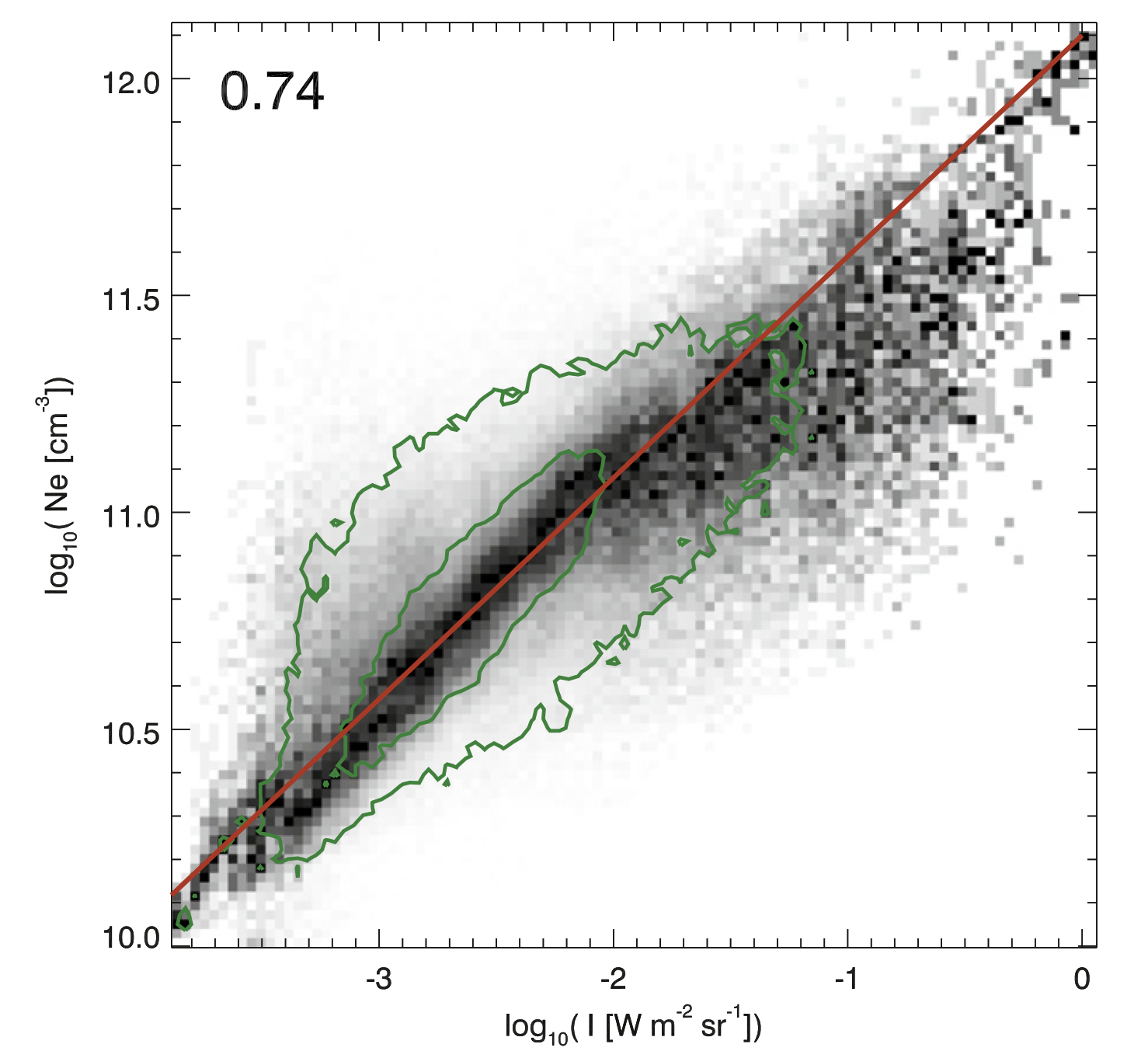}
   \caption{\small
     Analysis of publicly available advanced simulations and synthetic observables has provided the community with tools to derive physical information from the optically thick spectra, such as the upper chromospheric velocity which is well correlated with the Mg II k3 (central reversal) Doppler shift \citep[left panel, ][]{Leenaarts2013b}, temperatures in the middle chromosphere from the radiation temperature or brightness of the \ion{Mg}{ii} k2 peaks \citep[middle panel, ][]{Pereira2013}, and the electron density in the chromosphere from the intensity of the optically thin \ion{O}{i} 1355.6\AA\ line \citep[right panel, ][]{Lin2015}. }
       \label{fig_diag_phch}
\end{figure}

In the first approach, forward modeling of synthetic observables from advanced numerical simulations \citep[using the Bifrost code,][]{Gudiksen2011} has been used to study correlations between physical variables in the chromosphere and properties of the spectral lines (e.g., Doppler shifts, intensity, central reversal). These correlations are described in a series of nine papers and have provided us with new diagnostics that (approximately) map the extensive spectroscopic \iris\ observables to physical properties of the low solar atmosphere \citep[see also][]{Carlsson2019}. Here we briefly summarize the results of these studies. The many photospheric lines in the wings of the \ion{Mg}{ii} h and k lines in the \iris\ NUV passband provide velocity diagnostics over a wide range of heights above the photosphere, through measurements of Doppler shifts of absorption lines \citep{Pereira2013}. Analysis of the line formation of the \ion{Mg}{ii} h and k lines has shown that these lines have great 
diagnostic value for the chromosphere. They form over a wide range of
heights and the spectral line parameters (e.g., Doppler shift and intensity of k2v, k2r, k3, see Fig.~\ref{fig_diag_phch}) can be used to estimate physical parameters such as the middle and upper chromospheric velocity, chromospheric velocity gradients, and temperatures in the middle chromosphere \citep{Leenaarts2013a,Leenaarts2013b, Pereira2013}.
Software is available to determine these spectral parameters in the \iris\ tree of SolarSoft (see \iris\ data analysis guides for IDL and Python at \url{https://iris.lmsal.com/analysis.html}). Heating in the low 
chromosphere can be identified from emission in the 
\ion{Mg}{ii} triplet lines \citep{Pereira2015}, while upper chromospheric velocities can be estimated from the \ion{C}{ii} lines \citep{Rathore2015a,Rathore2015b}. Velocities in the middle chromosphere can be estimated from Doppler shifts of the \ion{C}{i} 1355.8 \AA\ \citep{Lin2017}. One of the most interesting diagnostics is the \ion{O}{i} 1355.6 \AA\ line, which is formed over a wide range of chromospheric heights under optically thin conditions \citep{Lin2015}. This means that it is uniquely sensitive to non-thermal motions in the chromosphere (through its broadening). In addition, the ratio of \ion{C}{i} and \ion{O}{i} intensities is inversely correlated with the electron density in the middle chromosphere. Velocity differences between the Doppler shifts of these lines can also be used to estimate the velocity gradient in the middle chromosphere \citep{Lin2017}.

\begin{figure}[tp]
  \includegraphics[width=0.99\textwidth]{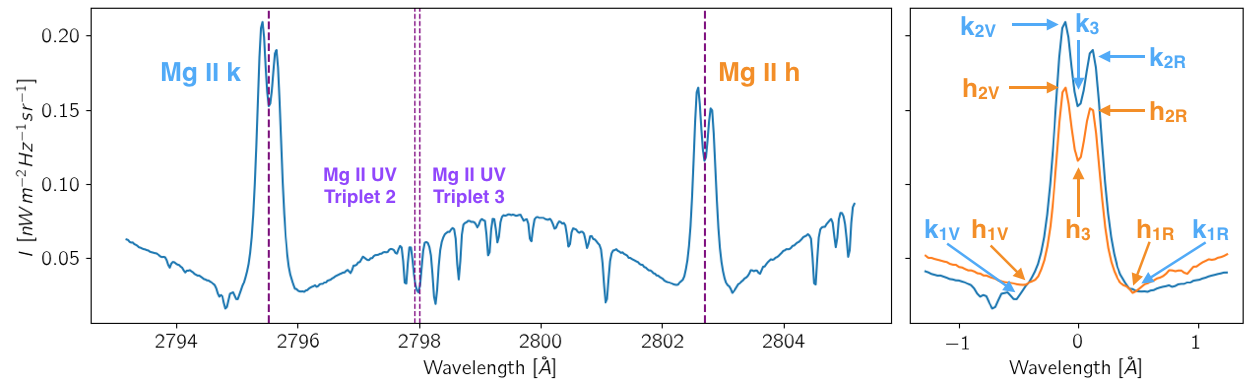}
  \includegraphics[width=0.99\textwidth]{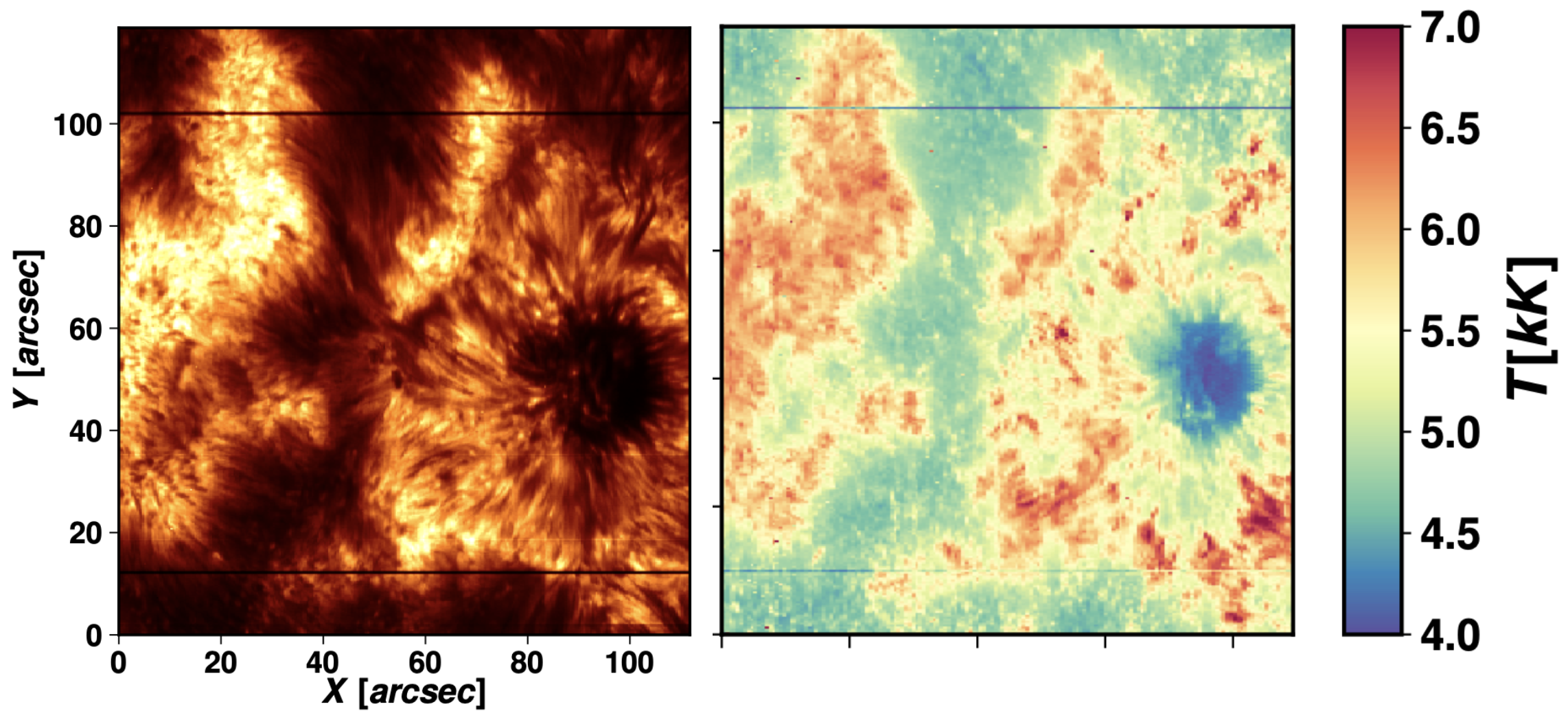}
     \caption{\small The \ion{Mg}{ii} h and k spectral range contains a wealth of spectroscopic features (top panel) that encode information about the physical conditions in the chromosphere, where this spectral range originates. Decoding this information requires sophisticated and complex inversion codes that include non-LTE radiative transfer, a process that can be computationally cumbersome. The machine learning-based \alb\ database allows rapid
       non-LTE-based inversions of \iris\ \ion{Mg}{ii} h/k spectra
       (\ion{Mg}{ii} k3 spectroheliogram, bottom left) into physical parameters (e.g., temperature, bottom right, as well as velocity, electron density, microturbulence) as a function of optical depth (which is related to height) $10^5-10^6$ times faster than classical approaches. Bottom panel adapted from \citet{Sainz-Dalda2019}.}
       
      \label{fig_alb}
\end{figure}

These results have been exploited to address a variety of phenomena, such as spicules \citep{Bose2019b}, flare evolution \citep{Hannah2019, Huang2019a}, reconnection-driven jets \citep{Cai2019}, and heating resulting from interactions between emerging and pre-existing magnetic fields \citep{Guglielmino2019a}.

An alternative approach to forward modeling is to exploit non-LTE
inversion codes such as the Stockholm Inversion Code \citep[STiC,][]{de-la-Cruz-Rodriguez2016} which allows quantitative determination of chromospheric conditions based on the extensive  spectroscopic diagnostics of \iris\ from the photosphere to the transition region, including the sensitive \ion{Mg}{ii} lines \citep[e.g., to estimate heating from cancelling granular-scale internetwork fields,][]{Gosic2018}. STiC inversions of \iris\ \ion{Mg}{ii} h and k spectra combined with \alma\ mm radiation are particularly promising in terms of constraining the chromospheric temperature and turbulent motions over a wide range of heights \citep{da-Silva-Santos2018,da-Silva-Santos2020}. Another promising avenue is to perform inversions of \ion{Mg}{ii} and \ion{C}{ii} lines simultaneously, given the different sensitivity to local conditions of both lines \citep{Leenaarts2013b,Rathore2015a,Rathore2015b}. 

These types of inversions are not readily
accessible to the broader community and are 
computationally cumbersome with inversion of a large \iris\ raster map
requiring many CPU-years. During the past few years, the \iris\ team has
exploited machine learning techniques and the STiC inversion code
to build a database (Fig.~\ref{fig_alb}) of representative spectral profiles and associated model atmospheres (i.e., temperature, density, velocity, and turbulent motions as a function of height). This novel approach drastically decreases the required computational effort (by a factor of 10$^5$ to 10$^6$) for chromospheric diagnostics while retaining similar accuracy as those provided by full STiC inversions: any \iris\ raster can be ``inverted'' to physical parameters in minutes using a laptop (rather than weeks on a large server). This \alb\ database
is publicly available \citep[\url{https://iris.lmsal.com/iris2}, ][]{Sainz-Dalda2019} and provides the
community with over 50,000 different model atmospheres that
capture the spatio-temporal complexity and diversity of the
solar atmosphere across a wide range of phenomena. It transforms the \iris\
data archive into a previously unavailable diagnostic goldmine.
The \alb\ inversions are of course subject to limitations related to
the underlying assumptions \citep[see,
e.g.,][]{de-la-Cruz-Rodriguez2016} and techniques. They do not
necessarily provide unique solutions, and the sensitivity of the
\ion{Mg}{ii} h and k lines to local thermodynamic conditions varies,
depending on the height in the atmosphere. This can lead to
significant uncertainties on the derived thermodynamic parameters at
some heights. A very conservative estimate for those uncertainties is
currently provided by the \alb\ database, while a more realistic
approach to estimating uncertainties, based on Monte Carlo
simulations, will be outlined in future work. Uncertainties can also
be introduced by a lack of diversity in the spectral profiles in the
database. All of these issues will be addressed in future developments of this database, which will expand on the number of profiles, include more photospheric lines as well as the \ion{C}{ii} 1334/1335 \AA\ lines, as well as include neural networks to improve the computational efficiency even more. 
  More information about using optically thick lines to diagnose physical conditions in the low solar atmosphere can be found on the \iris\ website: \url{https://iris.lmsal.com/itn39/}.

\subsection{Transition region and high temperature lines}
\label{diagnostics_thin}
In addition to providing crucial diagnostics in the photosphere and chromosphere, \iris\ observes several spectral lines which are formed over a higher temperature range,
from transition region (TR) to coronal and all the way up to flaring temperatures. Some of the strongest transitions in this range are the resonance lines of \ion{Si}{iv} at 1393.75\AA~and 1402.77\AA, formed at around T~$\approx$~10$^{4.9}$K (under ionization equilibrium conditions), which offer excellent diagnostics of plasma dynamics for a large variety of physical mechanisms, including spicules, jets, prominences, flares and UV bursts, as described in the following sections.  While it is often assumed that the \ion{Si}{iv} lines are formed under optically thin conditions, this might not always be the case, in particular during highly energetic events such as flares \citep{Kerr2019a}. In fact, the \ion{Si}{iv} 1393.75/1402.77$\lambda$~intensity ratio itself can be used as a diagnostic for opacity effects. In particular, a ratio 
different than 2 may indicate that the lines are optically thick \citep{Peter2014}.
Even when optically thin and in equilibrium, it is important to consider the effect of charge exchange, as well as photoionization, on the temperature of formation for \ion{Si}{iv}, as electron capture by ions from the dominant neutral species affects ionization balance in the TR \citep{Kerr2019a}. The \iris\ FUV bandpass also includes TR lines of the density-sensitive \ion{O}{iv} and \ion{S}{iv} multiplets, the coronal line \ion{Fe}{xii} 1349.4\AA, and the flare line \ion{Fe}{xxi} 1354.08\AA, as discussed below. 

Close in wavelength to the \ion{Si}{iv} 1402.77\AA\ line are the semi-forbidden transitions of \ion{S}{iv} (1404.81\AA~and 1406.02\AA) and \ion{O}{iv} (1399.78\AA, 1401.16\AA~and 1404.81\AA), which are formed at T~$\approx$~10$^{5}$K and $\approx$~10$^{5.15}$K respectively. The intensity ratios of these \ion{O}{iv} and \ion{S}{iv} lines provide excellent diagnostics of electron number densities in the $\approx$~10$^{9}$--10$^{12}$ cm$^{-3}$ and $\approx$~10$^{10}$--10$^{13}$ cm$^{-3}$ range respectively, which can be particularly useful during flares and other energetic events, when the plasma reaches high densities \citep[e.g.,][]{Polito2016b,Bradshaw2019}. 
One complication arises from the fact that the \ion{O}{iv} 1404.81\AA~and \ion{S}{iv} 1404.85\AA~lines are very close in wavelength and form a blend around 1404.82\AA, in which the two transitions are virtually indistinguishable. However, \citet[e.g.][]{Polito2016b} showed that the lines can be distinguished using the density diagnostic provided by the \ion{O}{iv} 1399.78\AA~and 1401.16\AA~ratio, and taking into account that the relative contribution of \ion{S}{iv} and \ion{O}{iv} to the blend varies with the plasma density and temperature. 

The \ion{Si}{iv} to \ion{O}{iv} ratio has also sometimes been used to provide density estimates \citep[e.g.,][]{Peter2014,Young2018}. However, the validity of this ratio to estimate densities has been highly debated based on a number of issues, including the fact that the ratio depends on the plasma temperature and density, and on the chemical abundances of O and Si, which are not known with great accuracy \citep[e.g.,][]{Judge2015}. In addition, these ions show a very different response to transient ionization because of their different formation processes, and their ratio can actually be used as a direct diagnostic of whether the observed plasma is in a non-equilibrium ionization state \citep[NEI;][e.g.]{Doyle2013,Bahauddin2020}. 
Nevertheless, \citet{Doschek2016} and \citet{Young2018} have recently discussed how the ratio can sometimes be used to provide some estimates for the electron density, keeping in mind the limitations described above.
\begin{figure}[tp]
   \includegraphics[width=\textwidth]{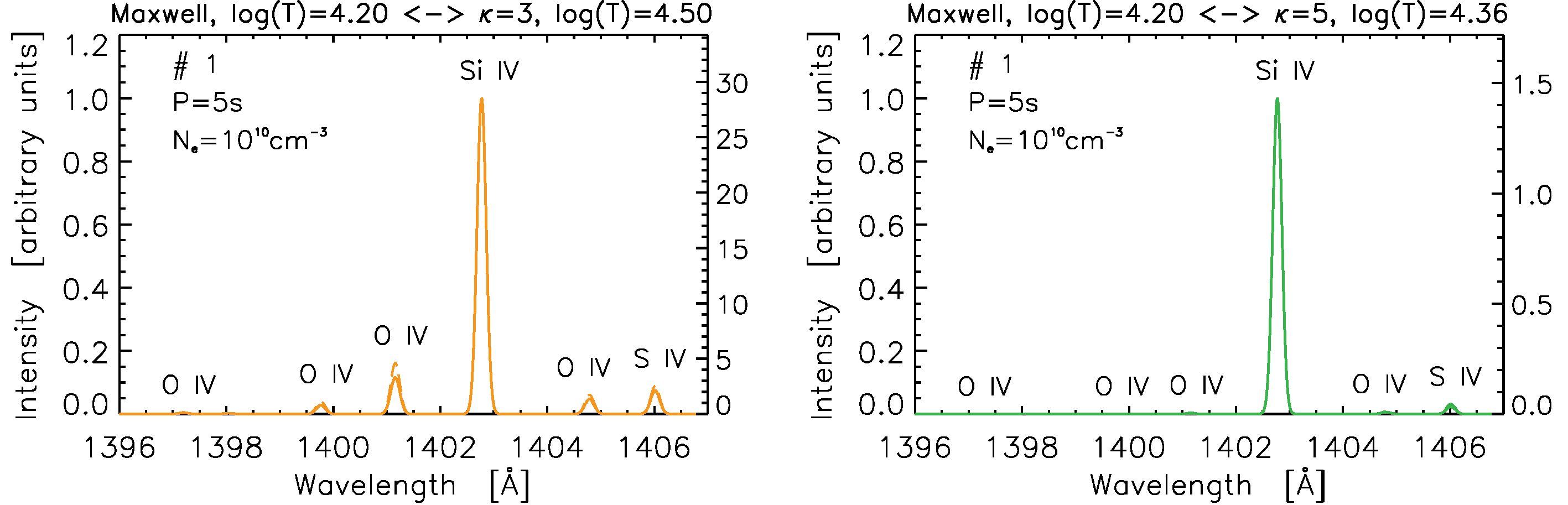}
   \caption{Synthetic \iris\ TR spectra assuming both non-equilibrium ionization and non-Maxwellian distributions using a periodic electron beam model, represented by a $\kappa$-distribution that recurs at periods of several seconds, approximating the effect of bursty energy releases from accelerated particles. In contrast to the equilibrium spectra from CHIANTI, such synthetic spectra are similar to those typically observed by \iris. Adapted from \citet{Dzifcakova2018}.}
   \label{fig:non_mawx}
   \end{figure}
   
\iris\ observations of these TR lines also have significant
potential diagnostic value for a variety of wide-ranging problems from
the oxygen abundance problem \citep[e.g.,][]{Asplund2009} to non-Maxwellian distributions. For
example, the \ion{Si}{iv} to \ion{O}{iv} brightness ratio observed with \iris\
deviates significantly from the theoretically predicted value for
ionization equilibrium 
 under coronal conditions. \citet{Olluri2015} and \citet{Martinez-Sykora2016a} compared  MHD simulations of a self-consistently heated low solar atmosphere with \iris\ observations and found that NEI effects are important for these lines, shifting the formation temperature of the ions to lower temperatures, and may remove some of the discrepancy between \ion{Si}{iv} and \ion{O}{iv} intensities. 
 \citet{Dudik2014} also found that
non-Maxwellian $\kappa$ distributions for the electron energy in the transition region can in principle bring the \ion{Si}{iv} and
\ion{O}{iv} intensities in line with the \iris\ observations. In a more recent study, \citet{Dzifcakova2018} (Fig.~\ref{fig:non_mawx}) investigated the effects of including both NEI and $\kappa$ distributions self-consistently in an impulsive heating model. Their calculations show that the combination of NEI with non-Maxwellian distributions can reproduce the \iris\ observations with a lower number of high-energetic particles than that needed in the time-independent scenario investigated by \citet{Dudik2014}. Evidence supporting the presence of such high-energy non-Maxwellian particles has been reported in other studies, as discussed in \S~\ref{nanoflares}.

Finally, \citet{Polito2016b} pointed out the importance of including the density dependency of dielectronic recombination when calculating the ionization state of the \iris\ TR lines. When such dependency is included in the calculations, the temperature of formation of the ions is shifted towards lower values for high densities as compared to the low density case. The influence of density-dependence effects and NEI on the formation of the \iris\ TR lines was also recently investigated by \citet{Bradshaw2019}, who could reproduce some of the observed AR spectra assuming heating by weak nanoflares in combination with these effects. 

The work summarized above highlights the need for advanced models, which take into account a variety of physical processes, and accurate atomic data. Further comparisons between these models and \iris\ observations will help determine the role of different non-equilibrium mechanisms in energetic events occurring in the TR. 

\iris\ also observes the weak forbidden \ion{Fe}{xii} 1349.4\AA~transition which is formed in the upper TR at around T~$\approx$~1.5MK. This line can be best observed using appropriate observation strategies (e.g., longer exposures, lossless compression) and in denser plasma (e.g., plage). When observed, the \ion{Fe}{xii} line can provide useful diagnostics of plasma dynamics in the TR \citep{Testa2016}. In addition, the ratio of the \iris\ \ion{Fe}{xii} 1349.4\AA~line to the \hinodee\ \ion{Fe}{xii}~lines provide in principle good diagnostics of temperature and non-Maxwellians, when the \ion{Fe}{xii} 1349.4\AA~line can be reliably observed and 
taking into account the radiometric calibration of both instruments \citep[see ][for the latest \iris\ calibration]{Wuelser2018}. 

Finally, the highest temperature line included in the \iris\ spectra is the \ion{Fe}{xxi}~1354.08\AA~line, which is formed at T~$\gtrsim$~10~MK. This line has mostly been used as a diagnostic of flows throughout the flaring region, from the reconnection site to the ribbons and flare loops, and has provided many new insights into the understanding of these energetic events, as highlighted in \S \ref{flares}, \S \ref{flare_reco},  and \S \ref{trigger}. It should be noted that measurements of \ion{Fe}{xxi} redshifts, especially when the redshifted component is very faint, can be complicated by the presence of a hot \ion{Mn}{xviii} line at 1355.02~\AA, which is formed at around 8MK. While this line is predicted to be more than~10$^{3}$ times weaker than the \ion{Fe}{xxi} line at the peak formation temperature of \ion{Fe}{xxi} ($\approx$~12~MK), the ratio between the \ion{Fe}{xxi} and \ion{Mn}{xviii} can become much smaller if the plasma temperature is closer to the peak formation temperature of \ion{Mn}{xviii} ($\approx$~8~MK).

More information and useful codes for calculating plasma diagnostics using these lines can be found on the \iris\ website: \url{https://iris.lmsal.com/itn38/}.

\section{Numerical models}
\label{models}

As described above, the interpretation of chromospheric observables can be highly complex because of the crucial role of non-LTE radiative transfer and non-equilibrium ionization. Numerical models are thus key for the interpretation of \iris\ observations. Conversely, observations have
been vital in improving the numerical models. Discrepancies between synthetic observables calculated from the models and the observations provide clues to what physical processes are missing in the models. 

Over the last eight years numerical simulations from the \iris\ project have been publicly released\footnote{\url{http://sdc.uio.no/search/simulations}} together with tools, guidelines, and documentation\footnote{\url{https://iris.lmsal.com/modeling.html}}. New models will be added to the existing ones as soon as they are created and validated. The new models will expand the parameter range of different regions and magnetic field scenarios as well as physical processes included in the simulations.  
The broader community has also made models publicly available. These can be of great assistance in interpreting observations under specific scenarios. One example is the large grid of 1D radiative HD flare models\footnote{\url{https://star.pst.qub.ac.uk/wiki/doku.php/public/solarmodels/start}}
produced with the RADYN code \citep[e.g.,][]{Carlsson+Stein1994,Allred2015} in the 
F-CHROMA\footnote{collaborative project funded under the EU programme FP7-SPACE-2013-1} project. 

As part of the \iris\ project, several self-consistent radiative MHD models have been publicly released. Most of these models have been created with the Bifrost code \citep{Gudiksen2011}. The Bifrost code aims to address the most relevant physical processes in the lower atmosphere, i.e., photosphere, chromosphere, TR, and lower corona. The Bifrost code can include: 1) radiative transfer with scattering in the photosphere and lower chromosphere \citep{Skartlien2000,Hayek:2010ac}; 2) radiative losses and gains in the upper chromospheric and TR through recipes derived from detailed non-LTE calculations \citep{Carlsson:2012uq}, 3) optically thin radiative losses in the corona, 4) thermal conduction along the magnetic field, 5) ion-neutral interaction effects using the generalized Ohm's law  \citep[GOL, ][]{Martinez-Sykora2017b,Nobrega-Siverio:2020tec}, 6) and ionization balance in non-equilibrium for hydrogen and helium \citep{Leenaarts:2007sf,Golding:2016wq}, 7) non-equilibrium ionization of minority species \citep{Olluri2013}. 

\citet{Cheung2019} provide access to a self-consistent 3D Radiative MHD simulation\footnote{Supported by NASA’s Heliophysics Grand Challenges Research grant on "Physics and Diagnostics of the Drivers of Solar Eruptions" (NNX14AI14G)} of a flaring active region using the MURaM code \citep{Rempel:2014sf}. Recently, this model has been added to the \iris\ publicly available models with the same FITS format as the models from the Bifrost code. This addition expands not only the number of targets but also adds variety to the numerical codes of the simulations. In contrast to the publicly available Bifrost models, this MURaM model does not include a detailed treatment of chromospheric radiative transfer. 

Table~\ref{tab:simlist} lists the publicly available simulations and, while all these simulations include radiative transfer and thermal conduction; GOL and non-equilibrium ionization are not always included. The fourth column describes if any of these two physical processes are included. The table summarizes the name, references, number of snapshots available, numerical domain size, grid size, resolution and properties of the simulation such as field configuration and type of target on the Sun that the model aims to represent. 

\begin{table*}
    \centering
	\begin{tabular}{|l|l|l|l|l|}
		\hline
	Name  &  Size (Mm) &  Res.  & Extra & General  \\
	Reference  &  \# points & (km) & physics & properties  \\
	\# snapshots  & ver. ext. (Mm) &  &  &  \\ \hline
	\hline
	en024048\_hion & ($24 \times 24 \times 17$) &  $ 48$ & H in  & $<|B_{ph}|>\sim 50$~G   \\ 
     MC2016 & $504\times504\times496$ &  &  NEI &  polarities 8 Mm apart \\
    157 & (-2.5,14) &  &  &  enhanced network. \\ \hline
 	ch024031\_by200bz005 & ($24 \times 24 \times 17$) &  $31$ & LTE  & $<|B_{ph}|>\sim 40$~G  \\ 
    MC2016 & $768\times768\times768$ &  &  &  coronal hole  \\ 
    619 &  (-2.5,14) &  &  &   \\ \hline
  	en096014\_gol & ($96 \times 43$) &  $14$ & LTE & $<|B_{ph}|>\sim 190$~G   \\ 
    JMS2018 & $6930\times1554$ &  & GOL &  polarities 40 Mm apart \\ 
    91 &  (-2.5,40) &  &  & plage \\ \hline
  	en096014\_nongol & ($96 \times 43$) &  $14$ & LTE & $<|B_{ph}|>\sim 190$~G   \\ 
    JMS2018 & $6930\times1554$ &  & &  polarities 40 Mm apart \\ 
    1 &  (-2.5,40) &  &  & plage  \\ \hline
  	qs006005\_dyc & ($6 \times 6 \times 10.5$) &  $5$ & LTE & $<|B_{ph}|>\sim 56$~G  \\ 
    JMS2019 & $1200\times1200\times1736$ &  & &  local dynamo  \\ 
    25 &  (-2.5,8) &  &  & internetwork/CH  \\ \hline
    qs024048\_by3363 & ($24 \times 24 \times 17$) &  $48$ & LTE & variable $<|B_{ph}|>$   \\ 
   VH2017 & $504\times 504\times 496$ & & &  Flux emergence \\ 
    101 &  (-2.5,14.5) &  &  & into weak field corona  \\  \hline
 	HGCR Flare model & ($98 \times 49 \times 49$) &  $191$ & LTE & $<|B_{ph}|>\sim 3$~kG  \\ 
 	CR2019 & $512 \times 256 \times 768$ & & gray & quadrupolar AR  \\ 
    11 &  (-7.5,42.5) &  &  & with flux emergence  \\ \hline
	\end{tabular}
    \caption{ Summary of publicly available numerical models.  MC2016, JMS2018, JMS2019, VH2017 and CR2019 reference \citet{Carlsson2016}, \citet{Martinez-Sykora2018},  \citet{Martinez-Sykora2019}, \citet{Hansteen2017} and \citet{Cheung2019}, respectively}
	\label{tab:simlist}
\end{table*}

\begin{figure}[tp]
   \includegraphics[width=0.32\textwidth]{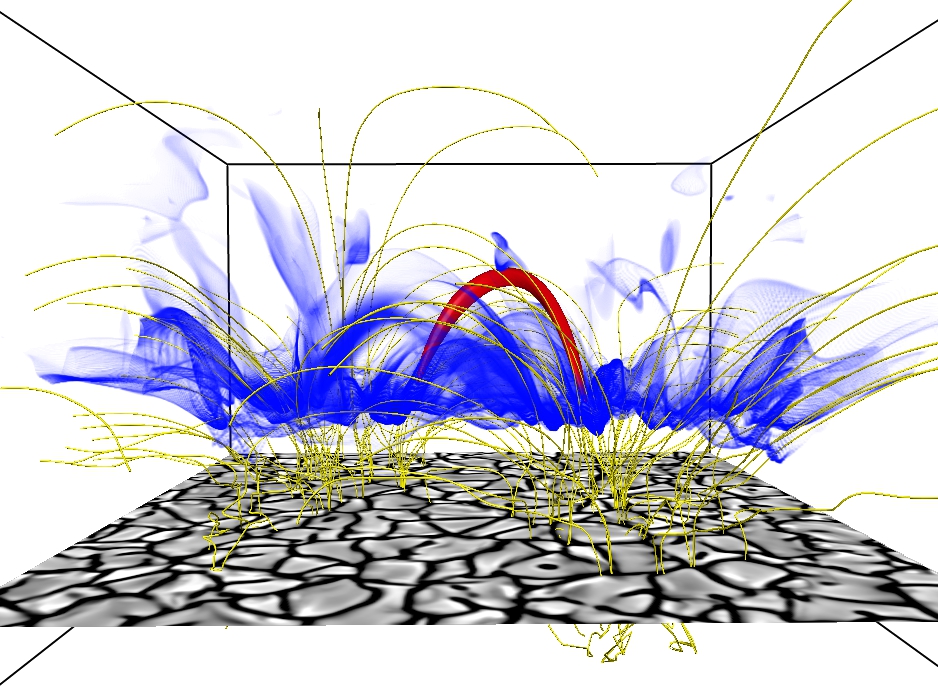}
   \includegraphics[width=0.32\textwidth]{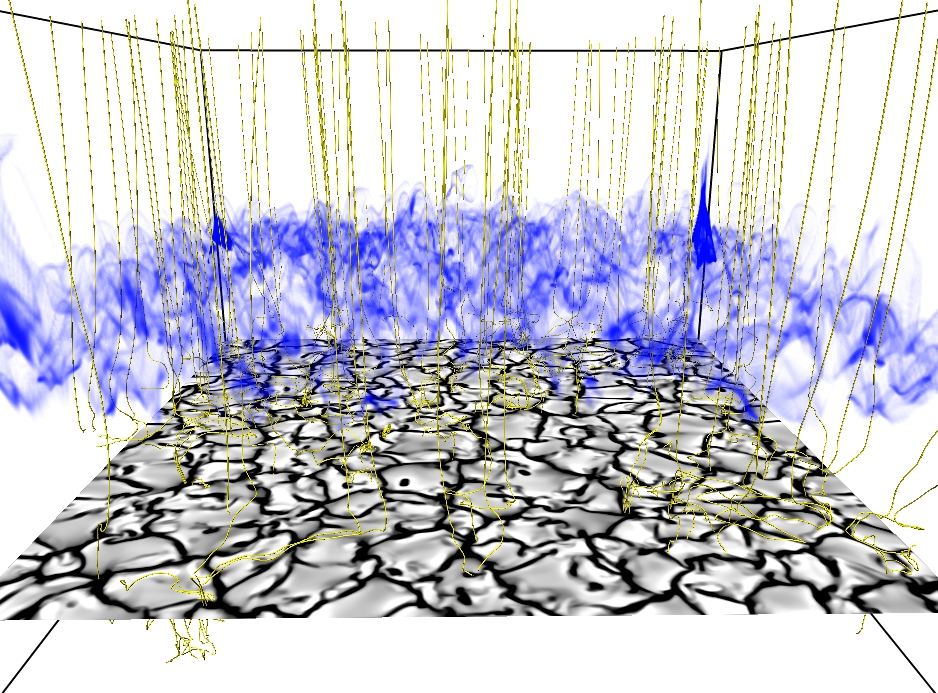}
   \includegraphics[width=0.31\textwidth]{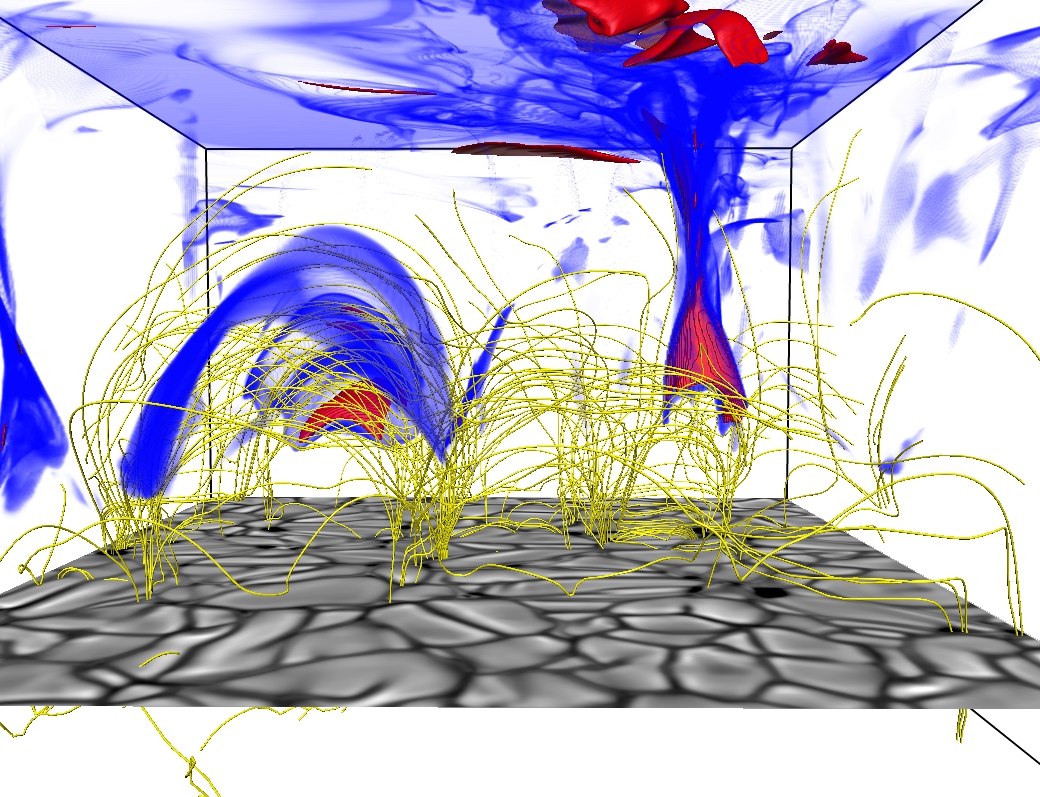}
   \includegraphics[width=0.16\textwidth]{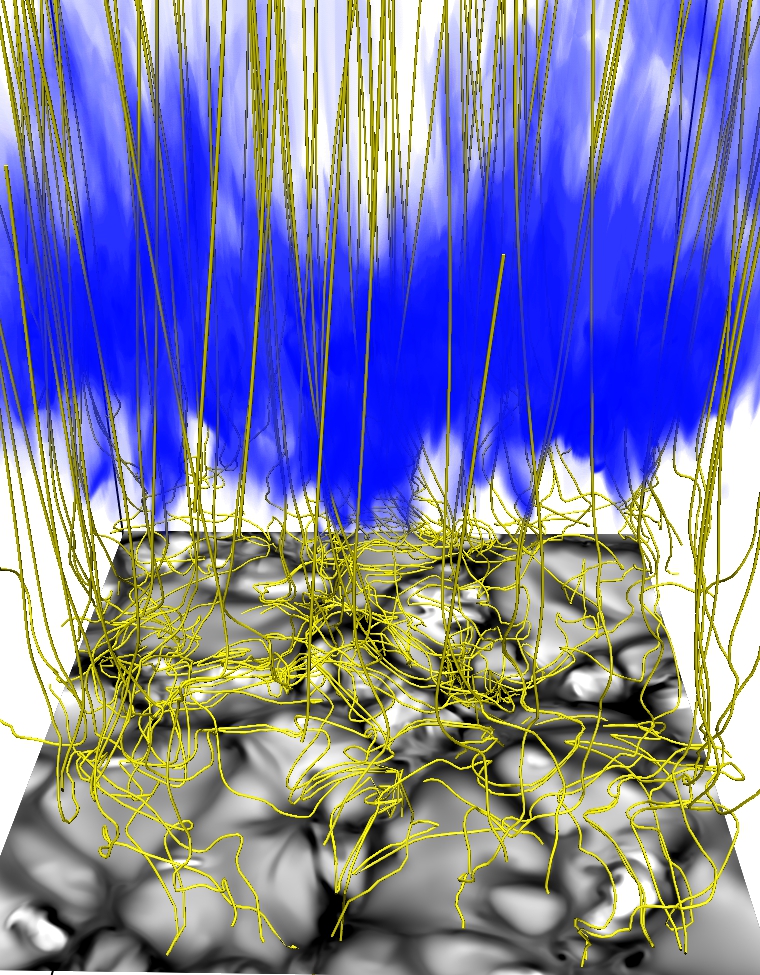}
   \includegraphics[width=0.50\textwidth]{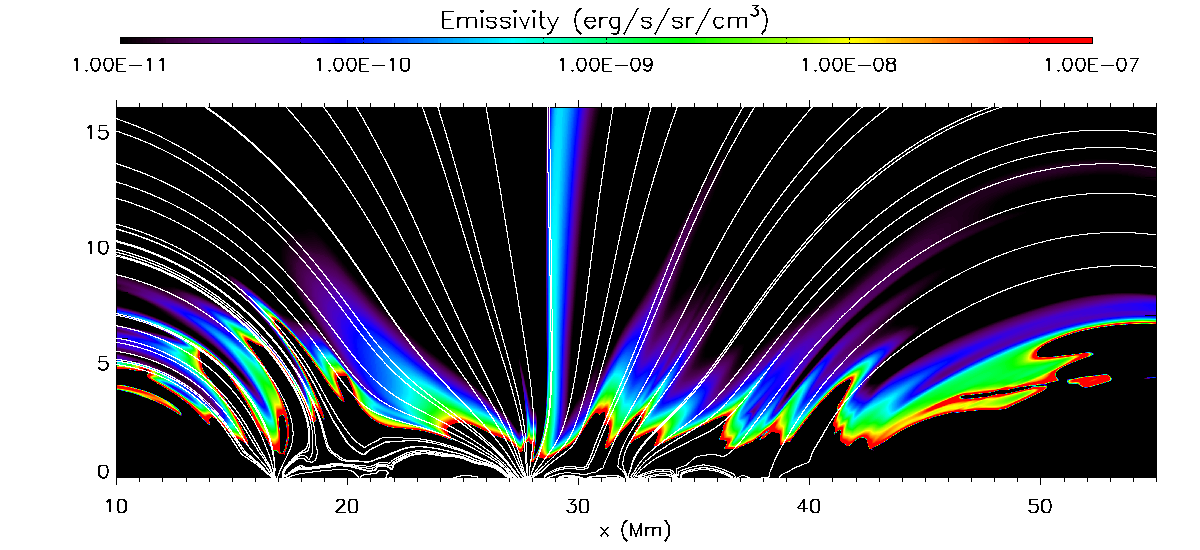}
   \includegraphics[width=0.32\textwidth]{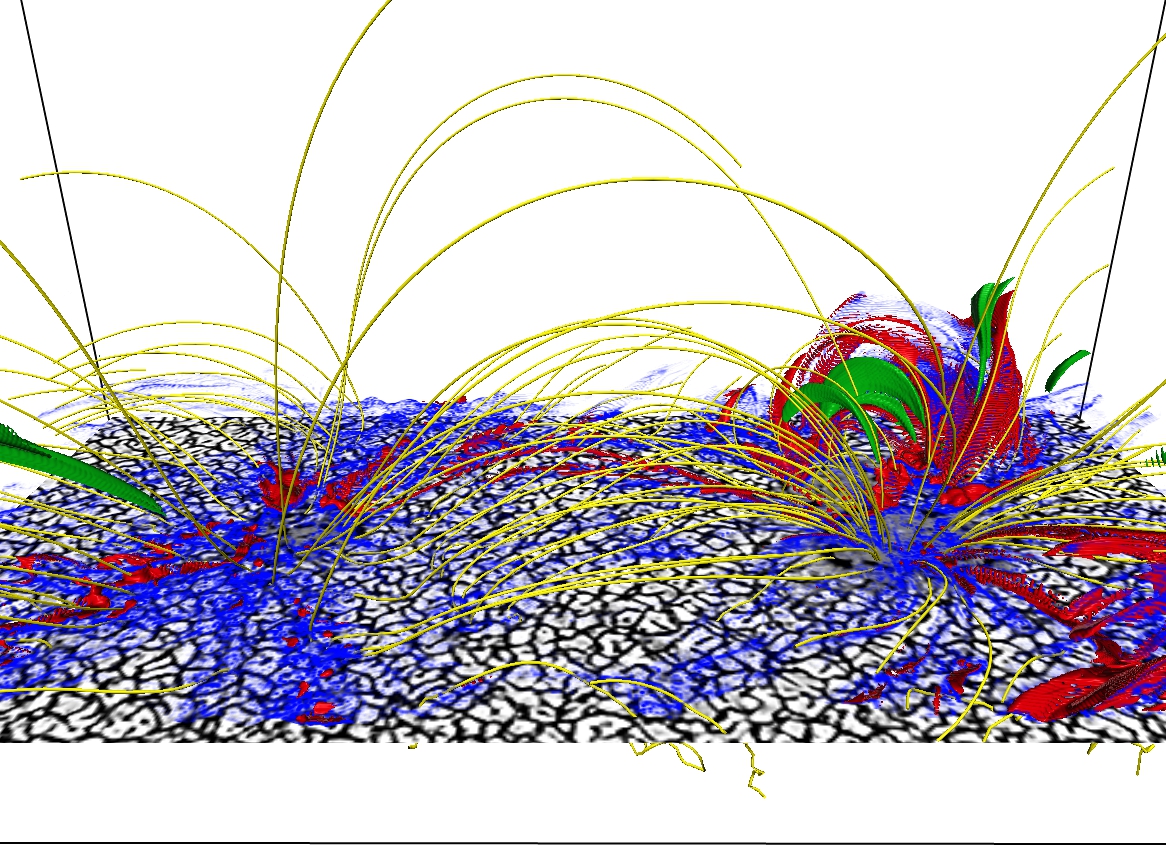}
   \caption{\small
     3D (2D) view of the 3D (2D) publicly available numerical models. The 3D images show the emission of \ion{Si}{iv}, \ion{Fe}{xii}, and \ion{Fe}{xxi} in blue, red and green iso-surfaces, respectively. The plots correspond to simulations  en024048\_hion, ch024031\_by200bz005, qs024048\_by3363, qs006005\_dyc, en096014\_gol and HGCR from left to right and top to bottom, respectively. The vertical velocity at the photosphere is shown in gray and the magnetic field lines in yellow. For the 2D en096014\_gol simulation (bottom middle) the map shows the emission in \ion{Si}{iv} and the magnetic field lines in white. }
       \label{fig_pub_sim}
\end{figure}

The existing publicly available numerical models cover various solar targets, from the quiet Sun to active regions with different spatial resolutions and various physical processes (see Fig.~\ref{fig_pub_sim}): 

\begin{itemize}
    \item Simulation en024048\_hion mimics an enhanced network with two main opposite magnetic field polarities separated by 8~Mm and connected with $\ge 8$~Mm long loops. The simulation shows fibrils but not type II spicules. The simulation includes non-equilibrium ionization for hydrogen. This simulation was the first to be publicly released and has been used in a large number of publications and is described in detail in \citet{Carlsson2016}.
    \item Simulation ch024031\_by200bz005 mimics a coronal hole with a coronal field strength of $\sim5$~G. Hydrogen ionization is in LTE, and the coronal temperature is relatively low ($5.8\,10^5$~K).
    \item Simulation qs024048\_by3363 models a strong magnetic region (3.4 kG) emerging into a quiescent pre-existing magnetic field \citep{Archontis2014, Hansteen2017}. The new emerging field pushes material out of the numerical domain as well as forms U-loops that reconnect and form UV-burst reconnection events with plasmoids and temperatures near 2~MK (See Figure~\ref{fig_pub_sim}). The resulting unsigned magnetic field in the photosphere after the emergence reaches an average of $\sim 230$~G. Hydrogen ionization is in LTE and ambipolar diffusion is not included.
    \item Simulation qs006005\_dyc \citep{Martinez-Sykora2019} mimics a coronal hole in which the corona has relatively low temperatures ($3\,10^5$~K) and 2.5~G unsigned magnetic field. Because of the high resolution of this model, it is possible to investigate at small scales the kinetic-magnetic energy conversion (and local dynamo) in the upper-convection zone, photosphere and chromosphere in the internetwork regions. Hydrogen ionization is in LTE and ambipolar diffusion is not included.
    \item Simulation en096014\_nongol \citep{Martinez-Sykora2018} is in 2.5D and mimics a plage region with two main polarities 40~Mm apart, connected with $\ge 40$~Mm long loops. Hydrogen ionization is in LTE and ambipolar diffusion is not included. There is only a single snapshot which is mostly used as a comparison to the next model. The 2.5 dimensions limit the braiding and types of waves present. 
    \item Simulation en096014\_gol \citep{Martinez-Sykora2018} has the same setup as en096014\_nongol, but includes ambipolar diffusion and the Hall term. This model reproduces many aspects of type II spicules despite its obvious limitations: it is 2.5D, and ionization is treated in LTE. It also shows an interesting thermal evolution of low-lying loops. 
    \item The HGCR model \citep{Cheung2019} is the first self-consistent 3D radiative MHD model\footnote{https://purl.stanford.edu/dv883vb9686} of a flare (equivalent to a GOES M class) driven by an emerging eruption. The setup is inspired by National Oceanic and Atmospheric Administration (NOAA) Active Region 12017 that appeared in late March and early April 2014. Contrary to the previous models, this one is computed with MURaM which assumes LTE and gray radiative transfer. Figure~\ref{fig_pub_sim} shows the hot ribbons in \ion{Fe}{xxi} (green). 
\end{itemize}

\section{Study of fundamental physical processes in the solar atmosphere}
\label{physical}

In this section we review how \iris\ observations and numerical modeling have contributed to a better understanding of several key fundamental physical processes that occur not only in the solar atmosphere but play a role throughout the heliosphere and astrophysical environments.

\subsection{Chromospheric Heating and Ion-Neutral Interactions}
\label{chromo_heating}

The solar chromosphere is sandwiched between the relatively cool surface, or
photosphere, and its million-degree outer atmosphere or corona. While
the chromosphere's temperature is only modestly increased over
that of the photosphere, the many scale heights of
dense plasma comprising the chromosphere imply that up to two orders of magnitude more non-thermal 
energy are required to sustain the chromosphere as compared to the
corona. Nevertheless, its dominant heating mechanisms remain
unknown and determining which physical processes dominate the heating of
the chromosphere is a major challenge in solar physics.

To tackle this long-standing problem, several different approaches are useful. It is important to properly diagnose thermodynamic conditions in the chromosphere, and to develop theoretical models for various heating mechanisms, including numerical simulations. Detailed comparisons between observations and numerical models play an important role in studying the various mechanisms that contribute to heating the chromosphere. Here we focus on \iris--related results, for a broader review, see \citet{Carlsson2019}.

\subsubsection{Observations}

Diagnosing chromospheric conditions is complex because many of the diagnostics are optically thick and formed under conditions that depart from local thermodynamic equilibrium (non-LTE). 
As discussed in \S~\ref{diagnostics_thick}, the development of the STiC inversion code allows much improved diagnostics of the thermodynamics in the chromosphere using the \ion{Mg}{ii} h and k lines, which are very sensitive to chromospheric conditions. The combination of the STiC inversion code and machine learning techniques has provided the community with the \alb\ database \citep{Sainz-Dalda2019}.  Exploitation in the near future of the \alb\ database (Fig.~\ref{fig_alb}) for a wide variety of solar targets and phenomena has the potential to drastically expand the quantitative constraints on numerical models, a key step for discriminating between various potential chromospheric heating mechanisms. 

An example of the inversion approach is the work by \citet{Gosic2018} who used STiC inversions to estimate the impact on chromospheric thermodynamics from the release of magnetic energy through cancellation of the pervasive weak magnetic fields in the internetwork. They found that while the local heating is significant, it likely does not play a dominant role in the average energy balance of the quiet Sun internetwork chromosphere. The spatio-temporal filling factor of such cancellations, at the sensitivity of the current (i.e., pre-\dkist) state-of-the-art magnetic field measurements, is not high enough. Recent coordinated \sst\ and \iris\ observations also highlight the possibility of significant heating from the interaction between pre-existing magnetic fields and recently emerged granular-scale internetwork fields \citep{Gosic2021}. Such heating would be compatible with statistical studies \citep{Schmit2016} that suggest a heating component in the internetwork that is unrelated to the magnetic network or the ubiquitous magneto-acoustic shocks (which are presumed to dominate the local energy balance).
To settle this issue, future coordinated \dkist/\iris\ observations are needed, given the sensitivity of the \iris\ diagnostics to the upper chromosphere and transition region.

\begin{figure}[tp]
   \includegraphics[width=0.95\textwidth]{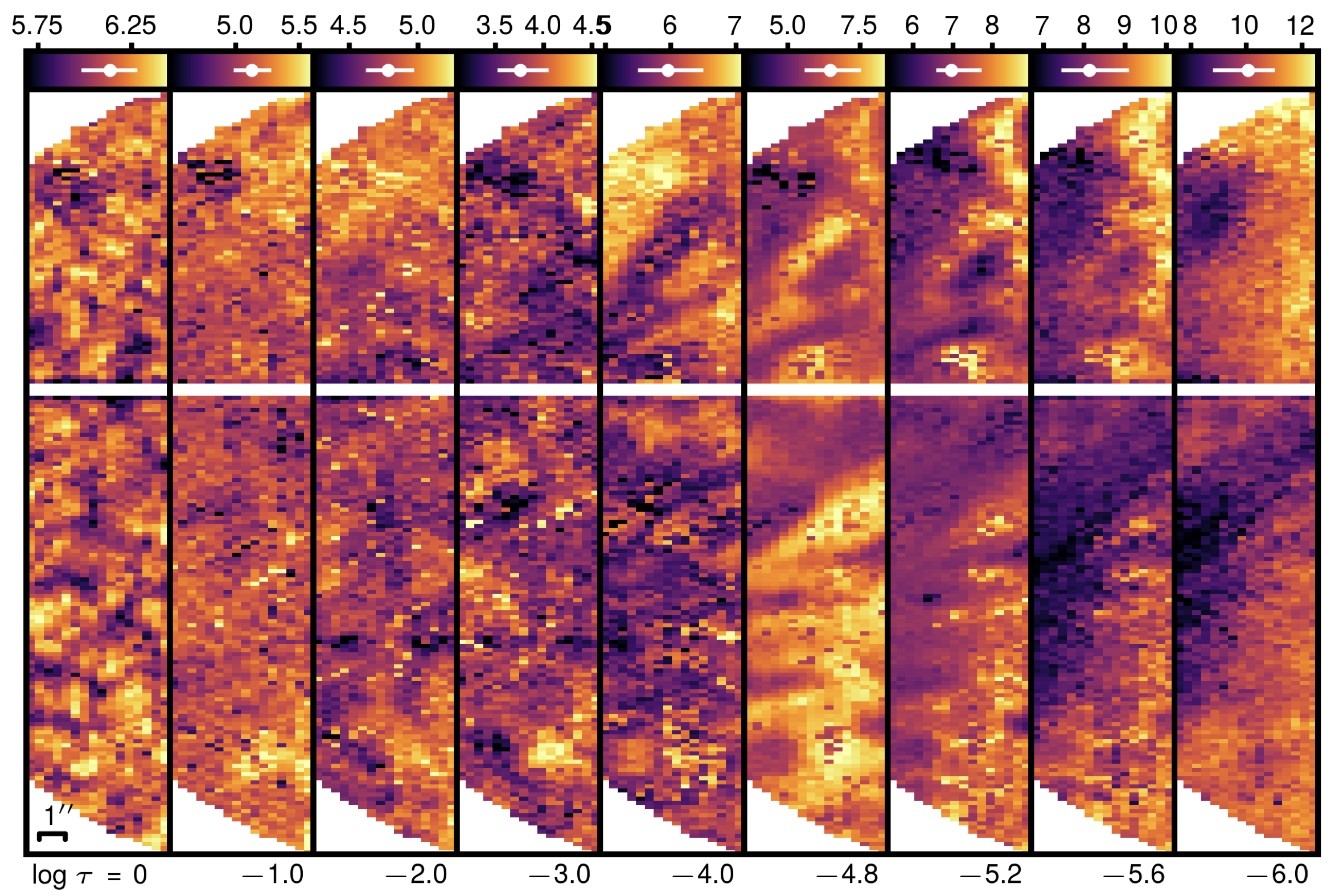}
   \caption{\small
Simultaneous inversions of \iris\ and \alma\ data provide unprecedented diagnostics of the chromospheric thermodynamic conditions, in this example temperature maps (kK) as a function of optical depth in a plage region. Such inversions show particular promise to help resolve the degeneracy between temperature and microturbulence \citep{da-Silva-Santos2020}. }
       \label{fig_alma_iris}
\end{figure}

The availability of chromospheric diagnostics from the \iris\ \ion{Mg}{ii} lines is particularly useful when combined with ground-based observations which often have spectral lines or diagnostics that are highly complementary. For example, the combination of \iris\ \ion{Mg}{ii} h and k spectra and \alma\ mm observations has been used (Fig.~\ref{fig_alma_iris}) to discriminate more accurately between chromospheric temperature and non-thermal motions \citep{da-Silva-Santos2018,da-Silva-Santos2020}. Similar synergies  exist with other spectral lines formed in the chromosphere \citep{Carlsson2019}. New advances in the inversion techniques that are focused on incorporating measurements from different observatories with a wide range of spatial resolutions \citep{delacruz2019}
allow the community to fully exploit combined \iris, \alma, or \dkist\
observations. 

Observational studies of heating in the magnetically dominated chromosphere have focused on a wide range of physical mechanisms. For example, magneto-acoustic shocks are ubiquitous not only in the internetwork but also in the magnetic network, plage, sunspots and penumbrae. They have a strong impact on the appearance of spectral lines wherever the shocks occur. But it is not clear whether they are responsible for the bulk of the heating in the chromosphere. \iris\ observations have been used to study their propagation and the evolution of non-linear harmonics as they travel upwards \citep{Chae2018}, as well as their frequency-dependent damping in sunspot umbrae \citep{Krishna-Prasad2017}. Shock waves have also been studied (using coordinated data from \dst\ or \hinode) in sunspot light bridges, with their provenance tied to magnetic reconnection and leakage of photospheric waves \citep{Tian2018a,Bai2019}. Similar processes have been invoked to explain the presence of high frequency signals discovered in active region plage \citep{Narang2019}. To determine whether these phenomena play a significant role in the local energy balance, novel techniques such as the \alb\ database and new methods to accurately determine shock wave properties \citep{Ruan2018a}, in combination with coordinated observations, will be very useful. Statistical approaches can also help elucidate the nature of the heating mechanism. For example, statistical analysis of the
correlation between the photospheric magnetic field (from \shmi) and \iris\ diagnostics shows an intriguing decrease in correlation around the temperature minimum, followed by an unexplained
increase towards the upper chromosphere \citep[in all solar targets, ][]{Barczynski2018}, a challenge for future models. 

Another approach to study chromospheric heating is based on exploiting the correlations between physical variables (such as temperature, velocity, density) and synthetic observables \citep{Rathore2015a,Rathore2015b,Leenaarts2016,Lin2017} derived from numerical simulations, primarily with the Bifrost code \citep{Gudiksen2011}. These types of correlations have been used to diagnose heating under a wide variety of conditions, from quiescent heating in the plage or quiet chromosphere \citep[e.g.,][]{Pereira2015, Park2016}, to the violent
conditions in reconnection-driven Ellerman bombs
\citep[e.g.,][]{Vissers2015b} and flares \citep[e.g.,][]{Tian2015}, thereby
providing constraints on the physical mechanisms (e.g., reconnection, Alfv\'en waves,
non-thermal electrons) responsible for the observed heating and dynamics. As more advanced numerical simulations become available for a wider range of solar targets (with the bulk of the results so far based on enhanced network simulations), the applicability of this approach will be enhanced. 

Several of these techniques have been applied to the study of heating in plage, regions with strong magnetic field outside of sunspots. Simplified forward models of \ion{Mg}{ii} k emission
combined with non-thermal broadening derived from \ion{O}{i}
\citep{Carlsson2015} suggest that plage regions show a step-like increase of heating in the low chromosphere and a transition region at high column mass, with significant non-thermal motions. 
More sophisticated inversions using STiC allow the height-dependent determination of these properties in plage, as shown by \citet{delacruz2016} using \iris\ and \sst\ observations. STiC inversions of combined \iris\ \ion{Mg}{ii} h/k spectra and \alma\ mm radiation (Fig.~\ref{fig_alma_iris}) above plage provide even more stringent constraints on how the temperature and turbulent motions depend on height \citep{da-Silva-Santos2018,da-Silva-Santos2020}. Such inversions have also provided the first direct measurements of plage heating associated with magneto-acoustic shocks \citep[see also, ][]{Chintzoglou2021a,Chintzoglou2021b}. The excellent synergies between \iris\ and \alma\ observations also highlight a possible path to further
improvements of the inversion approach from \iris\ data alone: preliminary studies indicate that the \ion{C}{ii} 1335\AA\ intensities are well correlated with \alma\ band 6 intensities (which are a good
proxy for temperature in the upper chromosphere), suggesting that \alb\ inversions that include \ion{C}{ii} would enhance the fidelity of temperature in the upper chromosphere, a region where the \ion{Mg}{ii} h and k lines are less sensitive \citep{Jafarzadeh2019,da-Silva-Santos2020}.

\subsubsection{Numerical modeling}

One of the key approaches of the \iris\ investigation has been the comparison between synthetic observables from advanced numerical simulations and \iris\ observations. In the Bifrost simulations, the bulk of the magnetic atmosphere is heated by dissipation of currents generated through magnetic field line braiding \citep[e.g.,][]{Hansteen2015}. These numerical models of the chromosphere are highly complex because of the wide range of physical processes that may play a role in this dynamic region. Because of the computational cost, it is currently not possible to include into a numerical model all of the candidate physical processes suspected of playing a role in the chromosphere. The \iris\ modeling approach has been to gradually include more diverse magnetic field conditions (the main free parameter in the Bifrost models), as well as more complex processes, both guided by comparisons with the observations. A key driver for this approach has been the finding that synthetic observations of \ion{Mg}{ii} h and k lines from Bifrost models are typically too narrow and often too faint compared to the high resolution \iris\ spectra \citep{Carlsson2019}. Various studies have been performed to understand this discrepancy. The lack of broadening in synthetic spectra could, in principle, be caused by a lack of turbulent motions in the numerical models. However,
microturbulence can be
estimated from the optically thin \ion{O}{i} 1356\AA\ line in the
\iris\ FUV bandpass \citep{Carlsson2015} and is found to be insufficient to fully explain the discrepancy. This result suggests that in the Bifrost models there is a lack of opacity in the \ion{Mg}{ii} lines, which could be caused by a lack of heating, mass loading, or density at chromospheric heights. One possibility is that the turbulent motions that are observed with \iris\ are directly associated with a heating process that is missing from the models, such as heating from turbulence driven by the Farley-Buhnemann instability \citep{Madsen2014} or Alfv\'en waves \citep{Arber2016}. Observational studies of the correlation between \ion{O}{i} broadening and chromospheric heating using \alb\ inversions would be interesting to help settle this issue. 

While it may not be easy to include physical mechanisms that occur on microscopic plasma scales into the Bifrost or other MHD models, it should be more straightforward to investigate the role of several other possible effects. For example, it is possible that numerical simulations at higher spatial resolution may lead to locally higher dissipation rates and subsequent heating, or even the generation of higher-frequency Alfv\'en or other waves \citep[e.g., through vortical motions, ][]{Moll2011}, both of which could contribute to more heating at chromospheric heights. Similarly, the interaction of small-scale, granular-scale fields with pre-existing network or plage fields is a possible candidate for explaining some of the heating in the magnetic chromosphere, also given the observational suggestions that cancellation of magnetic flux may play a role in heating the atmosphere \citep{Chitta2018}. Numerical simulations with more realistic magnetic field distributions that attempt to capture the properties of granular-scale fields are important to investigate this issue. Recent simulations also suggest that small-scale magnetic fields may also be generated at chromospheric heights \citep{Martinez-Sykora2019}. 

\begin{figure}
    \centering
    \includegraphics[width=0.95\textwidth]{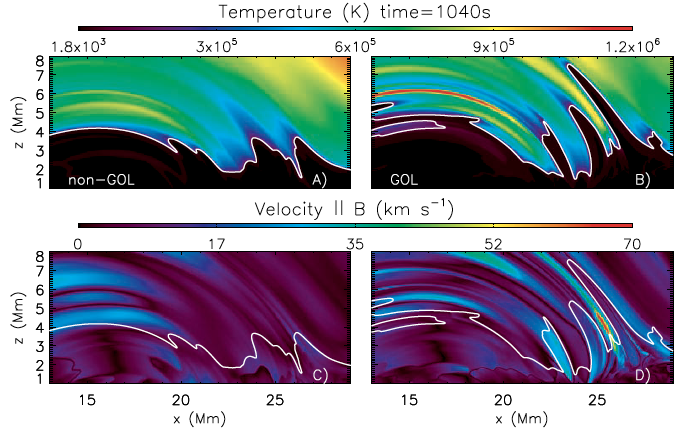}
    \caption{Temperature and velocity parallel to the magnetic field in two models, one in which ion-neutral effects are neglected (panels A, C) and one which implements these effects using a generalized Ohm's law (panels B, D). In the latter we find much longer fast spicules, similar to those observed, as a consequence of the increased role of ion-neutral effects such as ambipolar diffusion (adapted from \citep{Martinez-Sykora2017a}).  }
    \label{fig:ion_neutral}
\end{figure}

\subsubsection{Ion-Neutral Interactions}

An important aspect of the chromosphere is that it is partially ionized, like the Earth's ionosphere. This can lead to dissipation of magnetic energy from interactions and slippage between the electrically charged ions and neutral particles. These effects have long been
suspected of playing a significant role in the momentum and energy balance
of the chromosphere. \iris\ related numerical simulations have produced significant advances in our understanding of the role of ion-neutral interactions: ambipolar diffusion can lead to heating and significantly affect the dynamics of the chromosphere \citep{Martinez-Sykora2015,Martinez-Sykora2016b,Martinez-Sykora2017a}. In particular, these studies have provided new insights into our understanding of the formation and
evolution of chromospheric spicules, the most common jets in the solar atmosphere \citep{Martinez-Sykora2017a}. Advanced numerical simulations show that ambipolar diffusion fundamentally
changes the interaction between the strong network or plage magnetic
fields and the ubiquitous granular scale weak fields. It allows these
tangled weak fields to diffuse into the middle and upper chromosphere
where the violent release of magnetic tension (introduced in the
subsurface convection zone) leads to rapid acceleration of plasma to
drive supersonic jets with speeds of 50-100 km/s that are heated while
they expand upwards through the diffusion of ambipolar currents. Synthetic observables from these
simulations show good agreement with \iris\ and \sst\ observations
of spicules, including the significant heating to TR temperatures
observed with \iris\ and \hinodes\ \citep{Skogsrud2015, De-Pontieu2017b}.  Modeling and observational results also suggest a significant impact of spicules on the corona \citep{De-Pontieu2017a,Martinez-Sykora2018}. During the past few years it has become clear that other mechanisms may also produce jet-like features, including magneto-acoustic shocks \citep[e.g.,][]{Hansteen2006,Matsumoto2010}, vorticity, and twisted magnetic fields \citep[e.g.,][]{Iijima2017}. It remains unclear which of these is the dominant formation mechanism in the solar atmosphere. To settle this issue will require several advances on the modeling side: the identification of these jets as spicules is often based on physical variables in the model rather than synthetic observables. Also, many of these models lack a more sophisticated treatment for the significant radiative losses in the chromosphere. And finally, observations of spicules now extend into the transition region, so a full comparison across all wavelengths is key. In the Bifrost models at least, the introduction of ambipolar diffusion appears to have a significant effect on the strength, ubiquity, and thermal evolution of spicule-like features in the models.

Ion-neutral interactions can also produce damping of Alfv\'enic waves, especially at high frequencies \citep[e.g., ][]{De-Pontieu2001,Ballester2020}, further enhanced  by interaction between different species \citep[e.g., ][]{Zaqarashvili2011,Popescu2019,Martinez-Sykora2020b}. Detailed comparisons between numerical models and high-resolution observations with \iris\ and \dkist\ of spicules, which are known to carry Alfv\'enic waves \citep{De-Pontieu2007b,Okamoto2011,De-Pontieu2014}, could provide evidence of ion-neutral damping of Alfv\'en waves and possible associated heating. 

Recent  simulations also highlight the importance of non-equilibrium ionization in the chromosphere, which is of importance for both hydrogen and helium \citep[see, e.g.,][]{Golding2014}. The combination of ambipolar diffusion and non-equilibrium ionization appears to lead to enhanced electron densities in the chromosphere. This causes increased opacity at least in the \alma\ bands, with the effect on \ion{Mg}{ii} opacity not yet known. Comparisons with \iris--\alma\ observations \citep{Martinez-Sykora2020b} suggest that the increased \alma\ opacity may explain puzzling observations of chromospheric ``holes'', regions of unusually low
temperatures ($<4,000$~K) recently discovered with \alma\
\citep{Loukitcheva2019b}. These may occur as a result of low temperatures in the wake of magneto-acoustic shocks that provide mass to canopy loops. In addition, these simulations show that low-lying transition region loops could be heated by ambipolar dissipation of electrical currents \citep{Martinez-Sykora2020a}.

Further developments of the numerical models (and comparison with observations) thus hold promise to address many unresolved issues, such as the role of the Farley-Buhnemann instability, thermal instabilities \citep{Oppenheim2020}, backwarming from Lyman-$\alpha$, vorticity and Alfv\'en waves, interactions between small-scale weak fields and strong network or plage fields, and the interactions between multiple fluids and species \citep[e.g.,][]{Martinez-Sykora2020c}.

\subsection{Alfv\'en waves}
\label{alfven_waves}

The strongly inhomogeneous solar atmosphere is a rich environment for wave processes. Magneto-convection, density stratification and 
magnetic field emergence leads to large complexity in the physics of wave propagation. Because of the pervasive character of MHD waves 
and their substantial energy content, understanding this complexity is necessary for understanding the solar atmosphere, including coronal heating and solar wind acceleration.

Among MHD waves, Alfv\'en waves and, more generally, transverse MHD waves (defined here as having magnetic pressure and/or tension as one of the main restoring forces) are the ones to consider, due to their unique ability to carry enough energy to generate a corona \citep{Uchida_1974SoPh...35..451U, 
Cranmer_2005ApJS..156..265C}.

The study of transverse MHD wave generation and dissipation in the solar atmosphere is a field that has been significantly 
advanced by \iris. Indeed, a high resolution imaging spectrometer has a unique advantage to capture the 3D
flow associated with a wave, thereby more accurately identifying the nature of the wave. Furthermore, the temperature coverage of \iris\ allows to track the propagation of the wave 
across the chromosphere and transition region, where many wave processes are expected to occur. 

\begin{figure}
    \centering
    \includegraphics[width=\textwidth]{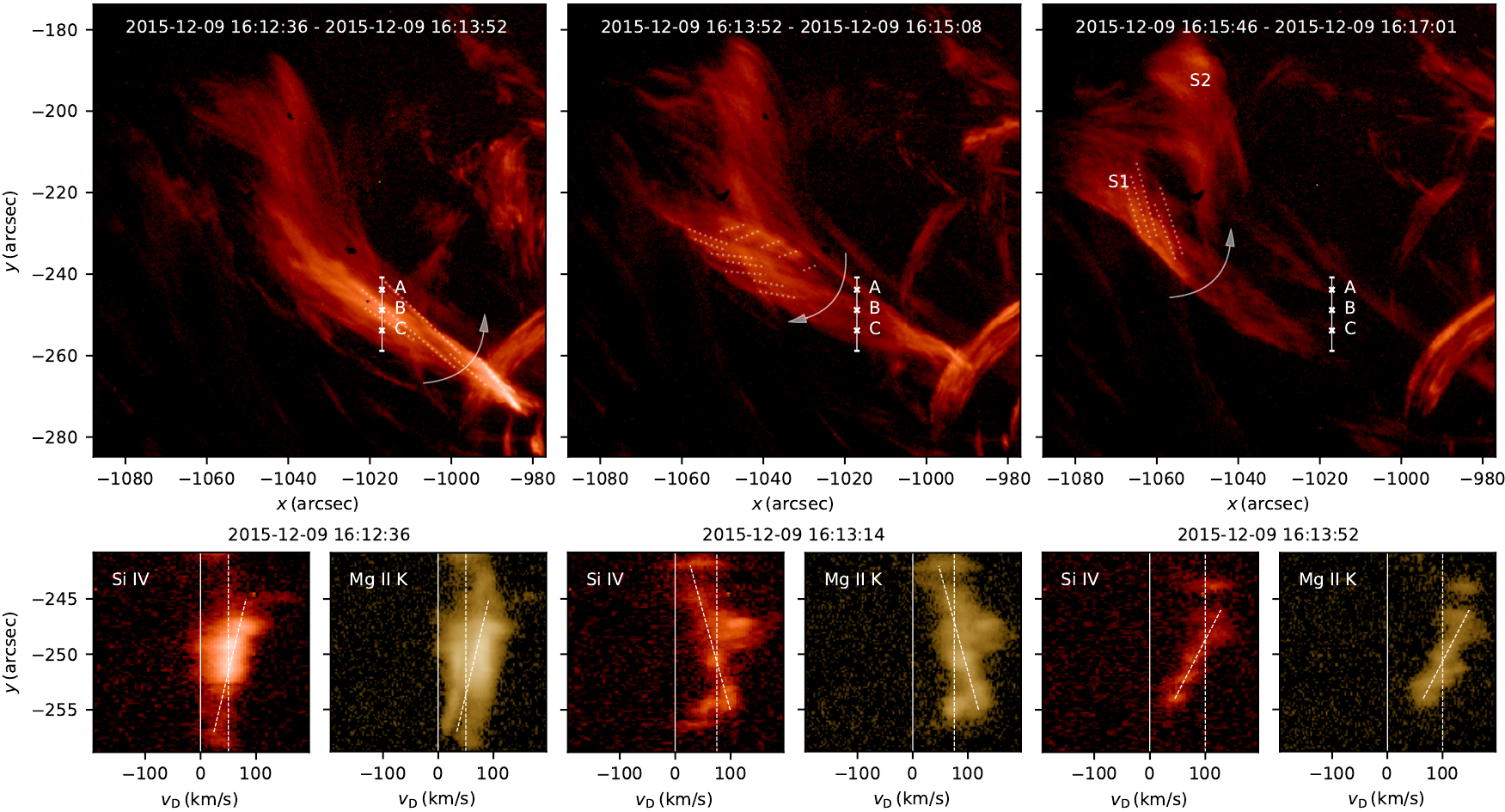}
    \caption{The first direct detection of a torsional Alfv\'en wave with SJI 1400 observations (top 3-set panel) and \ion{Si}{iv}~1394\AA, and \ion{Mg}{ii}~2796~\AA\ spectra(bottom panels). Top panels show 3 different phases of the torsional oscillation (each consisting of superimposed images at the times shown); the white dotted lines outline the helical trajectories (arrow shows direction of rotation). White vertical line indicate the \iris\ slit crossing the motion. The spectral profiles covered by this slit portion are shown at 3 different times in the bottom panels. Vertical dashed lines indicate the mean Doppler velocity. The tilted dashed line indicates the torsional motion. Adapted from \citet{Kohutova2020}.}
    \label{fig:torsional}
\end{figure}

For coronal heating purposes, any wave generated in the lower atmosphere needs to first propagate through the chromosphere and transition region before reaching the corona. This interface region presents sharp density and magnetic field gradients that lead to a significant loss of wave energy through processes such as mode conversion, shock heating, refraction and reflection \citep{Bogdan_2003ApJ...599..626B}, as well as ion-neutral interactions. In-situ generation of transverse MHD waves by-passes these obstacles and is therefore an interesting path to explore. Coordinated \iris--\sot\ coronal rain observations have provided a new mechanism of in-situ transverse MHD wave generation in the corona through colliding flows, leading to energy fluxes of $10^7-10^8$~erg~cm$^{-2}$~s$^{-1}$. Similarly, theory expects that a large fraction of the energy released through magnetic reconnection is in the form of Alfv\'en waves. 
\iris\ has now observed this process, with the first fully resolved torsional Alfv\'en wave in the corona \citep[][ see Fig.~\ref{fig:torsional}]{Kohutova2020}.

The main wave generator in the solar atmosphere is magneto-convection and related  processes such as magnetic buffeting \citep{Kato_2016ApJ...827....7K}, from which ample power is observed \citep[e.g., ][]{Oba_2020ApJ...890..141O}. 
Horizontal displacement of magnetic bright points is expected to be an efficient generator of transverse MHD waves \citep[e.g.][]{Jafarzadeh_2017ApJS..229....9J}. 
Recent statistical analysis of transverse waves on spicules using \iris\ \ion{Mg}{ii} spectra constrain the transverse wave amplitudes to 25~km~s$^{-1}$ \citep{Tei2020}, indicating significant wave energy fluxes at those heights on the order of $0.1-3\times10^{6}~$erg~cm$^{-2}$~s$^{-1}$. A $3-6$\% transmission into the corona leads to $1-2\times10^5$~erg~cm$^{-2}$~s$^{-1}$, which is sufficient for the quiet Sun corona and acceleration of the solar wind, in line with previous measurements from Hinode/SOT \citep{De-Pontieu2007b}. 


Recently, vortical motions in the form of swirls in the photosphere have been ubiquitously detected \citep[e.g., ][]{LiuJ:2019aa}. 
These structures are particularly interesting since they are expected to naturally generate torsional Alfv\'en and fast kink MHD waves, referred to as ``Alfv\'enic'' waves due to magnetic tension force being the main restoring force 
\citep{Bonet_2008ApJ...687L.131B,Wedemeyer_2012Natur.486..505W}. An important question 
is whether such vortical structures carry through to the upper atmosphere and whether they represent any major energy and mass conduit for the corona. First simultaneous observations of a vortex at photospheric and chromospheric levels have been obtained through coordinated \sst--\iris\ observations \citep{Park2016}, indicating high-speed upflows and a potentially important role in the chromosphere-corona mass and energy cycle. This has further been supported by the detection of propagating Alfv\'en pulses from swirls with a factor of $10-80$ the required energy power for locally heating the chromosphere \citep{LiuJ_2019ApJ...872...22L}. Chromospheric signatures of Alfv\'en waves from swirls have further been observed in case studies  
\citep{Tsiropoula_2020A&A...643A.166T}, while coordinated \sst--\iris\ observations have established the ubiquity of torsional motions on sub-arcsecond scales, occurring in active regions, quiet Sun and coronal holes alike \citep{DePontieu2014}. 
These observations show how these waves propagate upward and are associated with heating to transition region temperatures, as evidenced by \iris\ \ion{Si}{iv} 1402\AA\ images and spectra. 

Perhaps the most important but also most elusive wave process in the solar atmosphere is mode conversion, which involves the change from a specific wave mode into another, and therefore a change in the 3D flow of the wave, as well as its compressive properties and thus its ability to dissipate.  
Several types of mode conversion are expected to occur, such as longitudinal to transverse mode at the Alfv\'en - acoustic equipartition layer in the chromosphere \citep{Schunker_2006MNRAS.372..551S}, and fast mode to Alfv\'en mode in the transition region where a strong gradient in the Alfv\'en speed is expected \citep{Cally_2011ApJ...738..119C}. 
These processes are yet to be directly observed, but works such as \citet{Kanoh2016} combining \iris\ and \sot\ clearly show the enormous change in wave energy flux (from $10^7$ to $\approx 10^5$~erg~cm$^{-2}$~s$^{-1}$) in the longitudinal modes during propagation from the photosphere to transition region heights, suggesting either mode conversion or dissipation. If the former, such observations suggest a possible explanation to the ubiquity of Alfv\'enic waves in the corona \citep{McIntosh2011b}, part of which show frequencies characteristic of p-modes \citep{Morton_2019NatAs.tmp..196M}. If the latter, they provide evidence of an important chromospheric heating contribution, as also supported by other work \citep{DePontieu07, Okamoto2011, Morton_1012NatCommun...3...1315, LiuJ_2019ApJ...872...22L}.

Besides the above linear mode conversion processes, non-linear mode conversion is also expected, and can involve Alfv\'en to slow/fast modes or longitudinal to transverse modes. 
It is driven by terms such as the ponderomotive force, 
flux tube expansion, 
or wave-to-wave interaction. A characteristic signature of this process is a doubling of the wave frequency and therefore the generation of high-frequency waves \citep{Shoda_2018ApJ...854....9S}, which  are commonly detected in spicules \citep[e.g.,][]{Okamoto2011, Srivastava_2017NatSR...743147S, Tavabi2019}.



Mode conversion is also expected to occur ubiquitously in the corona because of its strongly inhomogeneous structure. Analogous to the fast to Alfv\'en mode conversion in the transition region, kink waves are expected to efficiently transfer their energy to azimuthal Alfv\'en waves \citep[resembling the torsional $m=1$ Alfv\'en wave,][]{Goossens_2020AA...641A.106G}, wherever the kink speed matches the Alfv\'en speed and in particular at the boundaries of coronal loops. Known as resonant absorption (or mode coupling), this process has long been hypothesised to be the leading explanation behind the observed strong damping of large amplitude kink oscillations \citep{Nakariakov_2020ARA&A..58..441N} and, together with the line-of-sight (LOS) superposition effect, is thought to be the main reason for the apparently small wave energy flux in the corona. 
Indeed, this idea has received support from numerical simulations, showing that 10\% or less of the wave energy is recovered from the POS motions detected with imaging instrumentation alone \citep{De-Moortel2012, Antolin2017}. On the other hand, a large fraction of the wave energy is expected to hide behind the observed large spectral line broadening resulting from these processes \citep{Pant_2019ApJ...881...95P}, naturally explaining the observed correlation between Doppler velocities and line widths \citep{McIntosh2012}. However, this interpretation of non-thermal motions
as evidence for waves has recently been questioned by interesting \iris\
results. 
\citet{Li2019a} captured the injection
of plasma into a loop, associated with the injection of helicity,
strong helical motions,
and (possibly) the subsequent development
of turbulence (from line broadening). While such broadening is
typically interpreted  in terms of waves, the observed scenario may also
be compatible with recent braiding models. To settle this interesting
issue, more observations and advanced models are required. Coordinated \iris\ and \solos\ measurements of line broadening in the same structure but from different vantage points will help distinguish between various mechanisms (\S~\ref{Sect:braiding}).

\begin{figure}
    \centering
    \includegraphics[scale=0.4]{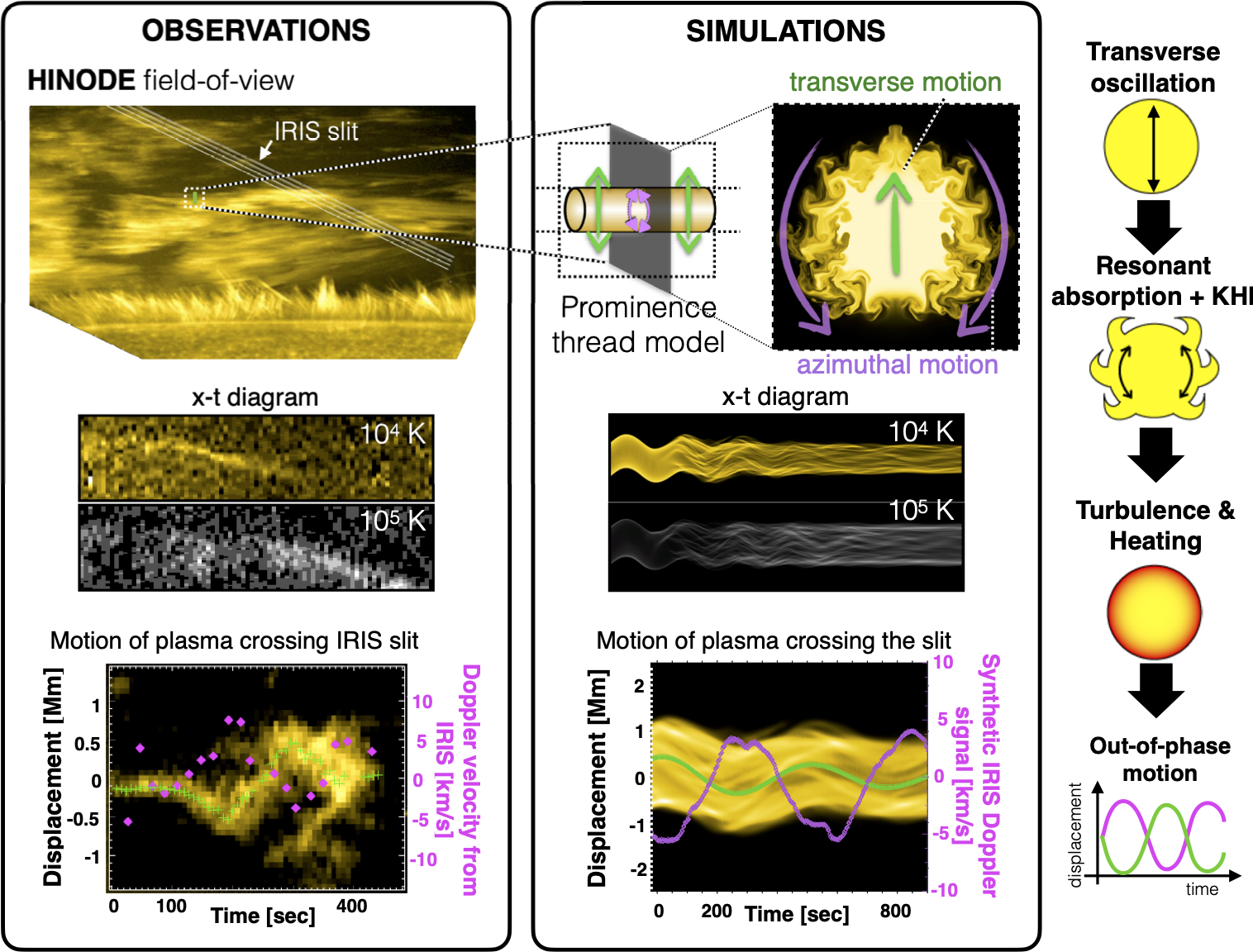}
    \caption{Coordinated \iris\ and \hinodes\ observations of an active prominence at the limb, reported by \citep{Okamoto2015}. Heating signatures were captured, evidenced by fading of the prominence threads in the \sot\ \ion{Ca}{ii}~H line (yellow) accompanied with intensity increase in the hotter SJI~1400 passband (grey). The Doppler velocity captured by \iris\  in the \ion{Mg}{ii}~k line (purple dots in bottom-left time-distance panel) shows an out-of-phase motion with respect to the plane-of-the-sky motion captured by \sot. 3D MHD simulations  of an oscillating prominence thread with a kink mode performed by \citep{Antolin2015b} reproduce the heating signatures and phase relations, based on the combined effect from resonant absorption and the KHI, as shown in the sketch. Figure taken from \citet{VanDoorsselaere_2020SSRv..216..140V}. }
    \label{fig:resonantkhi}
\end{figure}

Despite the overwhelming theoretical support for resonant absorption, direct observational evidence was only possible through \iris--\sot\ coordinated observations of a prominence. An out-of-phase relation between the POS oscillatory motion (detected with \sot) 
and the LOS velocity (detected with \iris) 
and accompanying heating of the prominence threads from $10^4$~K to, at least, $10^5~$K was observed \citep{Okamoto2015}. Through numerical 
simulations, 
such signatures were shown to be telltale signatures of resonant absorption \citep[][see Fig.~\ref{fig:resonantkhi}]{Antolin2015b}. Furthermore, the Kelvin-Helmholtz instability (KHI) produced by the velocity shear from the transverse displacement and the resonance, was shown to play a major role, allowing the enlargement of the resonant layer (to the observable scale) and the establishment of turbulence leading to wave dissipation and heating. These works have opened a very active research avenue for wave-based coronal heating. Further coordinated observations and modelling are required to determine how common this phenomenon is 
in other, more prevalent, structures in the solar atmosphere. 

The Transverse Wave-Induced Kelvin-Helmholtz (TWIKH) rolls predicted by the KHI-resonant absorption model have received further support in coordinated \sst--\sot\ observations of spicules \citep{Antolin2018a}. A very good match was found between the model and the observed characteristic strand-like structure in intensity \citep{Skogsrud_2014ApJ...795L..23S} and Doppler velocity, 
the very fast apparent upward motion \citep{De-Pontieu2017b} 
and ragged Doppler shift sign changes at maximum transverse displacement. 
On the other hand, the model has so far only been found to lead to mild temperature increase, at odds with \iris\ observations combined with \sdo\  and \sst\ showing the TR to coronal heating impact of spicules \citep{DePontieu2014b, Skogsrud2015, Skogsrud2016, De-Pontieu2017a}. More advanced simulations suggest that the observed heating and Alfv\'enic waves (and thus potentially the TWIKH rolls as well) may be a consequence of a whip-lash effect from the release of magnetic tension through ambipolar diffusion \citep{Martinez-Sykora2017a}. 

Several heating models based on Alfv\'enic waves have been proposed \citep{VanDoorsselaere_2020SSRv..216..140V}. Besides the combined KHI-resonant absorption model \citep[e.g.,][]{Karampelas_10.3389/fspas.2019.00038}, the long established Alfv\'en wave turbulence model \citep[based on wave-to-wave interaction, e.g.,][]{vanBalle2011,Matsumoto_2018MNRAS.476.3328M} and more recently, the generalised phase mixing model \citep{Magyar_2017NatSR...714820M}, all rely on the establishment of turbulence for wave dissipation to occur. It remains unclear whether the observable signatures from these models and, in particular, the observed non-thermal line broadening at both chromospheric and coronal levels match observational constraints. Measurements of non-thermal broadening of the
optically thin \ion{O}{i} line at the chromospheric
footpoints of coronal loops \citep{Carlsson2015} place strict
constraints on the available wave energy flux. At the coronal level, constraints come from measurements of Alfv\'en waves in coronal rain  
\citep{Kohutova2016}. 
Coronal rain investigation can further strongly constrain the wave energy flux in the corona since it uniquely combines the possibility of high resolution observations of coronal plasma \citep[catastrophically cooling and thus becoming observable in the \iris\ temperature range,][]{Antolin_2020PPCF...62a4016A}, 
with a strong reduction of the LOS superposition effect (due to the increased optical thickness). Combined observations of prominences and coronal rain between \solo/HRI and \iris\ may not only provide the phase relations between velocity, line width and intensity characteristic of resonant absorption and KHI processes, but may also directly observe the turbulent heating events that result. Furthermore, the high resolution provided by \dkist\ may also allow to distinguish the nature of the turbulence by directly observing TWIKH rolls. 
For example, the double mode conversion process, through which p-modes may mode convert into Alfvénic waves in the transition region and corona, may be detectable through coordinated \iris--\dkist\ observations.

\subsection{Shock waves}
\label{shock}

Oscillatory signals are common in the chromosphere. In many
locations these lead to the formation of shocks as they propagate
upward from the photosphere into the rarified chromosphere. 
Shock waves have long been known to pervade the chromosphere, from the
quietest internetwork regions to stronger magnetic field regions such as
network, plage and sunspots. There have been extensive numerical modeling efforts aimed at understanding the role of shock waves for several decades \citep{Carlsson1997,Bogdan_2003ApJ...599..626B}, both in weak and stronger magnetic field environments.  Studying shock waves with \iris\ has given
insight into how such waves behave in a complex magnetic environment
where they often cross the plasma $\beta=1$ surface(s), and also how they
impact the energy balance of the chromosphere, transition region or corona. 

\iris\ observations of shock waves in internetwork regions show that these waves only rarely impact the transition region emission \citep{MartinezSykora2015}. This is quite different in stronger field regions as evidenced by \iris\ observations
of sunspots \citep{Tian2014a,Yurchyshin2015} which reveal the
impact of shocks on the transition region above sunspots. 
Coordinated \hinodesp\ and \iris\ observations of sunspots have allowed
determination of photospheric and upper chromospheric energy fluxes
carried by slow-mode magneto-acoustic shock waves, revealing that these
shocks play an important role in heating the
sunspot chromosphere \citep{Kanoh2016}. Furthermore, recent \iris\ observations and numerical modeling have now uncovered the existence of pseudo-shocks (i.e., exhibiting only density discontinuity) in \ion{C}{ii} spectra above sunspot
umbrae, carrying a large energy flux upward \citep{Srivastava2018}. The pseudo-shock represents a discontinuity in density akin to the entropy mode (cf. \S~\ref{section:rain}). It is still unclear what the drivers of pseudo-shocks are and how ubiquitous they are.

By applying the helioseismic time-distance analysis technique on measurements with \iris, \shmi,
\saia, and the 1.6 m New Solar Telescope at Big Bear Solar
Observatory (\bbso), \citet{Zhao2016} have traced the source of some shock waves
all the way to below the solar surface. Leakage of photospheric waves is also seen in sunspot light bridges, although reconnection in the lower atmosphere also appears to play a role in generating waves in light bridges \citep{Tian2018a}. This is further discussed in \S~\ref{reconnection}.

\begin{figure}[tp]
   \includegraphics[width=0.95\textwidth]{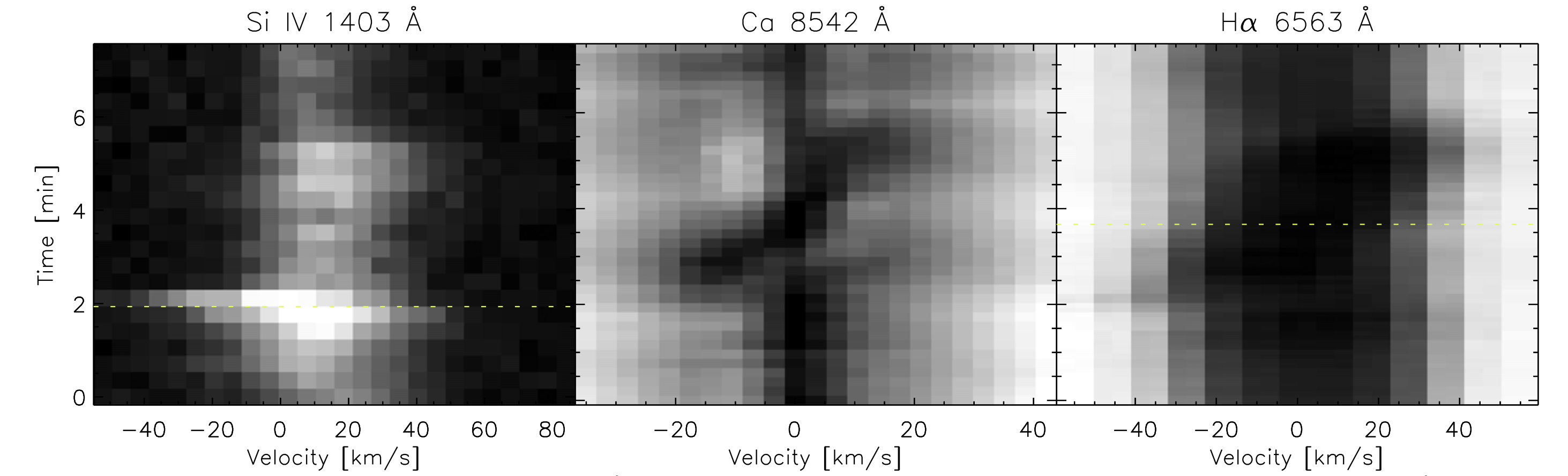}
   \caption{\small Slow-mode magneto-acoustic shocks are ubiquitous in active region plage, dominating the time evolution, as evidenced by sawtooth patterns in $\lambda$-time plots in chromospheric lines (middle and right panels) and the resulting formation of jets called dynamic fibrils. The impact of these jets on the transition region leads to a multitude of short-lived brightenings at the transition region footpoints of coronal loops (\ion{Si}{iv}, left panel)  
 \citep{Skogsrud2016}. }
       \label{fig_shocks_skogsrud}
\end{figure}

Magneto-acoustic shocks also play a
significant role in the dynamics and heating of the upper chromosphere and low transition region above plage, as evidenced by analysis of coordinated \sst\ and \iris\ observations \citep{Skogsrud2016} where they are seen to drive dynamic fibrils \citep{Hansteen2006} and strong brightenings in transition region lines (Fig.~\ref{fig_shocks_skogsrud}). In fact, many of the rapdily evolving brightenings in \iris\ slit-jaw images are caused by slow-mode magneto-acoustic shocks.

In strong field regions such as network and plage, \iris\
observations suggest a link between these chromospheric shocks and the long-elusive propagating coronal disturbances \citep{Bryans2016,De-Pontieu2017a} that continuously occur along coronal loops and that have been attributed to slow-mode waves or flows that feed the corona.  Other coordinated \iris--\aia\
observations have tied the shocks in umbrae to propagating
intensity disturbances in the corona, suggesting they may play
a role in the coronal energy balance \citep{Hou2018}. This is further discussed in \S~\ref{transfer}.

\iris\ observations have also shown that chromospheric shocks are not only driven
from below, as previously thought, but also occur when
strong coronal or TR downflows slam into the chromosphere. This has been observed both in
sunspots \citep[e.g.,][]{Straus2015} and arch filament systems created by
flux emergence \citep[see][who use \hinodes, \sdo\ and \iris\ data]{Toriumi2017}.

Given the ubuiquity of magneto-acoustic shocks, quantifying their contribution to the momentum and energy balance of the chromosphere is important. To determine whether shocks (or any other phenomenon) are energetically significant, typically the energy flux is estimated and compared to the average radiative losses expected from the chromosphere. The latter are most often based on spatio-temporal averages from semi-empirical models \citep[e.g.,][]{Withbroe1977}. This approach could be improved upon significantly if instead of averaged radiative losses the locally determined radiative losses are used \citep{Diaz2020}. In future work, this could be achieved by exploiting high-fidelity inversions either through STiC or \alb\ to determine the local thermodynamic conditions in the atmosphere and calculating the associated radiative losses.

\subsection{Fundamental MHD Instabilities}
\label{instabilities}

The solar atmosphere allows for the investigation of fundamental MHD instabilities in extreme conditions unattainable in the laboratory. Such instabilities can play a key role in major solar  questions, such as the dissipation of MHD waves or the onset and rate of magnetic reconnection, but can also provide seismological insight into the local conditions in which they form. With its high resolution and large temperature coverage, \iris\ has provided key new insight into these processes.

The Kelvin-Helmholtz instability (KHI) is an MHD shear flow instability characterised by vortex shaped, self-similar structures at the interface of  velocity shear regions. Importantly, in the high magnetic Reynolds number of the solar atmosphere, small-scale structures are generated in the turbulent cascade that results, allowing the kinetic and magnetic energy of the flow to be dissipated into heat, as well as momentum transfer due to the strong mixing of the plasma components across the shear boundary \citep{Fujimoto_1994JGR....99.8601F}. The KHI can also induce more efficient radiative cooling by enlarging the population of transition region plasmas in chromospheric-coronal interface regions \citep{Hillier_2019MNRAS.482.1143H, Fielding_2020ApJ...894L..24F}, whose temperatures can be detected with \iris.

In magnetized plasmas, magnetic tension can inhibit the KHI onset, and therefore shear flows misaligned to the magnetic field will more readily trigger the instability since the magnetic tension component opposing the unstable modes is effectively reduced \citep{Chandrasekhar_1961hhs..book.....C}. Periodic transverse shear flows, as in the boundary of coronal loops oscillating with kink modes, are always KHI unstable \citep[or unstable to a similar parametric instability,][]{Hillier_2019MNRAS.482.1143H}, and a large number of numerical studies supports such findings \citep[e.g.,][see \S\ref{alfven_waves}]{Terradas_2008ApJ...687L.115T, Antolin2014b}. However, the observations of the KHI have been scarce so far, either limited to quiescent prominences \citep[e.g.,][see also \S\ref{section:prominence}]{Berger_etal_2010ApJ...716.1288B} or to very energetic events, such as CME eruptions \citep{Foullon_etal_2011ApJ...729L...8F} or flares \citep{Brannon2015,Yuan_2019ApJ...884L..51Y}. Observations of the KHI in quiescent prominences with \iris\ suggest that only large field-aligned shear flows (with Alfv\'en Mach number larger than 2) are able to trigger the instability in such conditions \citep[][see also Fig.~\ref{fig_vanessa} and \S\ref{section:prominence}]{Hillier2018}. On the other hand, 3D MHD simulations of eruptions have shown that the characteristic KHI vortices (and Alfv\'enic vortex shedding at the wake of the eruption) are only visible under very specific LOS and at high resolution, and indicate that spectroscopic instruments are largely favoured over imaging instruments to readily detect the characteristic features \citep{Syntelis_2019ApJ...884L...4S}. Accordingly, broadened line profiles accompanied by vortex motions have been observed by \iris\ at the top of loop arcades with null point topologies \citep{Liu_2016SPD....47.0402L}, resembling the quiescent prominence dynamics. In an observation that is the first of its kind, the formation of the KHI produced by a blowout jet has been captured by \iris\ evidencing shear flow speeds of several 100s km~s$^{-1}$, a small-scale saw-tooth pattern at the shear flow boundary and a temperature increase of $\approx2$~MK during the event, indicating kinetic and magnetic energy dissipation \citep{Li2018nat}. 

Another fundamental MHD instability of magnetized plasmas observed in the solar atmosphere is the Rayleigh-Taylor instability (RTI). This instability forms at the interface of two fluids with large density variation, where magnetic tension works against gravity to support the densest fluid. 
In the solar atmosphere it has been observed in quiescent prominences, through the characteristic formation of tenuous but hotter plumes, which rise across the dense material. Because of the partial ionisation state of the prominence material, an interesting effect predicted by numerical simulations is the formation of localised large velocity drifts between the neutral and ion populations, due to the difference in the instability dynamics for both populations \citep{Khomenko_2014AA...565A..45K,Popescu2019}. The existence of such velocity drifts carries large importance due to its effect on physical quantities such as wave damping and the magnetic reconnection rate. While recent results provide support for such multi-fluid physics, more observations are needed to understand such plasma dynamics \citep{Anan_2017AA...601A.103A, Wiehr_2019ApJ...873..125W}.

Recently, the first coupled KHI-RT instability has been observed with \sot\ in a quiescent prominence \citep{Berger_2017ApJ...850...60B}. Large shear flow velocities of 100~km~s$^{-1}$ were inferred at the interface of prominence bubbles based on the observed phase velocity. However, reduced RT growth rates were measured, attributed to the existence of magnetic shear estimated to be on the order of 10~G at an angle of $70^{\circ}$ to the prominence plane in order to compensate the effect from the velocity shear flow. The existence of such large shear flows in prominence bubbles has been confirmed thanks to \iris\ observations \citep{2018cosp...42E.293B} and could lead to a better determination of the magnetic field in such bubbles, also in coordination with \dkist.

Thermal instability is yet another fundamental plasma instability whose understanding has been significantly advanced thanks to \iris\ observations. Thermal instability is discussed in \S\ref{section:rain}, together with its main observable feature (coronal rain). 

Finally, MHD instabilities can also play a role in reconnection (e.g., during flares) and the large-scale destabilization of the solar atmosphere (e.g., the torus instability). This is discussed in \S~\ref{flare_reco} and \S~\ref{trigger}.

\subsection{Dynamics of braiding}
\label{Sect:braiding}

Braiding of magnetic fields, resulting from the interaction between
the convective motions and magnetic fields, and subsequent reconnection
and dissipation of magnetic energy has been proposed as a dominant
coronal heating mechanism, an alternative to Alfv\'en wave
dissipation. 
The theory of braiding, or nano-flare heating, was first proposed by \citet{Levine74} but convincingly elaborated by \citet{Parker83,Parker88} and later developed further by \citet{Cargill94}. In summary, large scale photospheric motions drive the magnetic field to form large gradients in the chromosphere and corona as field lines of varying angles are forced together. These gradients are formed at small scales, eventually causing episodic dissipation at the same small scales. The dissipation events are known as nano-flares as they were first predicted to have a magnitude some $10^{-9}$ of a typical flare. These ideas were tested in early numerical simulations \citep{Galsgaard1996,Hendrix96} which showed that indeed small scale current sheets, the site of episodic dissipation, rapidly form as a result of the forcing of motions in a line tied magnetic field.   
It is not clear how ubiquitous or dominant braiding is in the solar atmosphere,
nor what the detailed properties of this mechanism are. However, several 3D numerical models, also in a realistic solar setting, show that braiding likely occurs as a result of coronal field lines being advected by photospheric flows \citep{Gudiksen2005,Hansteen2015,Rempel2017}.  Braided field lines, even at small angles to each other are shown to lead to episodic nanoflare dissipation of sufficient strength to heat the outer solar atmosphere to coronal temperatures. Thus braiding appears to provide a possible viable answer to the coronal heating problem. The unanswered questions concern the mode of dissipation (as current numerical models cannot reach kinetic dissipation scales) and the importance of heating through braiding as compared to wave or other forms of heating (as these may not be captured well by existing models). 

Several \iris\ results
during the past few years have found evidence that supports braiding
as a significant contributor to the energy balance of the outer
atmosphere. For example, coordinated \iris, \hinodex\ and \hinodee\
observations of moss, the upper transition region footpoints of hot
coronal loops, reveal evidence for small-scale heating events at
transition region heights. The sub-arcsecond spatio-temporal patterns
and center-to-limb variations of flows and non-thermal motions in these events, as derived from the highest resolution \ion{Fe}{xii}
measurements made, are compatible with numerical models of
coronal heating from braiding and appear in conflict with predictions
from Alfv\'en wave heating models \citep{Testa2016}. 

Similar signatures of small-scale reconnection are found in
low-lying cooler active region loops that only reach transition region
temperatures. These signatures are visible in the
sub-arcsecond resolution \iris\ \ion{Si}{iv} spectra as strong
brightenings with anomalously broadened profiles and are associated
with the formation of loops \citep{Huang2015,Huang2017}. Using \iris\ and \aia\ observations \citet{Bahauddin2020} suggest that brightenings in \ion{Si}{iv} and \ion{O}{iv} are consistent with magnetic-reconnection-mediated
impulsive heating at field-line braiding sites. The observations suggest that the impulsive heating occurs as a consequence of magnetic reconnection in multi-stranded transition-region loops and that the line profiles indicate the possible importance of ion cyclotron turbulence caused by strong currents at the reconnection sites. However, the large velocities (close to Alfv\'enic) found suggest that it is large angle reconnection that may be occurring, rather than the smaller angle ``component'' reconnection predicted by braiding theory and simulations. Further studies are needed to determine whether the heating is caused by reconnection as a result of flux emergence or as a result of braiding.

\citet{Pontin2020} investigate why the non-thermal broadening in most transition region and coronal lines is of the order of 15–30~km~s$^{-1}$, this seemingly independent of the instrument resolution \citep{De-Pontieu2015}, and further, why there is a correlation of line intensity and the non-thermal broadening \citep{Testa2016}, and why the line profiles are non-Gaussian with enhanced power in their wings. The authors construct a 3D numerical model of a coronal loop and suggest that line broadening with these characteristics is caused by the turbulent decay of an initially-braided magnetic field, and thus is evidence that braiding is a strong actor in heating the outer solar atmosphere. They do find that the line broadening is more pronounced perpendicular to the magnetic field than parallel to it, but claim that this finding may not hold up under coronal conditions where the perpendicular component of the field is expected to fall off more slowly than the parallel component with height. In either case, simultaneous observations of similar spectral lines with \iris\ and \solos\ from different vantage points would provide unique constraints on such models. Future studies of the properties of line broadening in various magnetic field topologies, and in active regions of varying ages, plage and quiet Sun at different locations on the solar disk, and comparisons of these with the synthetic observables derived from numerical models will clarify this issue further and constrain the importance of the braiding mechanism. 
\begin{figure}
    \centering
    \includegraphics[width=0.95\textwidth]{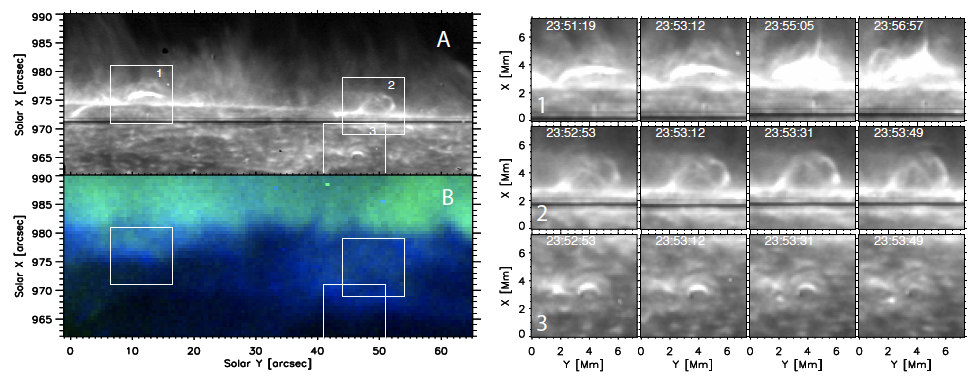}
    \caption{\iris\ \ion{Si}{iv} slit jaw images reveal highly dynamic, low-lying loops at transition region temperatures (adapted from \citet{Hansteen14}). (A) UFS-loops on the western solar limb. (B) the same field of view is shown, but with SDO/AIA images: the coronal 17.1~nm (blue) and 19.3~nm (green) filters, note that the UFS-loops are hidden by the high opacity of cool gas at these wavelengths. The loops are rapidly evolving as shown in three regions of interest in the small panels. Such loops are presumably heated episodically by braiding induced or flux emergence induced reconnection. In active regions low-lying loops are also common, connecting nearby plage regions of opposite polarity \citep{Huang2015}. These types of loops provide a unique sub-arcsecond resolution view of features that may have similar heating mechanism(s) to (some) coronal loops.}
    \label{fig:ufs}
\end{figure}

Low lying transition region loops are likely related to the so-called ``unresolved fine structure'' (UFS) loops (see Figure~\ref{fig:ufs}) whose existence had been
predicted from Skylab observations and which were discovered to be
ubiquitous in \iris\ slit-jaw images of quiet Sun regions
\citep{Hansteen2014}. Statistical studies of their properties and
comparisons with hydrodynamic simulations have now revealed that these
loops indeed appear to be resolved by \iris\ \citep{Brooks2016}. The
loops show collective behavior on cross-field spatial scales that are
of order 300 km, much larger than the spatial scales predicted from
theoretical models of magnetic energy dissipation in the solar
atmosphere. \citet{Brooks2016} suggest that this spatial scale could be caused by the
switch-on nature of the ``secondary instability'', one of the proposed
physical mechanisms involved in the dissipation of currents driven by
braiding. \citet{Pereira2015} suggest that flux emergence may play a role in forming some of these loops, although it remains unclear to what extent this is the case. As a result, it is not clear whether such low-lying loops are a good proxy for how coronal loops are heated or whether their low-lying nature implies a larger role for flux emergence than for typical loops in the solar atmosphere. Statistical studies of the properties of these loops and their relationship to the magnetic field could help address this open question.



\begin{figure*}[tp]


\centering
\includegraphics[width=0.95\textwidth]{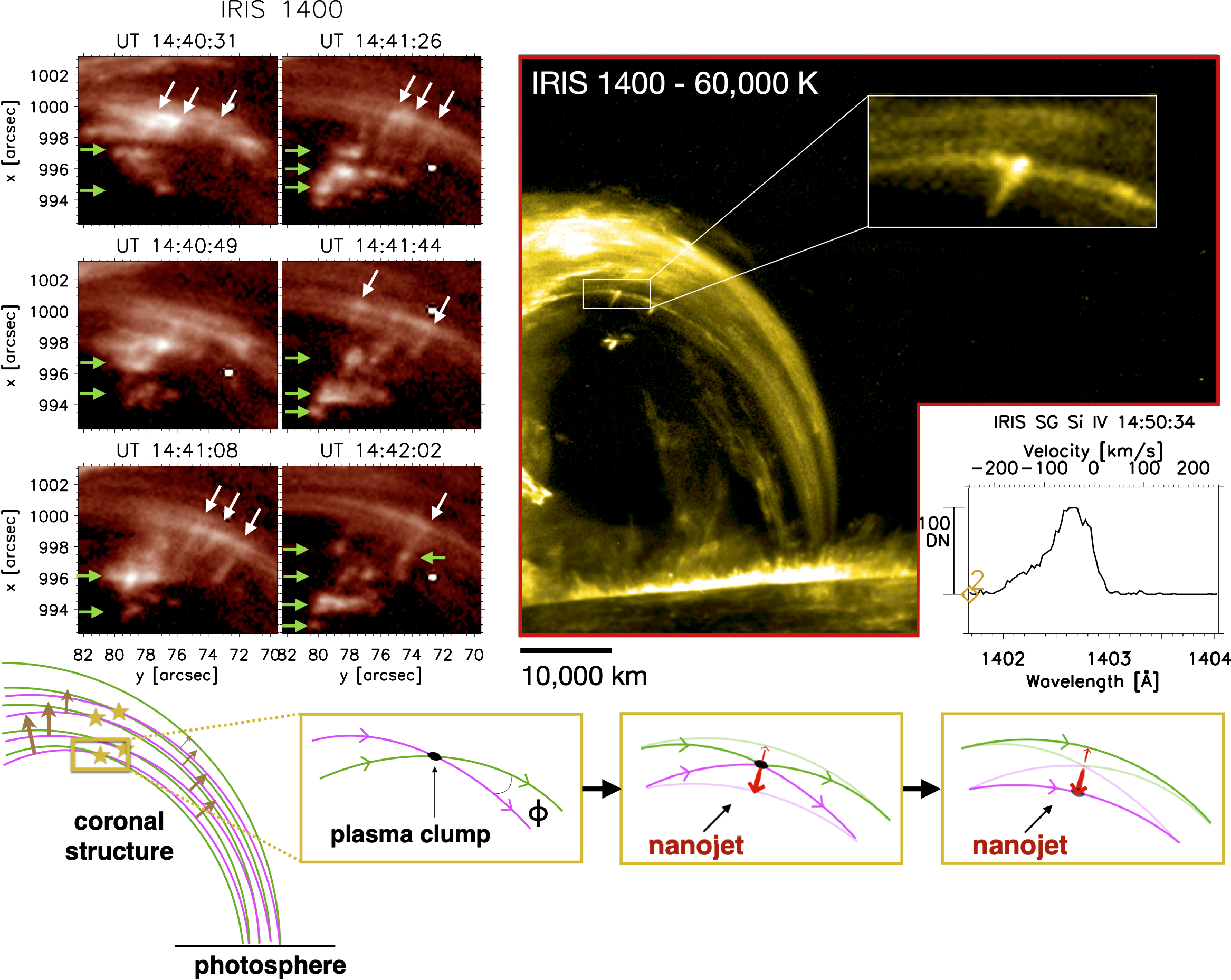}
     \caption{\small Examples of {\em nanojets} -- a  smoking gun of reconnection-based coronal heating -- as observed by \iris\ (adapted from \citealt{Antolin2021}; their Fig.2a, Fig.4d, and Fig.6). {\em Top left panels:} Time series of \iris\ 1400 SJI images showing a cluster of nanojets. {\em Top right panels:} Example of \iris\ \ion{Si}{iv} (bottom) spectra of the nanojets. {\em Bottom panels:} Sketch of the  reconnection process leading to a nanojet: Loop-like structures observed in the POS undergo a slow upward expansion (brown arrows); the stars denote the nanoflares, for which the evolution is shown in the other panels. Small misalignments between the green and magenta field lines lead to reconnection at small angles, and plasma is heated and advected transversely to the loop at large speeds due to magnetic tension, thereby creating the nanojet. The inward component is much larger than the outward component (red arrows) due to the curved topology of the field lines, thus leading to the singular nature of the jet. A final configuration is reached with reduced braiding and misalignment between field lines at a higher temperature. A coronal strand starts to form.
     }
      \label{fig_antolin2020}
\end{figure*}

Recently, exciting \iris\ results have revealed direct observational
signatures of fast ($> 100$~km/s), short lived ($\sim 15$~s), and small-scale ($<1$ \arcsec) jet-like features (Fig.~\ref{fig_antolin2020}), typically
perpendicular to a cool loop-like structure, and associated with nanoflare-sized heating
events \citep{Antolin2021}. Initially, the loop shows misaligned rain strands and complex rotational motions, compatible with the existence of braiding in the loop. The nanojets precede the formation of coronal strands, leading to a hot coronal loop. Comparison with numerical models suggest
that these are reconnection events predicted from braiding, with spatio-temporal properties characteristic of an MHD avalanche. This is supported by the expansion of the nanojets across and along the loop, their increase in frequency occurrence and the overall reduction in the misalignment of the rain strands over time.
These nanojets represent a key signature of magnetic reconnection leading to coronal heating.
It is not yet clear whether these features are related to the jet-like
threads and visible braiding seen with \iris\ in erupting loops
\citep{Huang2018a} and to other similar small-scale jet-like structures seen in tornadoes with \iris\ \citep{ChenH2017,Chen2020}. On the other hand, visible braiding or twisting at sub-arcsecond scales in non-erupting low-lying loops is not detected in
observations that exploit the high-resolution of \iris\ and of the High-resolution Coronal imager (\hic) 2.1 ($<0.5$ \arcsec; \citealt{Rachmeler2019}), suggesting that field relaxation may prevent build-up of highly braided structures with many twists, consistent with early numerical work on braiding \cite{Galsgaard1996} and compatible with recent MHD braiding models \citep{Peter2021}. 

Coordinated \iris-- \soloe\ observations are required to settle the issue of whether field relaxation is effective in 
hindering the build up of strongly twisted field or whether most braiding causes reconnection 
at small angles.
Despite the many observational clues for braiding in the low solar atmosphere, more extensive statistical
studies are needed and expected to provide key insight into the role
and mode of braiding in the solar atmosphere.

\subsection{Non-thermal particle acceleration and chromospheric response to coronal nanoflares}
\label{nanoflares}

Diagnosing physical processes in the corona, such as reconnection
or particle acceleration, can be extremely difficult because of the
optically thin nature of the corona and the long line-of-sight
integration. The transition region and chromosphere form 
boundaries to the corona and do not suffer as much from these
drawbacks. They can therefore be used to diagnose coronal heating properties by observing the impact of small heating events (nanoflares) on the lower atmosphere.

Recent \iris\ observations have demonstrated the potential for probing
the properties of coronal nanoflares even while only observing the chromospheric and transition response to such small-scale heating events \citep{Testa2014, Testa2020}. 
Coordinated \iris\ and \aia\ observations detected small-scale brightenings on timescales of $\sim 10-30$ seconds in chromospheric and transition region emission, at the footpoints of the hottest ($\sim 5$ MK) coronal loops in an active region (as observed with \aia). 
The \aia\ and \hinode\ coronal observations of these transient hot loops suggest that they are likely caused by relatively large angle ($\gtrsim 20^{\circ}$) reconnection \citep{Reale2019a,Testa2020b}, yielding short-lived ($<60$s) heating events leading to chromospheric and TR response.
Such rapid brightenings were previously seen with the \hic\ coronal imaging rocket
experiment, 
but the exact physical mechanism remained elusive \citep{Testa2013}. 
The addition of \iris\ spectra showed \ion{Si}{iv} blueshifted emission for several of these brightenings  \citep{Testa2014}. Comparisons with advanced radiation hydrodynamic modeling with RADYN, showed that such upflows are not compatible with energy release at coronal heights followed by thermalization and thermal conduction to the footpoints \citep{Testa2014,Polito2018}. Instead the observations are naturally explained by the generation of non-thermal electrons in the corona, which are thermalized when the beams reach the chromosphere. The wide thermal coverage of \iris\ and \aia\ combined with the \iris\ spectra appear to provide strict constraints for the properties of the electron beams, especially the low energy cutoff.

Hard X-ray observations are directly observing emission from non-thermal particles and are generally considered the measurement of choice for investigating accelerated particles. However, these new diagnostics with \iris\ are of particular interest for several reasons. First, they observe smaller (nano-flare sized) events compared to hard X-ray observations, which typically detect micro flares and larger flares \citep[e.g.,][]{ Christe2008, Hannah2011}. New observations of small events with \nustar\ are possible \citep[e.g.,][]{Glesener2020}, but still few and challenging, and derive NTE properties in agreement with the \iris\ diagnostics of \cite{Testa2014,Testa2020}. In addition, the \iris\ observations more easily constrain the important low-energy cutoff parameter, which is difficult to derive from hard X-ray observations because of the typical dominance of the thermal emission at lower energies.

\begin{figure*}[tp]
\centering
\includegraphics[width=1\textwidth]{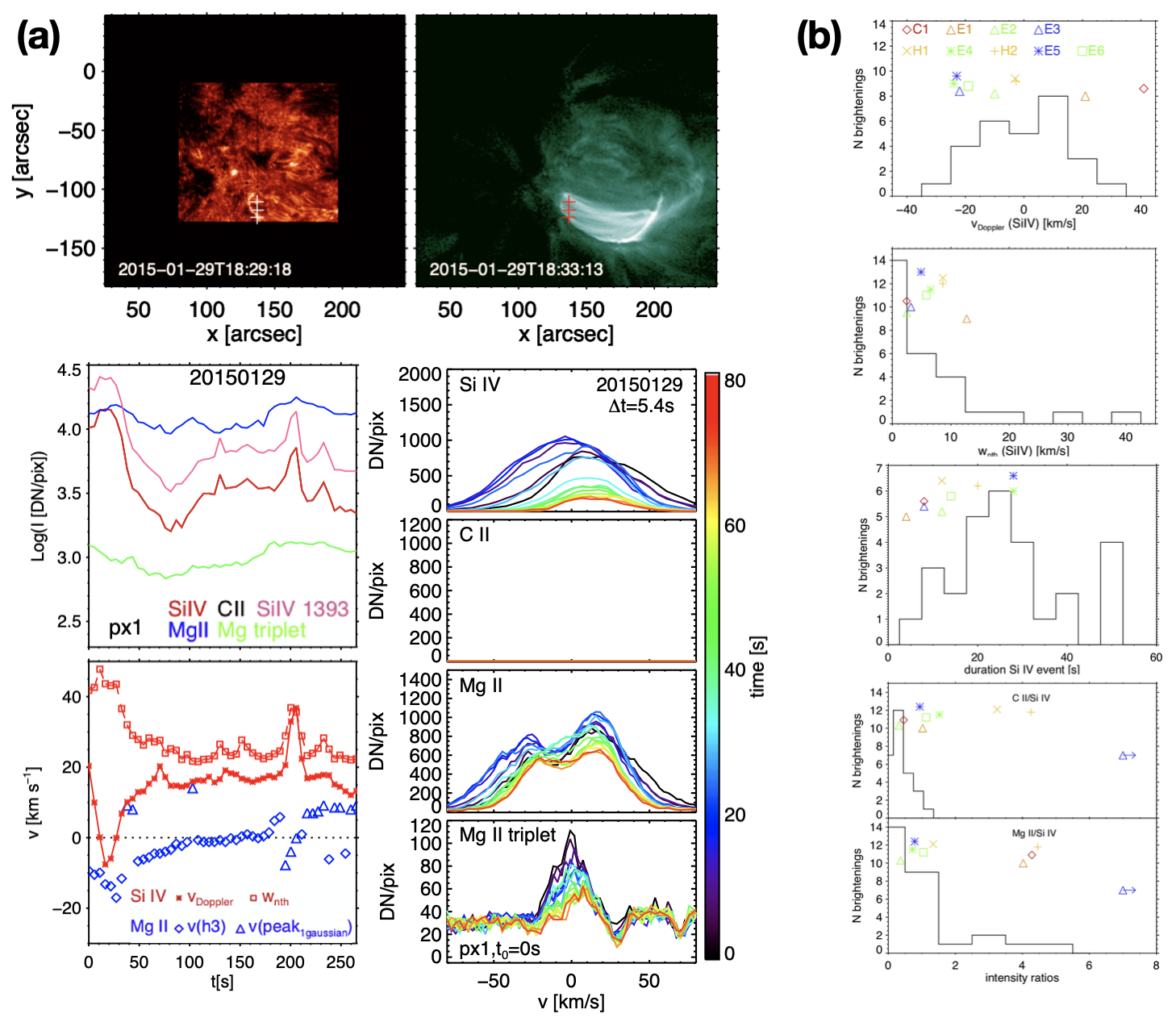}
     \caption{\small \iris\ observations of rapid variability ($\lesssim 60$~s) at the footpoints of transient hot coronal loops in active region cores  (adapted from \citealt{Testa2020}) provide spectral diagnostics of the properties of coronal heating and energy transport (thermal conduction or non-thermal electrons, NTEs).
     {\em Left panels (a):} \iris\  1400 SJI and \aia\ 94\AA\ images (top panels) showing the footpoint brightenings and the hot ($\sim 8-10$~MK) coronal emission respectively; temporal evolution of intensity in various \iris\ spectral lines (middle left),  and of the spectral properties of \ion{Si}{iv} and \ion{Mg}{ii} (bottom left); time series of \iris\ spectral observations during the brightenings (bottom right). 
     {\em Right panels (b):} Histograms of observed \iris\ properties of footpoint brightenings (from top): \ion{Si}{iv} Doppler velocity, \ion{Si}{iv} non-thermal line broadening, duration of the moss brightening in \ion{Si}{iv}, and ratio of intensity maxima for \ion{C}{ii}/\ion{Si}{iv} and \ion{Mg}{ii}/\ion{Si}{iv}. The colored symbols show the corresponding predictions of different RADYN models including heating exclusively by thermal conduction (C1), by a combination of thermal conduction and NTE (H1, H2), or by NTE only (E1-E6 with increasing energy [5, 10, 15~keV] from orange to green to blue; see \citealt{Testa2020} for details).
     }
      \label{fig_beams}
\end{figure*}

A follow-up statistical study to \citet{Testa2014}, including 10 different datasets, and a large grid of numerical models exploring a broad region of the parameter space, pointed  to an additional diagnostic of the presence of non-thermal particles, namely emission in the \ion{Mg}{ii} triplet (around $\sim 2798.8$\AA) observed by \iris, as well as to more stringent constraints on the NTE distributions (Fig.~\ref{fig_beams}) derived by using both chromospheric and TR diagnostics \citep{Testa2020}.  This study reveals that NTE appear to be present in a significant fraction ($\sim$ half) of the studied small heating events, including the smallest ones.
These findings not only provide constraints on particle acceleration mechanisms, but also highlight the presence of nanoflare-sized heating events in the corona and show that non-thermal particles play a role in heating at least the hottest coronal loops in AR cores. These results also show that \iris\ chromospheric and TR observations offer the opportunity to investigate the physics of electron beams with higher sensitivity, i.e., for smaller events than often possible with hard X-ray observations from instruments such as \rhessi\ and NuSTAR \citep[e.g.,][]{Testa2014,Hannah2016,Wright2017}.
Significant additional progress in the study of NTE outside large flares can be made with future investigations for instance by carrying out larger statistical studies of such heating events, and by combining \iris\ with additional chromospheric and/or hard X-ray coordinated observations to further pin down the properties of NTE and heating distributions.

Non-thermal electrons also appear to leave their mark in small-scale jets. \citet{Innes2015} find evidence for apparently super-Alfv\'enic ``jets'' in \iris\ slitjaw images that protrude from small-scale reconnection events. They suggest a scenario in which reconnection events (observed as \ion{Si}{iv} explosive
events) generate electron beams that propagate away from the source at 
speeds of thousands of km~s$^{-1}$, while the weak interaction with the
background loop plasma leads to thermalization of a fraction of the
electrons, thereby forming linear brightenings in the \iris\ slitjaw
images. These intriguing results clearly require more modeling and
analysis during the next few years.
Furthermore, \iris\ TR lines emitted by transition region ions in active regions often show evidence of significant wing emission, departing from a single Gaussian shape, and they  can be well fitted with non-Maxwellian k-distributions \citep{Dudik2017}. These non-Maxwellian distributions exhibit significant power-law tails at high velocities or energies, and can be caused by a variety of processes, including acceleration of particles due to magnetic reconnection, shocks, wave-particle interactions, or plasma turbulence. Their study, and modeling, can provide crucial information on the physical processes at work during flares and active region heating events.
Numerical modeling of such events also highlights the
diagnostic potential of \iris\ to detect effects of non-equilibrium
ionization 
\citep{Bradshaw2019}, or departures from Maxwellian distribution 
\citep{Dzifcakova2018}, key to a proper
interpretation of \iris\ data.

\subsection{Magnetic reconnection in small-scale events}
\label{reconnection}
Reconnection of magnetic field lines is thought to drive a variety of
energetic events, e.g., flares or CMEs, but also much smaller events further
down in the atmosphere. Reconnection occurring during flares is discussed in \S~\ref{flare_reco}. Here we focus on reconnection in smaller events occurring in the lower solar atmosphere. UV bursts, sudden and compact brightenings in UV light, provide an opportunity to study magnetic reconnection on the Sun \citep{Young2018}. A variety of phenomena such as transition region explosive events and \iris\ bombs are considered to be UV bursts. These events provide an opportunity for studying fundamental aspects of reconnection and its
impact on the solar atmosphere. 

%

\subsubsection{UV Bursts} 
\label{burst}
\begin{figure}
    \centering
    \includegraphics[width=0.95\textwidth]{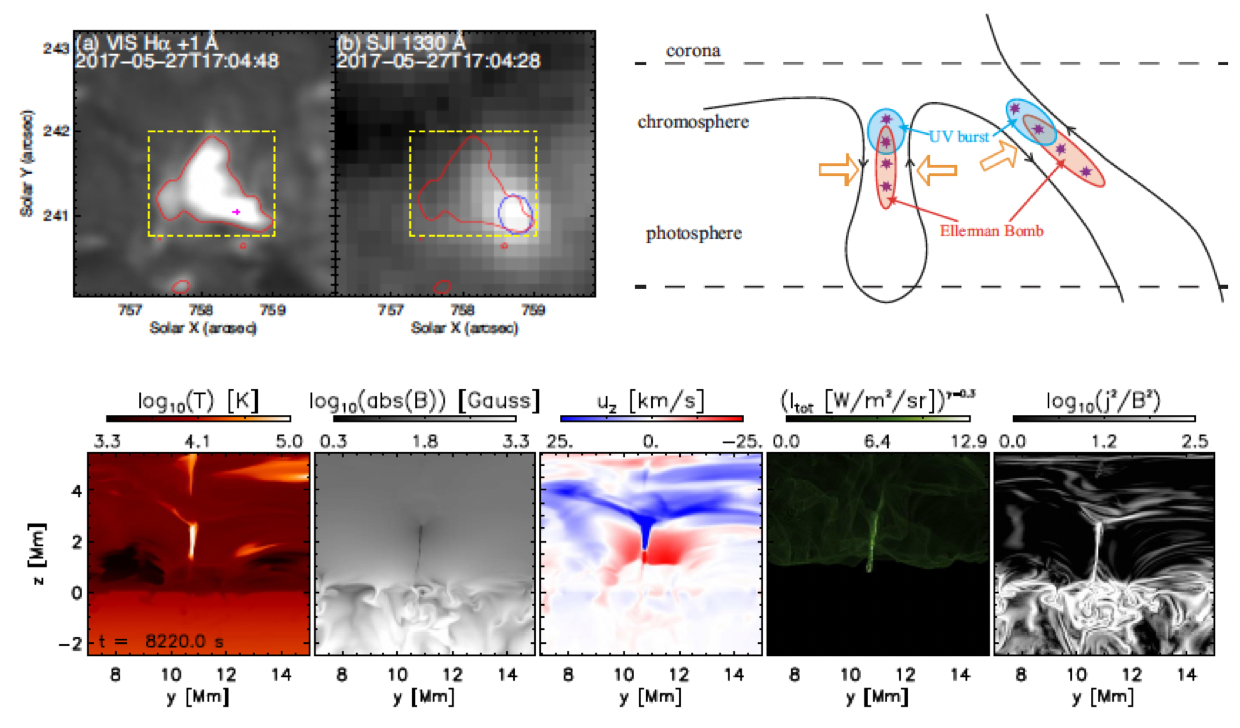}
    \caption{Simultaneous observations in the red wing of the H$\alpha$ line made at the {\em Goode Solar Telescope} and in \iris\ \ion{Si}{iv} slit-jaw images near the solar limb show a displacement (upper left panels) that is consistent with Ellerman Bombs and UV bursts forming at different heights along the same current sheet (upper right panel, adapted from \citet{Chen2019b}.) This scenario is consistent with numerical models of flux emergence in which several magnetic ``bubbles'' pierce the photosphere and reconnect along nearly the vertical current sheets formed by their expansion and interaction (lower panels, adapted from \citet{Hansteen2019}) }
    \label{fig:EBUVburst}
\end{figure}
Recent \iris\ results have capitalized
on the discovery of \iris\ bombs, or now more commonly ``UV bursts'' \citep{Peter2014}. 
UV bursts are short-lived \ion{Si}{iv} events driven by convergence and cancellation of opposite polarity flux \citep{Tian2018b, Wu2019}, in  which plasma is accelerated to $>100$ km/s and heated to what appear to be TR temperatures at low chromospheric heights.
These ``bombs'' or ``bursts'' are related to phenomena named ``explosive events'' as reported by earlier instruments such as High Resolution Telescope and Spectrograph (HRTS) \citep{Dere1989} or SOHO/SUMER \citep{Innes1997}.
Statistical studies using \iris\ and ground-based telescopes (e.g., SST and NST), show that some (but not all) UV-bursts appear to be closely related to the well-known Ellerman bombs \citep[e.g.,][]{Vissers2015b, Kim2015, Tian2016, Grubecka2016, Ortiz2020}. While at the same time some 20\% of observed Ellerman bombs have associated UV-bursts. Coordinated \iris\ and
ground-based observations show that the \iris\ bombs are associated with reconnection between granular scale oppositely directed field in the photosphere \citep{Gupta2015, Kim2015}, e.g., triggered by flux emergence \citep{Toriumi2017}; similar conditions to those found in Ellerman
bombs.

An analysis of a variety of 
spectral diagnostics shows that the hot UV-burst plasma is emitted at very high 
 densities \citep[as evidenced by
self-absorption in \ion{Si}{iv} lines, see][]{Yan2015} typical for upper
photospheric or low chromospheric conditions. It is also clear that the
temperatures are at least an order of magnitude higher than 1D non-LTE models for Ellerman bombs can produce \citep{Grubecka2016}. 
This argument is further strengthened by the existence in the \ion{Si}{iv} 1402\AA\
line of a very narrow \ion{Ni}{ii} absorption feature (as well as other lines of \ion{Fe}{ii} typically associated with photospheric temperatures).

Because of the rich diagnositic capabilities of \iris\ from the photosphere to the transition region, UV-bursts provide the most detailed views of reconnection events in the solar atmosphere. Coordinated \iris\ and SST observations show how the overlying canopy of chromospheric
fibrils hides the reconnection site from view in commonly used ground-based diagnostics like the H$\alpha$ core, while \iris\ observables such
as \ion{C}{ii} (upper chromosphere), \ion{Si}{iv}(transition region) and \ion{Mg}{ii} triplet (low chromosphere) lines see through this absorbing forest and unveil the ``bomb'' site \citep{Vissers2015b}. Along with the derived temperatures and densities described above, these direct views revealed
puzzling and hotly debated results about the formation heights and the temperatures to which
plasma is heated in UV bursts. Initially, some argued that the
\ion{Si}{iv} emission with superimposed cool absorption lines
\citep{Gupta2015} is caused by heating to
80,000 K \citep{Peter2014} at photospheric heights. This was strongly contested by \citet[e.g.,][]{Judge2015} based on radiative transfer considerations, or with alternative suggestions of, e.g., 10-20,000~K
temperatures based on LTE ionization \citep{Rutten2016}.

There is now consensus, based on observational evidence and numerical modeling (see Figure~\ref{fig:EBUVburst}) that the Ellerman bomb and UV burst, when they occur together, are formed at the opposite ends of the extended current sheet resulting from the interaction of recently emerged field with itself or with the pre-existing ambient field. The absorption seen in \ion{Ni}{ii} and \ion{Fe}{ii} is caused by cool gas carried high into the chromosphere, above the strongly emitting current sheet, as emerging flux breaks free of the photosphere and expands upwards into the corona \citep{Hansteen2019}.  Coordinated \iris\ and GBO observations have confirmed how Ellerman bombs and UV bursts (Fig.~\ref{fig:EBUVburst}), when they occur 
together, appear to be formed at different heights \citep{Chen2019b}, 
although still being part of the same reconnection system \citep{Ortiz2020}.

A statistical study of the properties of UV bursts (using \iris\ spectra and \shmi\ magnetograms), their formation environments, and the impact on the chromosphere would provide a robust set of
observational constraints for multi-fluid models of reconnection in the partially ionized plasma of the solar chromosphere. Such data will
also provide insight into the geometry and topology of current sheets in areas of newly emerging magnetic flux, a topic of great interest since this flux eventually fills and pervades the solar corona. 
A better understanding of this process can be sought by confronting numerical modeling \citep[e.g.,][]{Hansteen2019} with \iris\ observations, supplemented (when available) with \dkist, \sst, \sdo, \solo\ or \hinode\ data. 

Coordinated \iris\ and GBO data will also be
useful to study the intriguing reports of Ellerman bomb-like features in QS \citep{Nelson2017}. Using the \sst\ \citet{Joshi2020} found evidence for ubiquitous, but weaker, Ellerman bomb like events in the H$\beta$ line with potential implications for the heating of the quiet Sun chromosphere, and \citet{Rouppe2021} find similar events in sunspot penumbrae. This should be followed up with \iris\ studies of chromospheric and transition region observables.

\subsubsection{Magnetic reconnection mechanisms in the low solar atmosphere}

\iris\ observations have led to new results that have improved our understanding of how magnetic reconnection occurs in the solar atmosphere. For example, UV bursts and transition region explosive events have provided insight into the detailed mechanisms involved in facilitating fast reconnection. It has long been known that reconnection in the solar atmosphere, which leads to dynamic events that occur on timescales of seconds or minutes, appears to proceed at a much faster rate than classical reconnection \citep[e.g., Sweet-Parker, see][]{Priest2014} theories predict for solar conditions. 

Recent developments in reconnection theory suggest that under
certain conditions slow reconnection can transition to fast
reconnection when mediated by the plasmoid (or tearing mode)
instability. Evidence for reconnection mediated by plasmoids has now
been found by comparing recent numerical models with \iris\
observations of tell-tale ``triangular-shaped'' spectral line profiles
in transition region explosive events 
\citep{Innes2015}. Such profiles do not match predictions from the alternative fast
(Petschek-type) reconnection. The presence of small-scale plasmoids has been directly confirmed through analysis of coordinated \iris--\sst\ observations of
small-scale reconnection events in the low solar atmosphere and
comparison with numerical models \citep{Rouppe-van-der-Voort2017}. 
 Evidence for this mechanism has also been found with \iris\ in coronal jets
\citep{Zhang2019a}, flux cancellation events \citep[using \iris/\aia, ][]{Yang2018b}, as well as much larger flares to explain propagating sawtooth-like and quasi-periodic flare ribbon structures \citep{Brannon2015,Parker2017}. The reconnection-driven nanojets reported in the corona also exhibit plasmoid ejecta along their axis, particularly for the most energetic nanojet clusters \citep{Antolin2021}.
More indirect but similar constraints come from \iris\
observations of microflares which provide support for fractal-type
reconnection mediated by the secondary tearing instability
\citep{Reep2016}. 

\begin{figure}[tp]
   \includegraphics[width=0.45\textwidth]{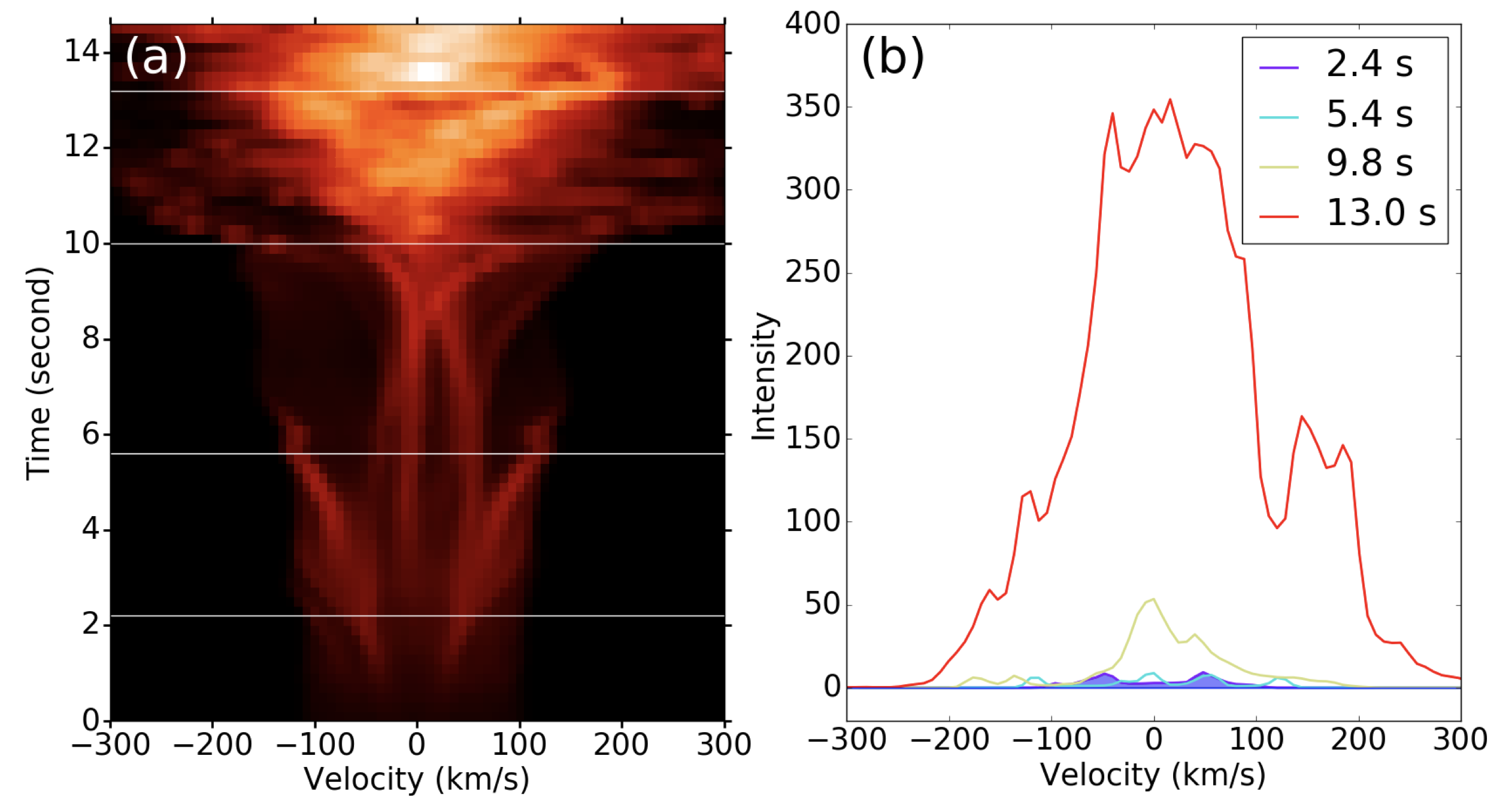}
    \includegraphics[width=0.45\textwidth]{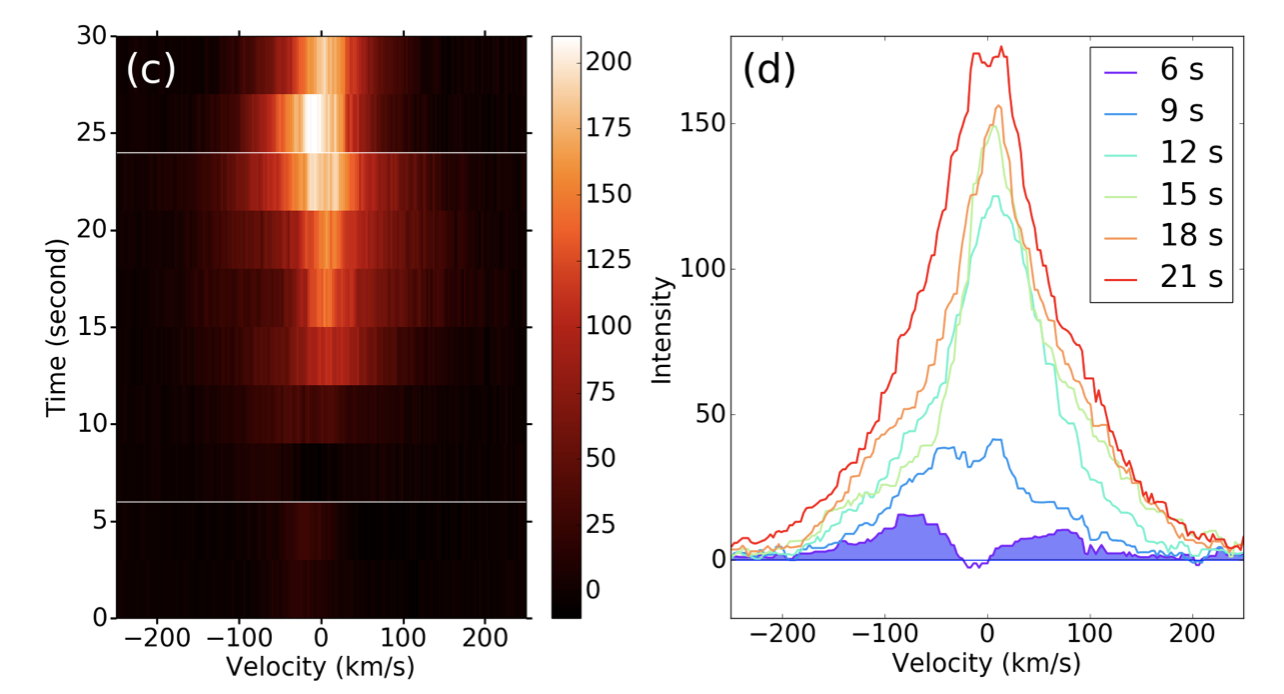}
   \caption{\small High-cadence observations of UV bursts or transition region explosive events have revealed a rapid transition from bi-directional flows to complex, often triangular-shaped profiles (panels c, d), which agree well with numerical simulations of the transition from slow Sweet-Parker like reconnection to fast reconnection mediated by the plasmoid instability (panels a, b) \citep{Guo2020}. }
       \label{fig_guo}
\end{figure}

A very recent interesting finding is the discovery of a transition between slow and fast reconnection in very high cadence \iris\ spectra. Comparison with numerical models shows that the observations are compatible with a transition from classical reconnection to fast reconnection mediated by plasmoids \citep[][ Fig.~\ref{fig_guo}]{Guo2020}. On the other hand, some recent studies suggest that another mechanism, turbulence, appears to drive fast reconnection in some microflares \citep{Chitta2020}. More observations and modeling are required to determine how common both of these mechanisms are and their relative importance. Such observational insights can come from \iris\ observations of the
Ellerman Bomb/UV burst phenomenon which can provide important clues
on the process of magnetic reconnection and the
formation and evolution of the AR magnetic field.
Unlike \saia\ passbands and GBO H$\alpha$ observations, the \iris\ spectral bands are not
opaque to the fibrils that overlie most bursts, allowing clear views of the reconnection process.

Novel insights into the conditions that lead to reconnection have been obtained from so-called ``light walls'' (also called peacock tails) that originate not only from sunspot light bridges \citep{Bharti2015}, but also around neutral lines in and around, sunspots, as evidenced by a statistical study of 6 months of \iris\ data \citep{Hou2016a}. Numerical simulations and \iris\ and \sdo\ observations have revealed the peculiar magnetic field configuration in
light bridges that leads to reconnection \citep{Toriumi2015a,Toriumi2015b}. \iris\
observations have also elucidated the cross-field coherence in dynamic evolution
of these lightwalls, which has been associated with slipping reconnection
\citep{Hou2016d,Bai2019}: the light wall dynamics appear to be intimately tied to and thus
help reveal the overall changes in magnetic topology during a flare. High resolution spatial and temporal data from \iris\ and \sdo\ has been used to confirm the observation of signatures of magnetic reconnection and accompanying flux cancellation \citep{Yang2018a} including two-loop interaction processes, plasma blob ejections, and sheet like structures above the flux cancellation sites. Further studies of the conditions that lead to reconnection using the rich \iris\ data archive are key to go beyond case studies and establish how common these mechanisms are. Coordinated \iris\ and \soloe\ observations of so-called campfires, small-scale and short-lived heating events in the EUV can help identify the role of reconnection in driving these recently discovered events \citep{Berghmans2021}.

\subsection{Effects of flux emergence on the chromosphere and beyond}
\label{flux}

As magnetic fields rise from the photosphere into the low atmosphere,
they interact; driving dynamic events (\S~\ref{reconnection}) and energizing plasma \citep[e.g.,][]{Cheung2014}.
Recent \iris\ results have shed light on the details and
importance of this process. Emergence of magnetic field into the atmosphere occurs on many spatial
scales, from active region size emergence down to small flux elements
that continuously emerge on granular scales. In sum the flux emergence process on all scales fills and replenishes the chromospheric and coronal field, countering the depletion caused by reconnection, losses through the solar wind, and submergence. Both the local and global topology of the emerging field play important roles in determining the heating rate in the outer atmosphere  and the mode of solar wind acceleration. Furthermore, understanding flux emergence will enable a deeper insight into the processes active in driving the solar dynamo. Coordinated \iris, \shmi\ and
\hinodes\ observations have led to new insights into the impact of this
new flux on the energetics of the atmosphere in a variety of regions.
For example, the early stages of AR flux
emergence can be tracked very well through the development of UV
bursts \citep{Tian2018b}, providing insight into how newly emerged flux
interacts with pre-existing fields. Such insights have also come from
coordinated \iris--\aia\ observations showing counterparts of
low-atmospheric UV-bursts in all coronal \aia\ passbands \citep{Guglielmino2019a},
suggesting long-lasting reconnection episodes between emerging and ambient fields. During the later phases of emergence, UV bursts may be less common, with energization higher up in the atmosphere when new loops interact with ambient coronal loops \citep{Huang2019b}.

Coordinated \iris\ and \sst\ observations of small-scale flux emergence
in active regions reveal the general mechanism through which coronal
magnetic fields are added into a new active region \citep{Ortiz2016}. These large scale
fields are built up through successive emergence of small-scale fields
on granular scales. Each emergence of granular scale
fields leads to cool bubbles that slowly rise into the atmosphere,
which leads, after many minutes, to reconnection and upper
chromospheric and TR heating if the incoming fields are not
well-aligned with the pre-existing fields. 

\begin{figure}
    \centering
    \includegraphics[width=0.95\textwidth]{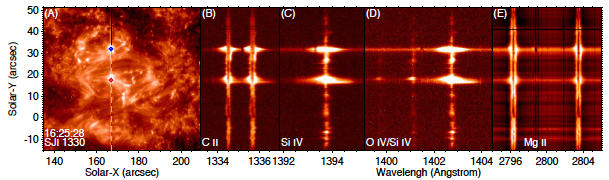}
    \caption{\iris\ 1330 SJI and spectra of the \ion{C}{ii}, \ion{Si}{iv}, \ion{O}{iv}, and \ion{Mg}{ii} h\&k lines in an active region in its earliest emerging phase (adapted from \citet{Tian2018b}). Strong heating and plasma acceleration, {\em i.e} UV bursts, occurs at sites where the emerging field interacts with itself or the pre-existing ambient coronal field.}
    \label{fig:UVburst}
\end{figure}

For example, flux emergence plays a key role in the formation of the
serpentine magnetic fields that drive reconnection events that lead to
strong acceleration and low atmospheric heating that is
visible as Ellerman or \iris\ bombs (see \S~\ref{burst} for more
details and Figure~\ref{fig:UVburst} for an example). \citet{Toriumi2017} show that flux emergence can also lead
to strong TR downflows along the arch filament systems formed as a
result of emergence. These supersonic downflows lead to shocks when
they hit the chromosphere, leading to significant heating (visible
in the \ion{Mg}{ii} triplet lines).


{\color{red}{

}}

Weak magnetic fields continuously emerge in the quiet Sun internetwork, and are
considered a potential contributor to the heating of the QS
chromosphere. \citet{Gosic2018} used \iris\ and \sst\ observations and
found that cancellations of such elements are unlikely to play
a significant role in bulk chromospheric heating at the
sensitivity levels of high cadence \sst\ measurements, although
coordination with deeper magnetograms from \sst\ or
\dkist\ are required to definitively settle this issue. On the other hand, the flux emerging in small magnetic loops may play an important role: Coordinated, high-resolution, multiwavelength observations obtained with the \sst\ and \iris\ 
of internetwork magnetic loops as they emerge into the photosphere reaching the chromosphere and transition region have recently been studied by \citet{Gosic2021}. In this case study, the footpoints of the emerging internetwork bipoles are clearly seen to appear
in the photosphere and to rise up through the solar atmosphere. Polarimetric measurements taken in the chromospheric \ion{Ca}{ii} 854.2~nm line provide direct observational evidence that internetwork fields are capable of reaching the chromosphere.
The concurrent \iris\ data show the effects of these weak fields on the local heating of the chromosphere and transition region. This work dovetails neatly with the discovery of H$\beta$ ``Ellerman bombs'' \citep{Joshi2020}. Future coordinated \iris\ and \sot, \sst\ or \dkist\ observations are needed still to determine whether emergence into the chromosphere plays an important role at a global level, i.e., whether the spatio-temporal filling factor of these events is large enough to impact the average energy balance of the internetwork chromosphere.

Flux emergence of course also plays a major role in driving flares and CMEs, the main exponents of space weather, as described in \S~\ref{trigger}.

\subsection{Energy and mass transfer between photosphere, chromosphere and corona}
\label{transfer}

All coronal plasma has its origins in the lower solar atmosphere, with
the coronal mass budget driven by a balance between upward mass
transport through, e.g., evaporative or eruptive flows, and the downward
mass transport through, e.g., coronal rain or gentle cooling
downflows. 
Although ultimately the coronal mass escapes into the heliosphere as 
the solar wind and to a lesser extent ($\sim$1\%) as CMEs or coronal jets, 
a fraction of coronal plasma returns to the chromosphere through downflows, forming a \emph{chromosphere\,--\,corona mass cycle}
\citep{Marsch2008, Berger2011, McIntosh2012}. 
Many processes appear to be involved and the outstanding
challenge is to determine which processes dominate in the various
solar regions (quiet Sun, coronal hole, active region, sunspots). 

\subsubsection{Solar Jets and Surges}
\label{jets}

The solar atmosphere exhibits a wide variety of ejective phenomena, ranging from coronal jets, at the largest scales, down to penumbral microjets, at subarcsecond scales. Observed in virtually all electromagnetic ranges, these ejective phenomena are of broad interest because of their implication in the energy and mass balance of the chromosphere, transition region, and corona. 

\begin{figure*}[t]
\centering
\includegraphics[width=0.6\textwidth]{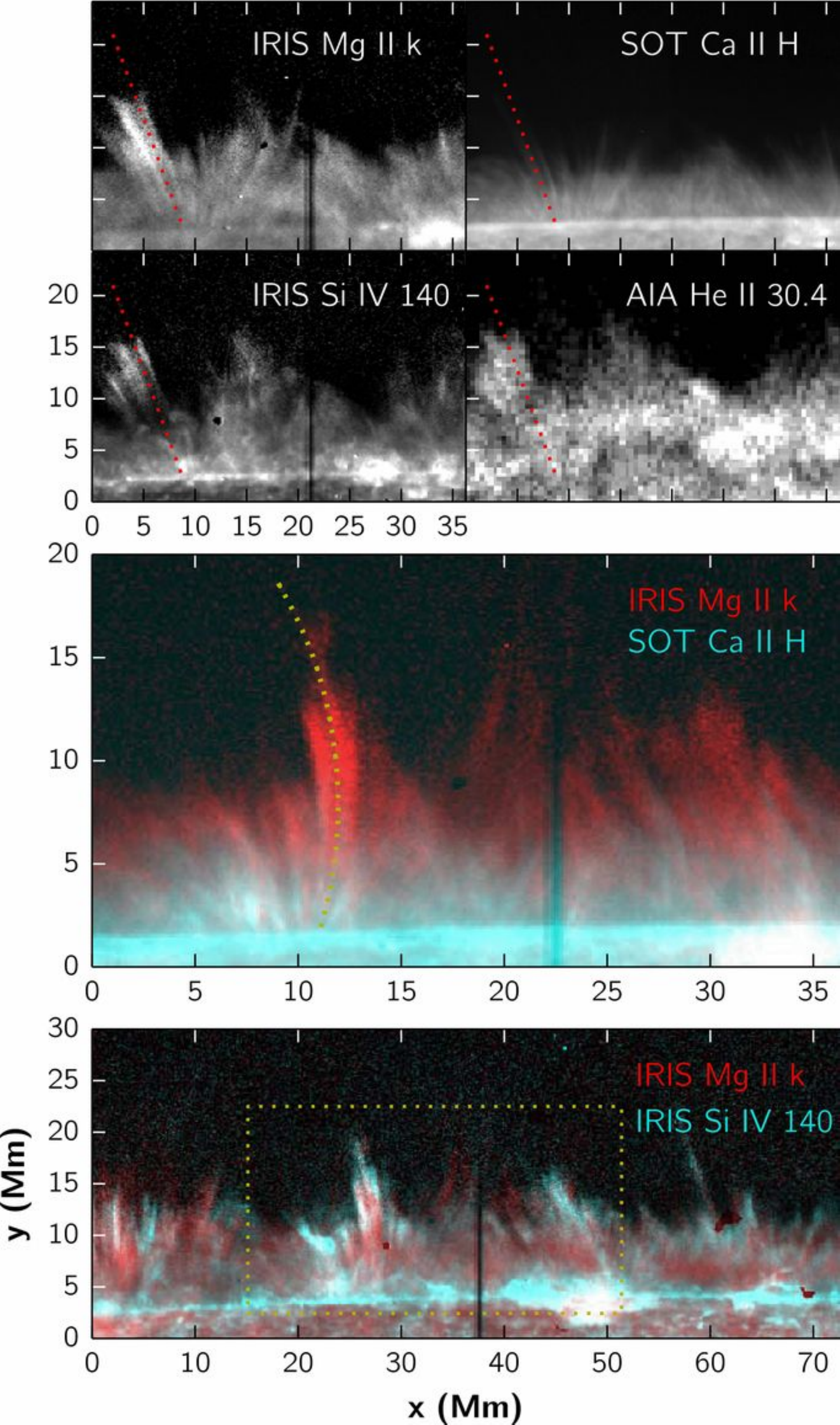}
\caption{Quiet-Sun spicule images extracted from \citet{Pereira2014a}. Top panel: two of the \iris\ slit-jaw filtergrams, SOT \ion{Ca}{ii} H filtergrams, and the AIA \ion{He}{II} 30.4 nm channel. Middle panel: color
composite of observations, where the red channel corresponds
to the IRIS \ion{Mg}{ii} k filtergram and the green and blue channels (cyan color) to
the SOT Ca ii H filtergram. Bottom panel: composite
image using the IRIS \ion{Mg}{ii} k and \ion{Si}{iv} filtergrams, for the same instant as the
middle panel and a larger field. The dotted rectangle denotes the field-of-view
of the middle panel.} 
\label{fig:spicules} 
\end{figure*}

The major improvements in observational and computational capabilities over the past decades have yielded new insight into the still open questions related to formation of solar ejections.
A representative example are spicules: the most ubiquitous jets in the solar atmosphere. Since their discovery, a huge effort has been performed to understand the generation mechanisms of spicules and their role to heat the solar corona \citep[see, e.g.,][]{Beckers1968, Hollweg1982, Sterling2000, DePontieu2004, Hansteen2006, Rouppe2009, Pereira2012}. Concerning the role of \iris\ in this progress,
\citet{Pereira2014a} exploited the broad thermal coverage provided by coordinated \iris\ and \hinode\ observations (see Figure \ref{fig:spicules}) to determine why spicules often fade rapidly from chromospheric passbands. They found that a significant fraction of chromospheric spicules undergo strong heating to at least transition region temperatures, a fact that was also found by \citet{Skogsrud2015} through \iris\ and \hinode\ data and \cite{Rouppe2015} with \iris\ and \sst\ measurements on the solar disk. These papers suggest a spicule model that consists of multiple threads with different heating. Some threads reach TR temperature while other remain at chromospheric temperatures. These observations also provide key constraints for theoretical models and considerable numerical effort has been carried out to unravel the spicule formation. For example, \cite{Martinez-Sykora2017a,Martinez-Sykora2017b} concluded that heating of spicular plasma to TR temperatures can be reproduced by their model, where ambipolar diffusion is a key mechanism in both the formation and heating of spicules. Later, it was shown that, when including non-equilibrium effects of hydrogen and helium, spicules have larger temperature variations and that ambipolar diffusion primarily acts in shock fronts and hot regions within the spicules \citep{Martinez-Sykora2020a}.  
Furthermore, \citet{Antolin2018a} used coordinated \iris\ and \sot\ observations and identified various types of Alfvénic waves with specific phase relationships in time and space, as well as a multi-stranded structure. These observations were found to be compatible with numerical models of transverse wave-induced Kelvin-Helmholtz rolls (see also \S~\ref{alfven_waves}) along spicules. More sophisticated modeling that includes both the dynamic formation of spicules, ambipolar diffusion and complex wave phenomena are key to further investigate the heating mechanism(s) in spicules. 

Other avenues have been recently opened to further explore spicules. For instance, combining \iris\ and radio observations obtained with \alma, \citet{Yokoyama2018} and \citet{Shimojo2020} showed that \alma\  spicular-like observations correspond well with the \iris\ counterpart, which is key for the diagnostics and interpretation of \alma\ data. Furthermore, theoretical support for such coordinated observations has been also provided by synthetic observables from numerical experiments \citep[e.g.,][]{Martinez-Sykora2020b,Chintzoglou2021b}. Another example is the use of machine learning techniques, where \cite{Bose2019b} applied \textit{k}-means to \sst\ \ion{Ca}{ii} and H$\alpha$ data to characterize the chromospheric spectral profiles of on-disk spicules, thereby leading to their unambiguous identification in \ion{Mg}{ii} k.

\iris\ has also been crucial to study other phenomena linked to spicules such as
propagating disturbances and network jets. For the former, \iris\ and \saia\ observations indicate that spicules appear to be the root of some coronal propagating disturbances \citep{Pant2015,Samanta2015b,Bryans2016}; however it remains unclear whether these propagating disturbances are caused only by waves or whether flows of heated material are also injected into the corona during such events. Concerning network jets, it has been argued that they are the TR counterpart of spicules \citep{Tian2014c,Narang2016,Chen2019a,Qi2019} with fast apparent motions ($>100$ km s$^{-1}$); nonetheless the combination of models and observations \citep[][]{De-Pontieu2017b,Chintzoglou2018} suggest that many of these jets may actually be rapidly propagating heating fronts along spicules, rather than caused by mass motions. Further modeling and observations are required to settle these issues, and thereby address the energy and mass flux associated with these events.

In addition to the step forward in the understanding of the spicules
and related events, \iris\ observations have revealed significant TR
emission associated with phenomena that were traditionally related to
the chromosphere such as surges \citep[][among others]{Yokoyama1995,Canfield1996, Chae1999,Brooks2007,Guglielmino2010}. By means of flux emergence experiments carried out with the Bifrost code, \citet{Nobrega-Siverio2016} showed that there is a significant part of the surge plasma that can be efficiently heated up to TR and even coronal temperatures, and then rapidly cools  back to chromospheric temperatures. This result implies that, during the ejection of surges, it is possible to detect TR emission. Evidence of such emission was independently found in the \iris\ \ion{Si}{iv} lines by \citet{Nobrega-Siverio2017} and \citet{Guglielmino2019a}. Other examples of chromospheric ejections with associated TR emission observed with \iris\ are fan-shaped jets originating from light bridges \citep{Yang2015,Bharti2015,Bai2019}, and penumbral microjets \citep{Vissers2015a,Humphries2020}, although penumbral microjets are not genuine jets with significant mass motions, but rather manifestations of heat fronts \citep{Esteban-Pozuelo2019,Buehler2019,Rouppe-van-der-Voort2019,Drews2020}. In order to properly understand the TR emission of the aforementioned phenomena, one has to take the highly dynamic nature of the chromosphere into account as well as the fact that some elements in the TR have long ionization and recombination timescales, leading to relevant non-equilibrium ionization effects \cite[e.g.,][]{Olluri2013,Olluri2015,DePontieu2015,Martinez-Sykora2016a}. For instance, in the particular case of surges, \cite{Nobrega-Siverio2018} showed that the sheath of these ejections suffers heating and cooling processes in short timescales due to the efficient action of mechanisms such as Joule heating, thermal conduction or optically thin losses, thus producing significant departures from the statistical ionization equilibrium in \ion{Si}{iv} and \ion{O}{iv} ions. The non-equilibrium ionization effects could moreover be a natural explanation for some of the puzzling correlations between TR intensity and non-thermal line broadening that had been discovered with Skylab in the 1970s.

Within the catalog of ejective phenomena, coronal jets have also received a noticeable development thanks to \iris\ observations. For instance, \citet{Cheung2015} used \iris\ and \sdo\ data to study recurrent coronal jets, which until now have been difficult to explain. They assimilated \shmi\ magnetograms into a data-driven simplified MHD model with the aim of comparing with \iris\ images and spectra. Thus, they were able to quantify the helical nature of the jets and found that emergence of current-carrying magnetic field supplies sufficient magnetic twist to drive recurrent helical jets.
 Signatures of transfer of twist to jets were also explored more recently by \cite{Liu2018} and \cite{JoshiR2020b}. \iris\ has also been key to studying the multi-temperature components of coronal jets \citep{Mulay2017, Guglielmino2018,Lu2019,Cai2019,Ruan2019a}; analyzing their fan-spine topology  \citep{Jiang2015}; obtaining evidences of Kelvin-Helmholtz instabilities associated with blowout coronal jets \citep{Li2018nat}; and
analyzing the subarsecond structure of multiple blobs, possibly produced by the tearing-mode instability, that are ejected together with coronal jets \citep{Zhang2019a}. In addition, through \iris\ and \sdo\ observations, \cite{JoshiR2020} found that the characteristics of the observed surges and kernels (plasmoids) that accompany the coronal jets, as well as the detected double-chambered structures show striking similarities with numerical jet models \citep[see, e.g.,][]{Moreno-Insertis2013,Nobrega-Siverio2016}.

\subsubsection{Coronal Rain}
\label{section:rain}

Coronal rain is a phenomenon that was initially considered only as a specific kind of prominence \citep{Levine1977} and it was not until the advent of high resolution instrumentation and multi-temperature coverage of the transition from chromosphere to corona that its unique characteristics were recognised. \iris\  played a fundamental role in this process, marking the unique links to multiple fundamental plasma processes. 

Coronal rain corresponds to cool and dense, clumpy plasma appearing seemingly out of nowhere in the solar corona on a timescale of minutes, and falling back towards the solar surface along coronal loops \citep{Antolin_2020PPCF...62a4016A}. It is a catastrophic cooling phenomenon from coronal to chromospheric temperatures and can thus, in a way, be considered as the opposite of coronal heating. This fascinating cooling problem, representing somewhat `snowflakes in the oven', holds many unsolved mysteries upon which \iris\ has shed new light.

The leading explanation for coronal rain is thermal instability in the solar atmosphere, which can occur when radiative losses locally overcome the heating gains. Due to the shape of the radiative loss function with respect to temperature, runaway cooling occurs, which can lead to the growth of unstable thermal modes (also known as entropy modes), that produce a local loss of pressure and a subsequent accretion of material to form a condensation \citep{Field_1965ApJ...142..531F,Mueller2004, Claes_Keppens_AA624_2019}. 

\begin{figure*}[t]
\centering
\includegraphics[width=1\textwidth]{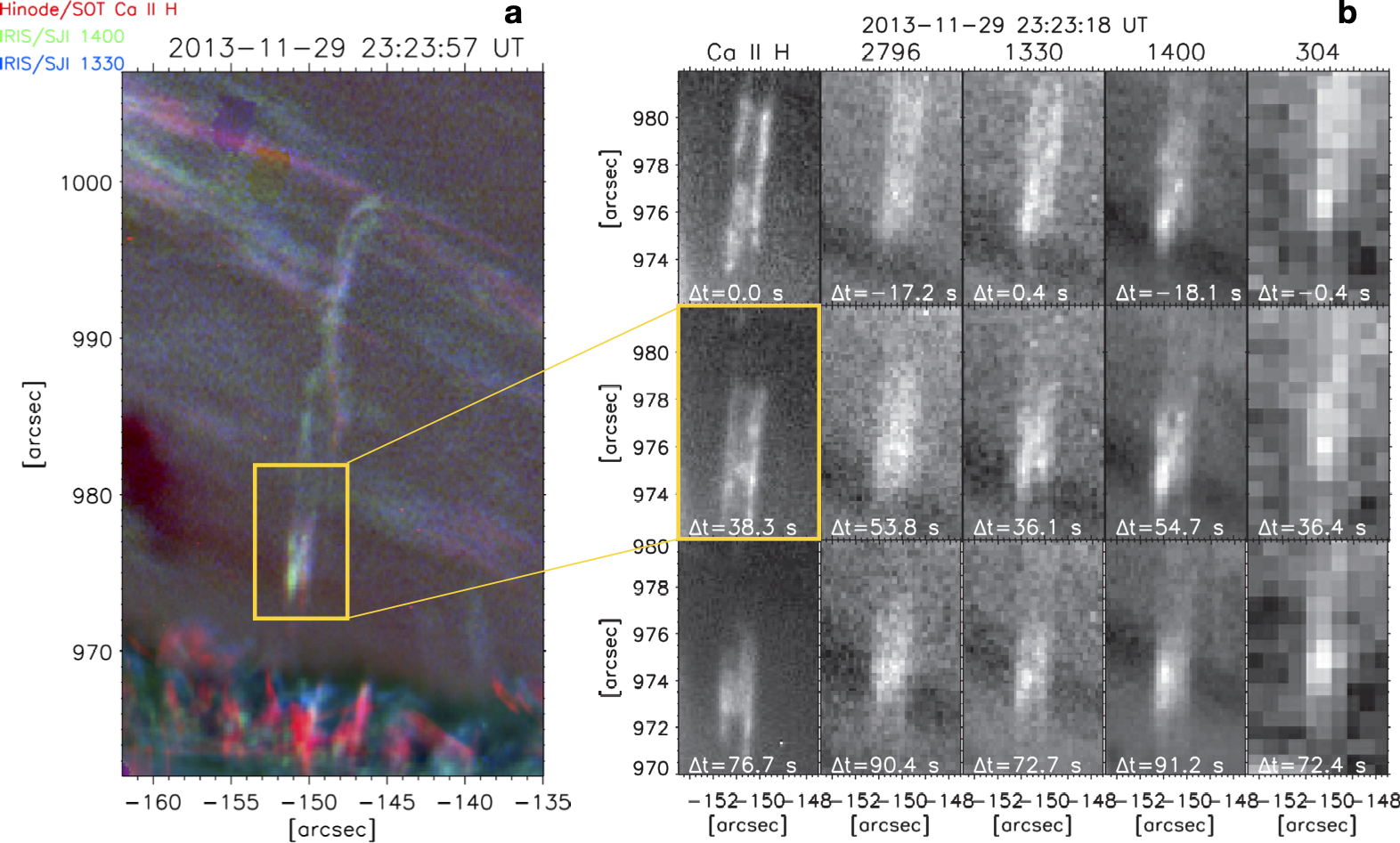}
\caption{(a) Coronal rain along an active region loop at the limb observed with \iris\  in the SJI 1400 (green) and SJI 1330 (blue), and with \sot\ in the \ion{Ca}{ii}~H line (red). (b) A zoomed-in view on the yellow rectangle shown in (a) of a coronal rain clump observed with the various filters, ordered in ascending temperature formation order from left to right, \ion{Ca}{ii}~H (\sot), SJI 2796, SJI 1330, SJI 1400 and AIA 304. Each row corresponds to 3 time instances, with each column showing the snapshot closest in time to the \sot\  snapshot. Figure adapted from \citet{Antolin_2015ApJ...806...81A}. } 
\label{fig:rain} 
\end{figure*}

The rapid recombination process on a timescale of minutes was first resolved through \iris--\sot\  observations in \citet{Antolin_2015ApJ...806...81A} by combining the SJI 1400, 1330 and 2796 passbands with the \ion{Ca}{ii}~H filter of \sot\ (Fig.~\ref{fig:rain}) \citep[see also][]{Liu2014,Antolin2014}. A fast-slow 2 step cooling process was observed suggesting either the transition from optically thin to optically thick cooling, or the presence of a heating source linked to the compression produced by the rain on the downstream plasma. Strong changes in the morphology are observed along the magnetic field direction (taken as the direction of the flow), with longer tails at the wake of the rain at transition region temperatures, thereby potentially playing a role in the characteristic filamentary morphology of coronal loops observed in transition region to low coronal temperatures, as seen in \aia\ 171~\AA. Prior to \iris\ and \sst\ observations, coronal rain was considered as a sporadic phenomenon of active regions, whereby catastrophic cooling would occur in a loop at most once every two days \citep{Schrijver2001}. Despite the large evidence for cool chromospheric and transition region downflows from coronal heights \citep{Levine1977,Kjeldseth-Moe1998}, and EUV downflow signatures (both bright from emission or dark from absorption, \citet[e.g.][]{Ugarte-Urra2006,O'Shea2007,Warren2007,Tripathi2009,Kamio2011}) there was no direct link between the two, and thereby no consensus for the physical mechanism of such multi-thermal downflows.

On the other hand, observations with \sst\  have revealed that these warm flows contain a large number of small cool chromospheric cores with densities on the order of $10^{11}~$cm$^{-3}$ or higher \citep[see Fig.~\ref{fig:rain},][]{Antolin_2015ApJ...806...81A, Froment_2020AA...633A..11F}, suggesting a tip-of-the-iceberg clump distribution in which the bulk at smaller sizes may still be unresolved. Combined, these chromospheric and transition region return flows to the solar surface show mass flux values on the order of $1-5\times10^9~$g~s$^{-1}$ (similar to prominences), suggesting that most of the material in the loop undergoes catastrophic cooling. Therefore, it is now clear that coronal rain plays an important role in the chromospheric-corona mass and energy cycle. 

In the direction perpendicular to the flow (the latter is expected to trace the magnetic field), the width of the interface from chromospheric to transition region temperatures is observed to be very sharp, below $0^{\prime\prime}.33$, evidencing the highly structured and multi-temperature corona at unparalleled spatial resolution \citep[Fig.~\ref{fig:rain} and][]{Antolin_2015ApJ...806...81A}. This fact also illustrates the highly anisotropic nature of thermal conduction in the solar corona. Another characteristic feature of coronal rain evidenced by these authors is its multi-stranded structure, particularly in the chromospheric lines, in the direction perpendicular to the field. Although still  an open question, it has been argued that such morphology should naturally result from the  spatial structure of the unstable thermal modes \citep{VanderLinden_1991SoPh..134..247V}.

The dynamics of such cool downflows further challenge our understanding of the solar atmosphere. Despite the high coronal origins, the falling speeds are on average on the order of $100$~km s$^{-1}$, with maxima close to $200$~km s$^{-1}$ \citep{Schad_2016ApJ...833....5S, Schad2017}. While the kinetic energy of these flows can be comparable to that of a microflare, and indeed lead to brightenings in the visible and UV spectrum upon impact \citep{Kleint2014, Deng2016, Hou2016c, Ishikawa2020,Nelson2020}, with termination shocks \citep{Straus2015, Chitta2016} and strong chromospheric oscillations \citep{Kwak2016}, the speeds of the downflows, 
and in particular the observed acceleration, are usually a factor of 3 lower than that expected from gravity, suggesting other forces at play \citep{Antolin_2010ApJ...716..154A, Antolin_Rouppe_2012ApJ...745..152A}. \citet{Oliver_2014ApJ...784...21O} have shown that the condensation process generates a restructuring of the gas pressure downstream of the rain. The acceleration process therefore occurs mainly in a timescale given by the sound speed, with heavier rain clumps experiencing stronger acceleration through a yet unidentified physical effect. 

Some of these downflows are observed to be persistent over a puzzlingly long time, much longer than the draining time of a loop \citep{Straus2015, Chitta2016} and have thus favored a siphon flow interpretation over that of coronal rain. 2D MHD simulations by \citet{Fang_2013ApJ...771L..29F} show, however, that coronal rain can have a siphon-like nature, with the siphon flows becoming thermally unstable along the way. The intermittent and clumpy nature of coronal rain is also observed to become more continuous and persistent just before impact due to a funnel effect from the expansion of the magnetic field with height \citep{Antolin_2015ApJ...806...81A}. These effects combined can therefore provide an explanation to the plume-like and filamentary structure of coronal loops often observed above sunspots \citep{Foukal_1974ApJ...193L.143F}.

The amount of thermally unstable material in the corona at any one time is still unclear. However, \iris\  has made it clear that coronal rain is a far more common and pervasive phenomenon than previously thought. This is supported by recent results from \citet{Samanta2018}. Using 60 large \iris\ rasters above active regions, it was found that 80~\% of the datasets present high-speed downflows at chromospheric to transition region temperatures. 

Coronal rain is not only an active region phenomenon. Indeed, new \iris\ and \aia\ observations (in the 304~\AA~ passband) indicate the common occurrence of this phenomenon in QS coronal loops associated with null point topologies, where the presence of magnetic dips are common and help accumulate material that subsequently becomes thermally unstable \citep{LiuW2014shin.conf, Mason_2019ApJ...874L..33M, LiL_2018ApJ...864L...4L, LiL_2019ApJ...884...34L, LiL_2020ApJ...905...26L}. In this scenario, interchange reconnection between open and closed loops is invoked to explain the presence of coronal rain within the closed loops, although it is yet unclear if reconnection itself plays a direct role in the generation of coronal rain. This seems indeed possible, as suggested by \iris\ observations from \citet{Kohutova2019}.

The link between coronal rain and coronal heating has been highlighted by the discovery of long period EUV intensity pulsations  \citep{Auchere_2014AA...563A...8A, Froment_2015ApJ...807..158F}. This phenomenon corresponds to highly periodic pulsations in most EUV passbands lasting up to a week in coronal structures in both active and quiet Sun regions. Simulations have shown that the driver mechanism is very likely thermal non-equilibrium \citep{Antiochos1999, Froment_2017ApJ...835..272F, Froment_2018ApJ...855...52F}. This scenario is presented by footpoint concentrated heating that is frequent enough (with a mean recurrent time less than the radiative cooling time), leading to cycles of heating and cooling (also known as evaporation and condensation cycles), with periodic EUV flows \citep{Pelouze_2020AA...634A..54P}. Although globally the loop is in a non-equilibrium state, during the cooling stage thermal instability can set in to produce coronal rain \citep{Antolin_2020PPCF...62a4016A}. 

Observations of coronal rain further allow a direct high-resolution window on the coronal topology and dynamics, as evidenced by, e.g., \citet{Antolin_Verwichte_2011ApJ...736..121A} and \citet{Nelson2019} for the detection of Alfvénic waves, and, e.g., \citet{Antolin2021}  fort the detection of braiding-induced reconnection.

The advent of new instruments on \solo\ and \dkist\ provide exciting new opportunities to address unresolved issues regarding coronal rain. For example, \soloe\ will provide an effective spatial resolution in the EUV of 200~km or less, which, together with \iris\ and DKIST/VBI observations in the chromospheric lines, will address whether the fine multi-stranded structure of coronal rain extrapolates to the entire coronal loop structure. If so, this would establish thermal instability as one of the main causes behind the filamentary morphology of the corona, and would more strongly constrain the energy transport scales. 


Combined observations with \iris\ and \dkist/VISP and DL-NIRSP will allow to observe the fast recombination process involved in catastrophic cooling leading to coronal rain. The details of this process - e.g. how cool coronal rain can get, how does the cooling rate change from the optically thin to thick regimes - may provide an insight into the plasma composition and processes beyond MHD such as ambipolar diffusion.

\subsubsection{Prominence Diagnostics}\label{section:prominence}


Another manifestation of `snowflakes in the oven' on the Sun is prominences
\citep{Tandberg-Hanssen1995, VialJ.Engvold.prom.book.2015ASSL..415.....V}, 
some mysteriously cool and dense material found at elevated heights in 
the hot and tenuous corona. Unlike coronal rain that is usually small in size
and transient in time, prominences are relatively large-scale, long-lived structures
that generally appear as bright emission at the limb and dark absorption on the solar disk
(thus called filaments).
Prominences play important roles, not only in the mass budget of the corona with 
prominence drainage serving as the return flow of the
chromosphere\,--\,corona mass cycle \citep[together with coronal rain, e.g., ][]{LiuW2014IAUS, LiuW2018cosp}, 
but also in solar activity with prominence eruptions forming the cores of CMEs
(see \S\ref{wind} and \S\ref{trigger}).
For example, about 10 large quiescent prominences of typically $3\times 10^{16} {\rm g}$ each
\citep[e.g.,][]{Gopalswamy1998} can hold
the mass equivalent to the entire solar corona.
Consisting of numerous downflow streams at typical speeds of $\sim$$10 \, {\rm km} \, {\rm s}^{-1}$,
a {\it single} large prominence is capable of 
draining the {\it entire} corona in just about one day,
at a rate 
twice that of the mass loss to the solar wind.	



\begin{figure}[t!]
\centering
\includegraphics[width=0.95\textwidth]{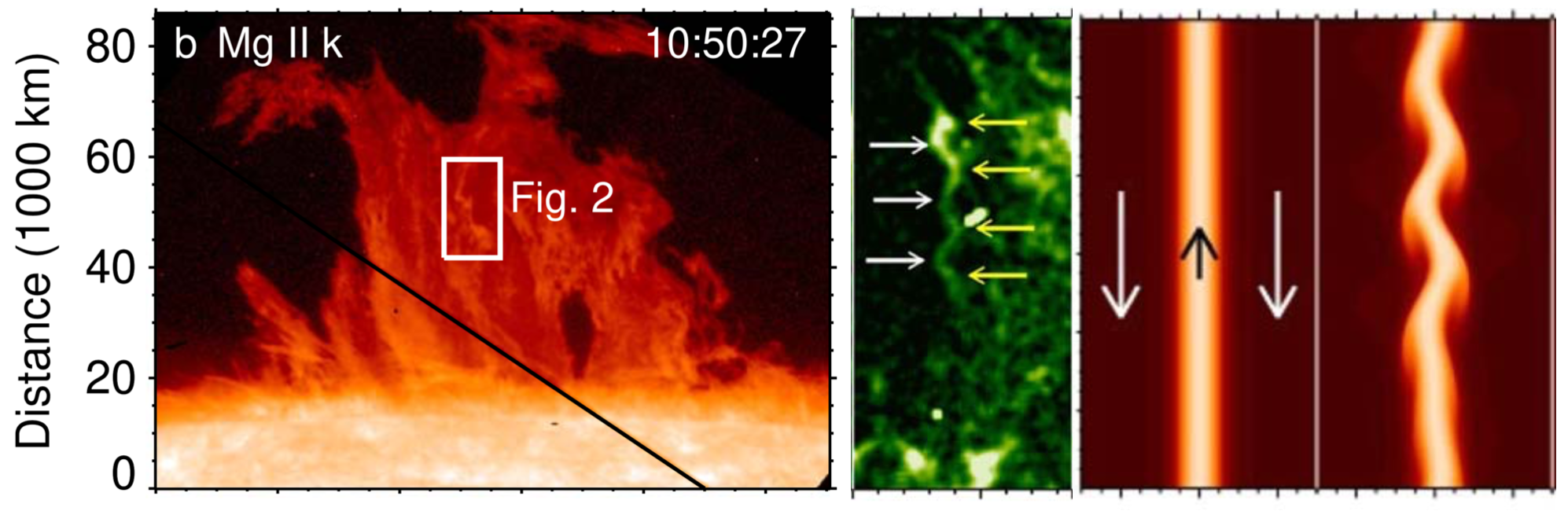}
\caption{\small High-resolution \iris\ observations (left panels)
  and simulations (right panels) of prominences have revealed
  details of fundamental physical processes such as the generation of turbulence as a result of Kelvin
  Helmholtz instability, caused by velocity gradients (right
  panels) \citep{Hillier2018}.
}\label{fig_vanessa} 
\end{figure}

The mass supply to prominences is a key aspect of the chromosphere\,--\,corona 
mass cycle and the transfer of mass into the outer atmosphere in general,
which has remained a topic of debate with coronal condensation, injection from the 
lower atmosphere, and flux emergence all considered plausible candidates. 
Coronal condensation, driven by thermal instability (see \S\ref{section:rain}), has found compelling support from \saia\ with
its multi-temperature coverage \citep[e.g.,][]{LiuW2012, Berger2012}.
The high spatio-temporal resolution of \iris\ has recently revealed
evidence for injections of plasma into prominences through quasi-periodic 
reconnection (possibly mediated by p-mode oscillations) that lead to the well-known counterstreaming flows in prominences \citep{LiT2016}. Similar
injection of plasma was observed by \citet{ZhaoJ2017} 
in association with reconnection at quasi-separatrix layers.  
The so-called prominence bubbles
\citep{Berger2008, Berger2010, Berger2011, DudikJ.bubble2012, LevensP.bubble.tornado.2016}
are dome-shaped, void-like structures intruding upward into prominences from below.
Some bubbles exhibit shearing flows at its interface with the prominence 
resulting in Kelvin--Helmholtz instability rolls and spawn Rayleigh-Taylor instability plumes rising into the overlying prominence threads \citep[][see also \S\ref{instabilities}]{Berger2017}.
Prominence bubbles and plumes have been hypothesized as signatures of flux emergence 
transporting mass as well as magnetic flux and helicity into the prominence
and its hosting coronal structure (e.g., a flux rope). 
The outcome of this is the accumulation of magnetic flux, helicity, and free 
energy in the flux rope, with its mass remaining largely unchanged or even reduced
due to drainage through the prominence. This could ultimately render the flux rope unstable and lead to its  eruption as a CME \citep[e.g.,][]{LowBC.CME-review.2001} supported by recent numerical simulations \citep{Fan2020}. 
\iris\ has detected about a dozen such bubbles and revealed preliminary,
yet important and unique spectroscopic evidence of such processes.

\iris\ observations have revealed the prevalence of turbulence in prominences \citep{Schmieder2014}. Further, \iris\ spectra and modeling have shown evidence for the role 
of the Kelvin Helmholtz instability (Fig.~\ref{fig_vanessa}) in generating such turbulence \citep{Hillier2018}.
Non-LTE modeling using \iris\ diagnostics has also
provided much needed constraints on the temperature, density, and
flows in prominences \citep{Levens2019}. This
type of modeling, coupled with coordinated 
data, 
has elucidated the nature of so-called solar tornadoes 
\citep[e.g.,][]{SuYang2012}
and identified that such helical motions are often apparent, caused by LOS
superposition, and thus perhaps not as good measure of helicity or eruptive potential as thought 
\citep{Panasenco.tornado2014, Schmieder2017}.
Meanwhile, as exceptions to this general trend, 
actual rotational motions were also revealed by Doppler measurements
\citep{Orozco2012}, including those in two tornadoes detected by \iris\ \citep{Yang2018a}. This suggests that there could be two separate classes of tornado involving either apparent or real rotational motions.
\begin{figure}[tp!]
   \includegraphics[width=0.95\textwidth]{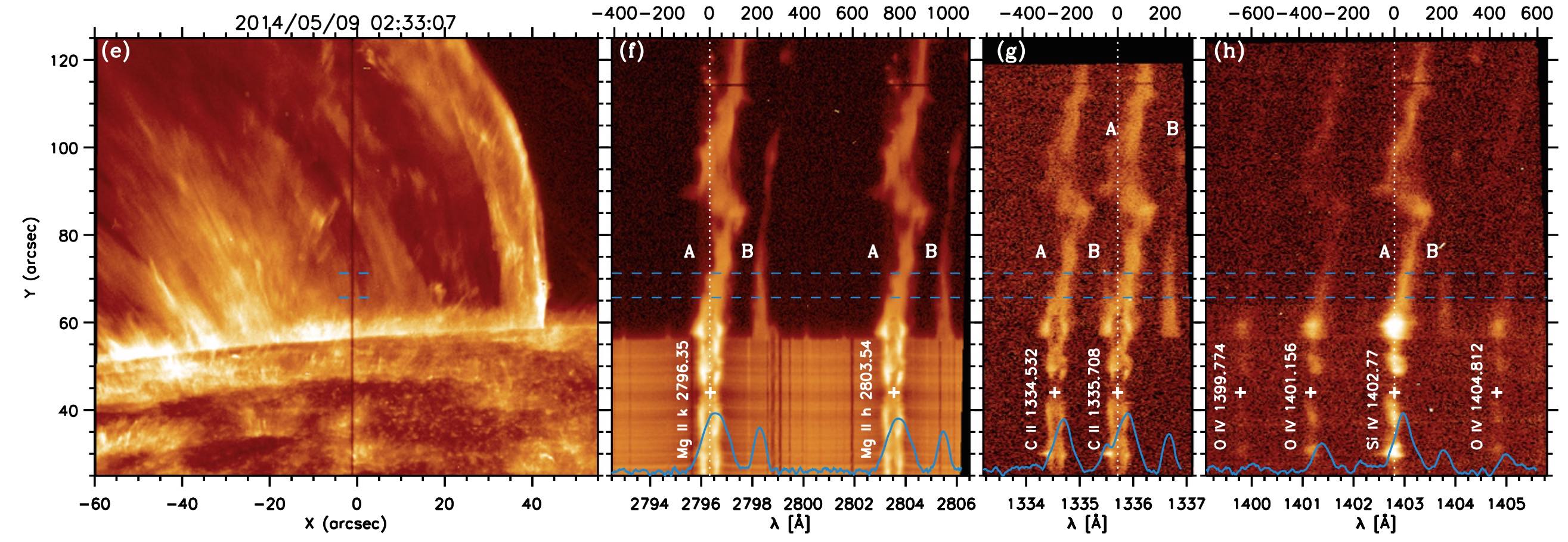}
   \caption{\small Coordinated \iris, \sdo\ and \stereo\ observations have provided unparallelled views of the 3D velocity structure and mass flow during the onset of CMEs (\iris\ slit-jaw, left), exploiting high cadence \iris\ spectra (right) that have revealed the first evidence of Doppler dimming in the \ion{Mg}{ii} lines \citep{LiuW2015}. }
       \label{fig_wei_cme}
\end{figure}

A unique plasma diagnostic for the prominence material that \iris\ can offer 
relies on the \ion{Mg}{ii} k and h lines, as alluded in \S\ref{diagnostics}. 
Unlike their on-disk or chromospheric counterparts, 
the off-limb \ion{Mg}{ii} k and h lines from prominences 
have no central reversals,  
suggestive of a low pressure or thickness and optically-thin regime.
In general, the integrated intensity ratio of the \ion{Mg}{ii} k and h lines 
is expected to be $\sim$2 for optically thin, 
collisionally excited (thermal) emission. Reported values include 2 for an active-region prominence \citep{VialJC.AR.prom.MgII.k/h.ratio=2.1979SoPh...61...39V},
1.7 for a quiescent prominence \citep{VialJC.quiescent.prom.MgII.k/h.ratio=1.7.1982ApJ...253..330V},
and 1.3 for quiescent prominences observed by \iris\
\citep{Heinzel2014a,SchmiederB.IRIS.prom2014AA...569A..85S}.
\citet{LiuW2015} found a surprisingly low value of 1.2 for the fallback material from
an eruptive prominence, which exhibits a positive linear correlation with the Doppler velocity
and \ion{Mg}{ii} k line intensity. They ascribed this behavior to the so-called Doppler dimming effect, which applies to radiatively excited emission from moving objects, for which the incident line emission from the solar surface onto a moving object 
is Doppler-shifted out of resonance. To fully exploit the diagnostic potential 
of the \ion{Mg}{ii} k and h lines requires detailed radiative transfer modeling for realistic
prominence geometry \citep[e.g.,][]{Heinzel2014a, HeinzelP.prom.MgII.2015ApJ...800L..13H, VialJC.IRIS.MgII.prom.model.2016SoPh..291...67V, VialJeanClaude.relation.prom.MgII.model.2019AA...624A..56V}, in comparison with \iris\ observations.
With such modeling, \iris\ can also help elucidate the distinction and relation between coronal rain and prominences.

\subsubsection{Mass supply to the solar wind}
\label{wind}

The connection between the low solar atmosphere and the solar wind has been studied extensively with previous instrumentation, focusing on the role of, e.g., jets \citep{Cirtain2007, Raouafi2016}, funnels \citep{Tu2005}, and active region outflows \citep{Doschek2007}. \iris\ is well suited to study the physical mechanisms at the roots of the fast and slow solar wind, key to interpreting the downstream properties as measured by \psp\ and \solo\ in-situ instruments. 

\begin{figure}[t!]
\centering
\includegraphics[width=\textwidth]{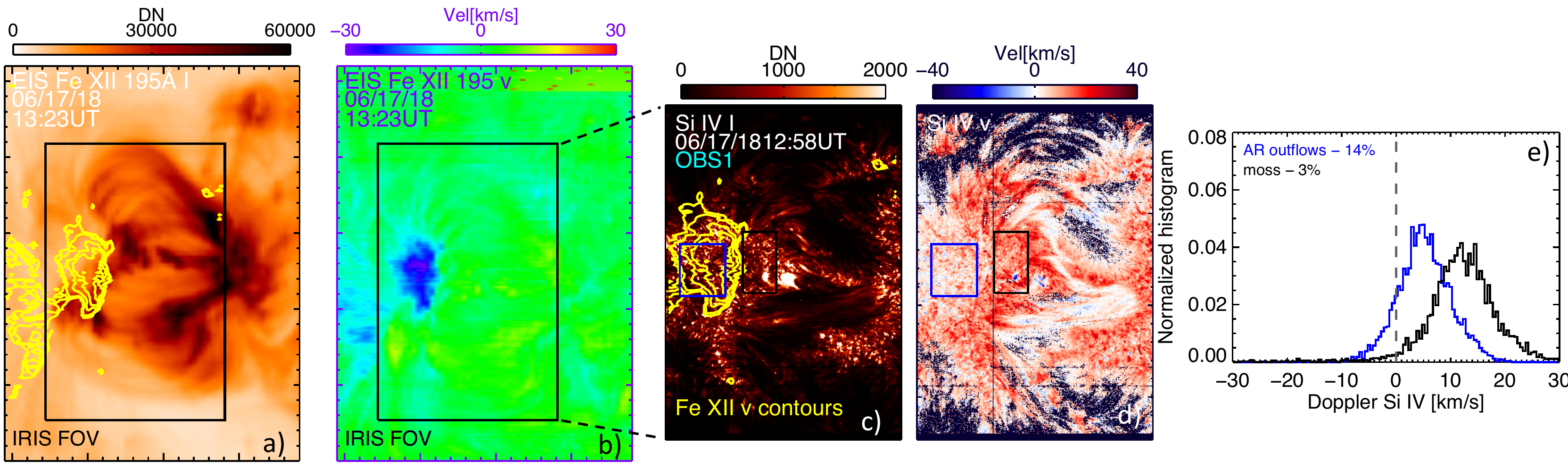}
\caption{\small Analysis of observations of AR outflow regions,
  thought to supply mass to the slow solar wind, with \eis\
 (a,b) and \iris\ (c,d,e) will build on preliminary results
  suggesting significant differences in the low TR, challenging
  current models. Adapted from \citet{Polito2020}.} \label{fig_vanessa2} 
\end{figure}

Jets of all types are common at the roots of the solar wind. \iris\ observations of such jets have been used to elucidate the mass supply to the solar wind. 
For example, \citet{Narang2016} studied the evolution and properties of jets in
coronal holes at the footpoints of the fast solar wind. They found that
the transition region counterparts of spicules
or so-called TR microjets appear faster and longer in
the open field configuration at the roots of the fast solar wind.  \iris\ results have also
provided new insights into the formation of larger-scale jets. For example,
coordinated \iris, \sdo\ and \hic\ observations
have established a clear relationship between magnetic cancellation
and the formation of so-called
jetlets \citep{Panesar2019}, larger than spicules but
smaller than coronal jets. Such cancellation and the resulting
formation and eruption of small-scale ("mini") filaments has been proposed as driving mechanism for many
coronal jets that feed into the solar wind \citep{Sterling2015}. Recent \iris/SJI and \sdo\ observations have been used to provide support for this mechanism \citep{Hong2019} through imaging of the
evolution of the predicted breakout current sheet that precedes the eruption of
a mini-filament and jet.

However, the formation mechanism and impact of such jets remains poorly constrained. Coordinated magnetograms from ground-based observatories could be used to help better understand the variety of magnetic field configurations that drive coronal jets and determine the relative roles of flux emergence and cancellation. Detailed studies using very high cadence \iris\ spectroheliograms may also allow diagnostics of the evolution of the mini-filament, if it is present, before it erupts.  Similarly, coordinated \iris--\aia--\psp\ observations during \psp\ perihelia with front-side connectivity can be used to study whether these jets are related to the magnetic field ``switchbacks'' \citep{Sterling2020b} reported by \psp\ \citep{Bale2019}.

Another potential mass source for the (slow) solar wind are the active region outflows \citep{Doschek2007}. Recently, \citet{Polito2020} studied the chromospheric and transition region signatures of these outflows. In the past these outflows have been detected in coronal lines as strong blueshifts. Preliminary analysis suggests that, surprisingly, these outflows leave distinct signatures in the spectral properties of chromospheric and transition region lines (Fig.~\ref{fig_vanessa2}). It is currently unclear what the physical cause is for these signatures. The canonical model for these outflows invokes reconnection at coronal heights and does not seem to provide an explanation for the \iris\ observations \citep[e.g.,][]{Baker2009,DelZanna2011}. Follow-up studies are needed to determine whether perhaps the reconnection also occurs at lower heights and temperatures, or whether the different properties of open field regions (e.g., spicules) perhaps play a role. 

Similar low-atmospheric signatures of the open fields connecting to the solar wind have been found through
analysis of part of the timeseries of more than 80 (almost monthly) \iris\ full-disk mosaics and large rasters of equatorial and polar coronal holes. This analysis suggests that the boundary of open and closed fields in coronal holes can (sometimes?) be detected in the shape and properties of the \ion{Mg}{ii} h/k line profiles \citep{Bryans2020} and their correlation with the underlying magnetic field properties \citep{Kayshap2018b}. Preliminary results suggest that the different properties of spicules in coronal holes may play a role in this intriguing detection. Further studies are needed to obtain a full understanding of the physical cause of this chromospheric signature, perhaps through investigating the physical parameters (temperature, density, microturbulence, and velocity) derived from \alb \citep{Sainz-Dalda2019}. 
In addition, high-cadence sit-and-stare \iris\ observations of the boundary
between quiet Sun and coronal holes could be used to help reveal chromospheric and
TR signatures of interchange reconnection, thought to play a key role in
energizing the fast solar wind. These studies would benefit from coordinated magnetic field observations from \shmi, \hinodes, \solop, or GBOs.  The \solop\ measurements, once they occur from a vantage point above or below the ecliptic, would be ideal to help study polar coronal holes from different vantage points.

One topic that is at the core of the solar wind connectivity to the low solar atmosphere is that of the enrichment of elements with low first ionization potential (FIP). While \iris\ cannot perform measurements of abundances, the roots of this FIP effect are thought to lie in the chromosphere where the ionization of these elements occur. Future coordinated observations with \iris, \eis, and \solos\ can help constrain theoretical models for this important physical phenomenon.

During space weather events, coronal mass ejections play a significant
role in the mass supply to the solar wind. Initial studies indicate
that coordinated \iris, \sdo\ and \stereo\ observations hold
promise in estimating the mass budget of coronal mass
ejections. For example, \citet{LiuW2015} exploited the line-of-sight
and plane-of-the-sky velocities of a CME measured with \iris\ to
reconstruct the 3D flow-field. They found rapid initial acceleration at very low heights
(constraining CME models) and obtained estimates of the returning mass from
the first observations of Doppler dimming in the \ion{Mg}{ii} lines.
Future observations with \iris, \sdo, and \solo\ could be used to perform similar studies for more CMEs, further elucidating the mass supply to the solar wind during these eruptions.

\section{Large-scale instabilities}
\label{instability}
\subsection{Energy deposition process during flares}
\label{flares}

Despite its relatively small
field-of-view (which covers $\approx$~5\% of the solar disk), \iris\ has so far observed 10 X-class, over 100 M-class
and more than 600 C-class flares since its launch in June 2013 \footnote{For a list of flares observed by \iris\, see: \url{https://iris.lmsal.com/data.html\#flares}}. Coupled
with the rich diagnostics that cover conditions from the photosphere
to the hottest parts of the flaring corona and the high
spatio-temporal resolution, \iris\ observations have provided many discoveries and a wealth of information that have expanded our knowledge of
how flares are triggered, and how non-thermal energy is released during
flares \citep[see e.g.][for a review of some earlier results]{Fletcher2011,Benz2017}.

\subsubsection{Chromospheric evaporation and condensation with \iris}
\label{flares_evaporation}
Unprecedented
observations of the \ion{Fe}{xxi} 1354\AA\ line (formed at $\approx$~10~MK during flares) have provided
new significant insights into how energy deposition in the chromosphere
drives hot plasma into the corona, i.e., the chromospheric evaporation
process. \citep[see also previous observations with the EIS and CDS spectrometers, e.g.][to name a few]{Brosius2003,DelZanna2006,Milligan2009,Young2013}
Several studies have shown that the line is observed to be fully blueshifted at the flare ribbon sites (without the presence of a rest component), with typical speeds ranging from $\approx$~100 to 300 km s$^{-1}$ \citep[e.g.,][to name a few]{Young2015, Polito2015, Tian2015, Graham2015, LiY2015, Brosius2015, Tian2018c}, in contrast to observations with previous lower resolution spectrometers, that most often showed
an additional stationary component \citep[e.g.][]{Polito2016a}.
The observation of fully blueshifted \ion{Fe}{xxi} profiles indicates that \iris\ is finally resolving the emission from the flare footpoints that contain the upflowing plasma, solving a long-standing problem and obviating the need for more complex
theoretical explanations \citep[e.g.][]{Doschek1986, Young2013}.

\begin{figure*}[t]
\centering
\includegraphics[width=\textwidth]{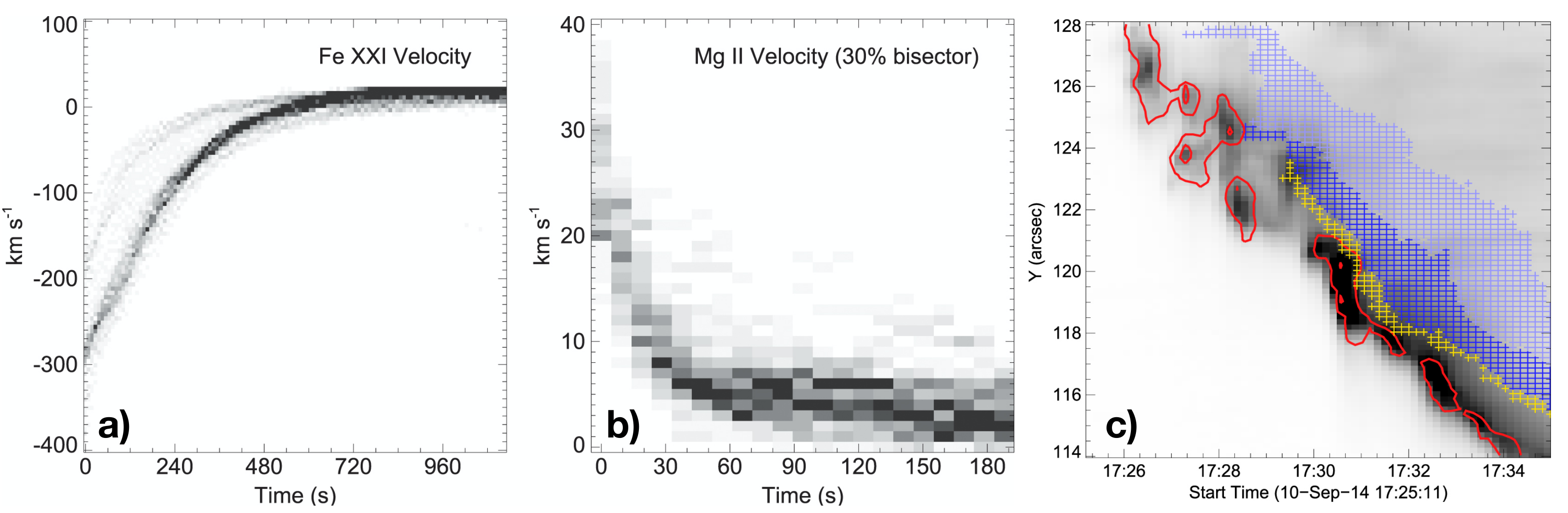}
\caption{Panels (a) and (b): Superposed epoch analysis of \ion{Fe}{xxi} and \ion{Mg}{ii} flows respectively for selected slit pixels at ribbon of the 2014 September 10 X-class flare  (negative/positive velocities showing upflows/downflows). The grayscale darkens with increasing occurrence within a given velocity interval. Panel (c): \ion{Mg}{ii} intensity spacetime map (inverted B/W color table) with overlays of: \ion{Mg}{ii} downflows at 15 and 30 $\mathrm{km}\ {{\rm{s}}}^{-1}$ (red contours), \ion{Fe}{xxi} upflow velocities above 270, 200, and 100 $\mathrm{km}\ {{\rm{s}}}^{-1}$ (yellow, dark blue, and light blue crosses, respectively). Adapted from \citet{Graham2015}. } 
\label{fig:flare_graham15} 
\end{figure*}

However, \iris\ has also provided several new results
that have challenged our current understanding of chromospheric evaporation. 
Most notable of this is the work of \citet{Graham2015} (Fig.~\ref{fig:flare_graham15}a), who exploited the high cadence
of \iris\ sit-and-stare observations of the 2014 September 10 flare to study the \ion{Fe}{xxi}
velocities in many tens of individual pixels in
flare ribbons.  They found that the upflow evolution
is strikingly similar among all of these pixels, possibly suggesting that either the energy input is remarkably uniform, or the evolution of the evaporation is more governed by the local conditions than the initial energy input.  The upflows are also observed to smoothly decrease over about 6 minutes (Fig.~\ref{fig:flare_graham15}a). 
It should be noted that some other studies \citep{LiD2015b,Brosius2015, Polito2016a, Dudik2016} have reported an initial increase in the blue shifted
\ion{Fe}{xxi} velocity followed by a decrease, a phenomenon that still has not been systematically investigated. 
Long and gradual decay times for the evaporation flows as those shown by \citet{Graham2015} can be hard to explain by single flare loop models, and seem to be best reproduced by models involving heating of multi-threaded flare loops \citep[e.g.,][]{Reep2016b, Reep2018, Reep2019}. 
Finally, multiple episodes of chromospheric evaporation in the \iris\ \ion{Fe}{xxi} line, which are suggestive of multiple individual energy injections into the atmosphere, have sometimes been observed \citep{Tian2018c, Brosius2018}. Such recurring energy injections have been interpreted as being due to a bursty type of reconnection (see also \S~\ref{flare_reco}).


The broadening of the \ion{Fe}{xxi} line profiles during the impulsive phase has also provided new diagnostics of flare heating. \citet{Young2015} examined the ``bar'' of enhanced emission that is
  typically found at the loop-tops in flare arcades.  They found that the
  \ion{Fe}{xxi} emission is completely resolved at the loop-top, and
  the line profile in this location is stationary and does not have
  excessive non-thermal widths, indicating a lack of turbulence. The
  cause of this enhanced emission at the loop-tops is still a mystery,
  but this observation rules out the idea that it
  is caused by turbulence in tangled magnetic fields at the loop tops
  \citep{Jakimiec1998}.  Other models for this emission have been
  proposed \citep[e.g.,][]{Reeves2007,Longcope2014} and further \iris\
  observations of flare loops will help clarify the
  mechanism at play.
  
\begin{figure*}[t]
\centering
\includegraphics[width=\textwidth]{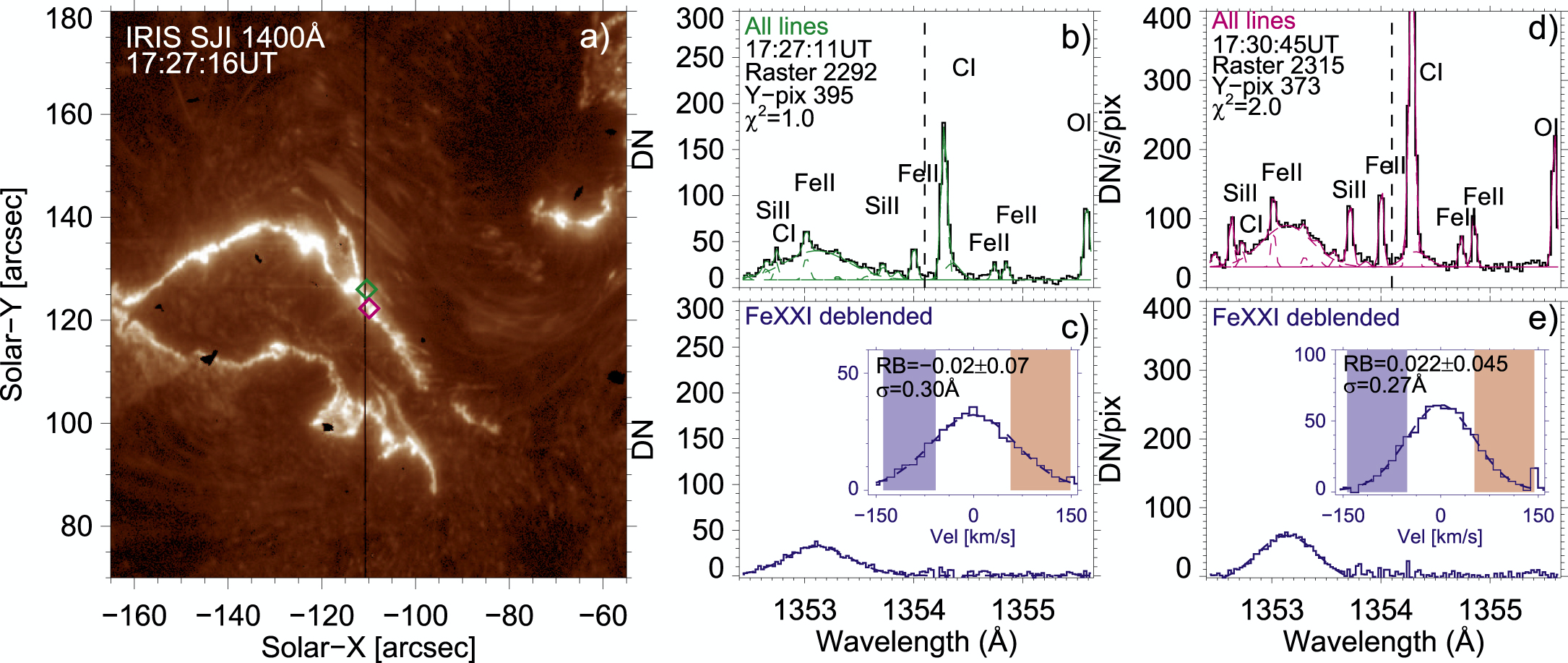}
\caption{Broad and symmetric \ion{Fe}{xxi} profiles observed by \iris\ during the 2014 September 10 flares. Panel (a): SJI 1400~\AA images of the flare ribbons and the location of the spectrograph slit (vertical line). Panels (b) and (d): \ion{Fe}{xxi} spectral window at the locations indicated by the corresponding colored symbols in panel (a). Panels (c) and (e): deblended \ion{Fe}{xxi} with a Gaussian fit (dotted curve) superimposed. The insets illustrate the red-blue asymmetry method showing that the line profiles are mostly symmetric. From \citet{Polito2019}.} 
\label{fig:flare_polito19} 
\end{figure*}

In addition, \citet{Polito2015} found that the Doppler shift and non-thermal width of the line gradually decrease as a function of time and are highly correlated, while no significant non-thermal line broadening was observed at the loop tops, in agreement with \citet{Young2015}. Superposition of evaporative upflows has long been a popular explanation for the large non-thermal broadening
in flare ribbons \citep[e.g.][]{Doschek1986,Milligan2011}. By using 1D hydrodynamic simulations of multithread flare loops mimicking a 3D geometry, \citet{Polito2019} found that this scenario may not be
compatible with the \iris\ \ion{Fe}{xxi} profiles, which appear to be broad but at the same time symmetric and well-fitted with a single Gaussian profile (Fig.~\ref{fig:flare_polito19}), and suggested that other mechanisms such as \alfvenic turbulence or non-equilibrium ionization might need to be invoked to explain the observations.

Preliminary results from the RADYN 
arcade model \citep{Kerr2020}, which is obtained by grafting 1D hydrodynamic simulations of flare loops heated by electron beams onto an observed AR loop arcade in a 3D domain, also show discrepancies with the \iris\ \ion{Fe}{xxi} observations (such as the long upflows decay and the large non-thermal broadenings), suggesting that important physical mechanisms might still be missing in these models \citep[see also][for another recent arcade model]{Reep2020}. For instance, hydrodynamic models including non-equilibrium ionization for highly charged iron atoms using the HYDRAD code \citep{Bradshaw2003} appear to reproduce some of the observed features, such as the broad and symmetric profiles in hot lines, assuming short flare loops and a combination of input parameters \citep{Mandage2020}. However, simulations of longer loops with lengths $\gtrsim$~50~Mm, such as the ones observed in the 2014 September 10th flare analysed by \citet{Polito2019}, still show profiles with blue wing asymmetries which are not seen in the observations. Further investigation into the parameter space of these models and inclusion of additional heating mechanisms, such Alfv\'en waves, might be needed to solve some of these discrepancies.  

Due to momentum conservation, the evaporation upflows are usually accompanied by downflows of chromospheric and transition region plasma (``chromospheric condensation”), which can be observed by \iris\ as redshifts of cooler lines such as \ion{Si}{iv}, \ion{C}{ii} and \ion{Mg}{ii} \citep[e.g.,][]{Tian2015,Brosius2016,Graham2015, Graham2020}. The \ion{Si}{iv} and \ion{C}{ii} spectra sometimes show persistent redshifts or red-asymmetries of a few tens of km s$^{-1}$ during the whole impulsive phase \citep{Warren2016,Yu2020}, which provide challenging constraints to the 1D loop models \citep[e.g.,][]{Reep2016b, Reep2018}. 
On the other hand, downflows in chromospheric lines such as \ion{Mg}{ii} are usually observed to be more short-lived and last $\approx$~30--60s \citep[e.g.,][Fig.~\ref{fig:flare_graham15}b]{Graham2015,Graham2020}, in agreement with early models of chromospheric condensation \citep{Fisher1985}.

In some cases, the \iris\ chromospheric lines (including \ion{Mg}{ii}, \ion{C}{i}, \ion{Fe}{i}, \ion{Fe}{ii}, \ion{Si}{ii}) also show an enhanced rest component superimposed with a broad and highly redshifted ($\approx$~25--50 km s$^{-1}$) secondary component, which rapidly decays in 30--60 s \citep{Graham2020}. Such double-component profiles can be explained by RADYN hydrodynamic simulations assuming that the most energetic electrons penetrate into the deep chromosphere, while the bulk of the electrons dissipate their energy higher up into the atmosphere \citep{Kowalski2017Mar29,Graham2020}. These studies also highlight the diagnostic potential of the \iris\ chromospheric lines such as \ion{Fe}{ii}, which are extremely sensitive to the model input parameters.

Further, several observations have shown significant delays (up to 2
minutes) between the evaporative upflows and chromospheric condensation flows observed in \ion{Mg}{ii} h and k
\citep[e.g.,][Fig.~\ref{fig:flare_graham15}c]{Graham2015, Sadykov2016}. Such delays are not predicted by the models. 
Clearly, more comparisons between targeted or tailored \iris\ observations and advanced models will be required to fully explain the diverse plasma dynamics observed during the flare impulsive phase and resolve some of these puzzles.

Sometimes upflows in cooler \iris\ lines are also observed.  For instance, \citet{LiY2019} recently observed blueshifts in the \ion{Si}{iv}, \ion{C}{ii} and \ion{Mg}{ii} \iris\ lines at the ribbons of a small GOES B-class flare. Blueshifts in the \ion{Si}{iv} have also been observed in smaller micro- or nano-flare size events \citep[][see \S~\ref{nanoflares}]{Testa2014, Testa2020}, suggesting that similar energy release mechanisms may be at play in energetic events of different size, from nano- or micro- to larger class flares. 
In addition, an earlier study by \citet{LiY2015} combined \iris\ and \eis\ spectra of flare footpoints to observe the temperature dependence of the evaporation and condensation in the same flare, and found some locations where  coronal and TR lines were blue shifted, and others  where the TR lines were redshifted, which had been interpreted by the authors as being suggestive of  ``gentle'' or ``explosive'' evaporation, respectively \citep[e.g.][]{Fisher1985,Milligan2009}. 
Recent theoretical studies have shown that the threshold between different trends of evaporations with temperature mainly depends on the location of the energy deposition, and is very sensitive to the details of the heating \citep[e.g.,][]{Reep2015,Polito2018}. 

After the flare peak phase, condensation downflows along the cooling flare loops are also observed (which are sometimes referred to as coronal rain). Such cooling loops have been observed in the \iris\ \ion{Mg}{ii} as TR lines, which exhibit strong redshifts and increased non-thermal widths \citep{Brannon2016, Mikula2017}. Another study exploited data from the New Solar Telescope at \bbso\ and \iris\ to reveal the evolution of coronal rain and its impact on the chromospheric ribbons \citep{Jing2016}. Importantly, the widths of the ribbon, the rain, and that of the observed footpoint brightenings were all in the same range of 80--200 km. The authors speculated  that these observations might be close to resolve the spatial scale of the energy transport during flares. Future coordinate observations between \iris\ and GBOs during the flare gradual phase may help investigate this further. 

\subsubsection{\iris\ lines as diagnostics of flare heating mechanisms}
\label{Flare_diagnostics}
Understanding what physical mechanisms contribute to what extent to
flare heating  remains an active topic of debate \citep[e.g. see review by][]{Fletcher2011}, to which \iris\ has provided new and challenging insights. 

Several recent studies
have provided support
for electron beams being the main carrier of energy from the flare
site to the ribbons \citep[e.g.,][]{Tian2015,LiD2015b, Sadykov2015, Polito2016a, Brosius2016, Brosius2017,Zhang2016a,Zhang2016b, LiD2017a, Reep2019}. These studies were based on correlations between evaporation/condensation flows observed with \iris\ and hard X-ray and radio fluxes (assumed to be associated with
non-thermal electrons), as well as comparison with hydrodynamic models.
In particular, a statistical work comparing \iris\ and HXR \rhessi\ observations 
with a grid of hydrodynamic models has shown that
non-thermal electrons can reproduce many aspects of the \iris\
observations, such as the correlation between the non-thermal energy flux and the \ion{C}{ii} Doppler shifts, although significant discrepancies remain,
including a puzzling lack of correlation with the \ion{Fe}{xxi} Doppler
shifts \citep{Sadykov2019}. 
Besides, evidence for both non-thermal electron and thermal heating has also been reported \citep[e.g.,][]{Yu2020}, sometimes within the same flare event \citep[e.g.,][]{Battaglia2015}.

In addition to particle beams and in-situ heating followed by thermal conduction, it has also been suggested that Alfv\'en waves may play a role in
transporting energy from flare sites and depositing or thermalizing this magnetic free
energy in the lower atmosphere \citep{Fletcher2008,Russell2013}, but observational
evidence has been sparse. Numerical work has shown that the heating from electron beams alone cannot account for  coronal rain, a common occurrence in flaring loops during their last cooling stage \citep{Reep_2020ApJ...890..100R}. Such work suggests the need for an additional, long lasting (radiative cooling timescale) and footpoint-concentrated heating mechanism triggered by the flare and future work needs to consider whether Alfv\'en waves fit the purpose. 
Other \iris\
observations find some suggestions that Alfv\'en waves may be at work
in driving sunquakes, photospheric perturbations in the aftermath of a
flare whose origin has remained difficult to explain since their discovery in
the 1990s \citep{Matthews2015}. Additionally, \iris\ observations at an unprecedented temporal resolution (1.7s) have shown increased and oscillatory \ion{Si}{iv} non-thermal broadening preceding significant brightening in flare ribbons \citep{Jeffrey2018}. 
These observations are compatible with a scenario in which turbulent dissipation of magnetic energy first occurs in the lower atmosphere footpoints, possibly as the result of interacting waves,
rather than thermalization of electrons that have been accelerated in
the corona.

\begin{figure*}[t]
\centering
\includegraphics[width=0.8\textwidth]{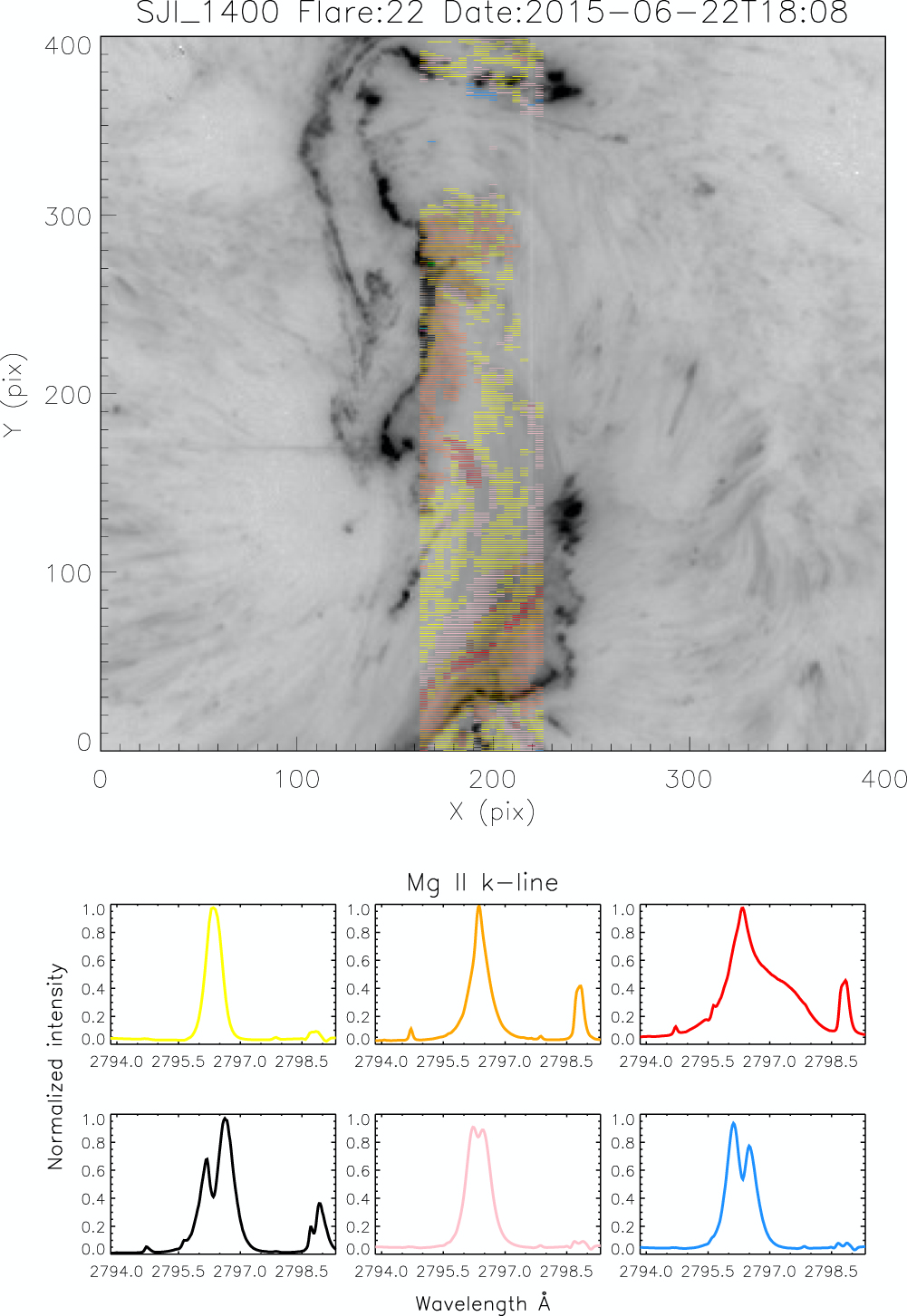}
\caption{Example of a k-means clustering of \iris\ \ion{Mg}{ii} profiles for the M6.5 flare on 2015 June 6, highlighting the potential in identifying spatio-temporal patterns in spectroscopic data. The line profiles are assigned to the representative groups in the bottom panel they most closely resemble, according to the colors. The black reversed profiles are those observed at the ribbon fronts. 
From \citet{Panos2018}.} 
\label{fig:flare_panos18} 
\end{figure*}

The \iris\ \ion{Mg}{ii} h and k lines in flare ribbons provide unique chromospheric diagnostics, which previously had
been seen only once before in 1984 with OSO-8. The complex and broad line
profiles observed in and around the flare ribbons \citep[e.g.,][]{Kerr2015,LiuWen2015} and loops \citep[e.g.][]{Mikula2017}
appear to be difficult to fully reconcile
with current numerical models \citep[e.g.,][]{Rubio-da-Costa2016}.  This is also due to the complex formation mechanisms of these lines, especially under flaring conditions. 
\citet{Kerr2016} performed
radiative hydrodynamic modeling and non-LTE radiative synthesis of the
\ion{Mg}{ii} h \& k lines in the aftermath of a flare and found that
the shape of these chromospheric lines appears to be very sensitive to the detailed physical
mechanism that transports and deposits energy in the
chromosphere. Their preliminary results appear to favor Alfv\'en waves
over electron beams, but much work remains to be done to provide
definitive evidence for a significant role of Alfv\'en waves in flares.

In addition to previous studies \citep{Rubio-da-Costa2017}, recent work by \citet{Zhu2019} has succeeded in reproducing some of these profiles assuming very large electron
energy fluxes (typical of a X-class flare) and increased Stark
widths. Nevertheless, it is still unclear how to reproduce such single
peaked profiles in the case of smaller flares, and how to fully
explain their increased broadening. \citet{Kerr2019b}
highlighted the importance of improving the realism of the codes
 that are currently used to forward model these \iris\ lines by taking
 into account all the relevant physics (such as non-LTE, PRD and
 opacity) for these transitions.

Puzzling blue-wing enhancements in the \ion{Mg}{ii} lines \citep{Tei2018,Huang2019a} have also been observed and could be
indicative of ``cool upflows'', caused by non-thermal electrons
penetrating much deeper into the lower atmosphere than expected and
causing an upward motion in the upper chromosphere \citep{Huang2019a,
  Zhu2019, Hong2020}. A recent study has also shown that the location where the cool upflows occur varies in different flare models \citep{Hong2020}.

These preliminary results suggest that the shape of the \ion{Mg}{ii} lines may hold a key to a better understanding of the energy deposition mechanisms in flares. Exciting advances in identifying these complex profiles through
machine learning has recently enabled a
statistical approach that captures the full
spatio-temporal information in \iris\ observations. In particular,
\citet[][Fig.~\ref{fig:flare_panos18}]{Panos2018} analyzed thousands of profiles and discovered a
class of broad profiles with a blueshifted reversal that occurs at the fast-moving leading
edges of ribbons (black profiles in Fig.~\ref{fig:flare_panos18}). Such peculiar profiles are similar to those observed by \citet{Xu2016} at the ribbon fronts of a M-class flare, in the same location as increased absorption (``negative ribbons") in the \ion{He}{i} 10830 chromospheric line was also observed with the \emph{Goode Solar Telescope} at \bbso. 
More observations from \iris, combined with ground-based and X-ray satellites may be needed to understand these puzzling spectral features. 

 
\iris\ observations of NUV continuum have also provided exciting diagnostics of energy deposition in white light flares (WLF).
 A significant fraction of the energy deposited in the atmosphere
during many flares is in fact observed in white light emission, which emanates
from the photosphere. Since energetic particles have difficulty
penetrating that deep into the atmosphere, some other mechanism is
thought to be responsible for the emission. \iris\ has now been able to
provide key new insights into this long-standing puzzle by exploiting observations
that were coordinated with ground-based telescopes, \hinode, \rhessi\ and \sdo. For example, extending the work of \citet{Heinzel2014},
\citet{Kleint2016} combined observations covering wavelengths in the UV, visible and infrared, as well as \rhessi\ HXR observations and numerical models, to show that the white-light
emission in the 2014 March 29 flare originated from temperature increases in
the photosphere due to back-warming from hydrogen recombination (e.g., Balmer continuum) in
the chromosphere, where the flare energy was dissipated. 
Moreover, the authors found that the energy contained in the high energy electrons observed by \rhessi\ is sufficient to explain the observations. 
Using data from the same flare, \citet{Kleint2017} also showed that enhanced spectral line emission can contribute significantly to the \iris\ SJI 2832\AA\ filter, whose emission cannot therefore be considered as being purely due to Balmer continuum during WLFs.

Further
 evidence for this process was reported by \citet{Kowalski2017Mar29}, who combined \iris\ and \rhessi\ data with electron-beam driven flare simulations and found that the NUV continuum intensity was consistent with hydrogen Balmer recombination radiation emitted in two separate atmospheric layers where the electron beams release their energy (see also \S~\ref{stellar}). 
These findings appear to support the hyphothesis that WLF emission is associated with energy deposition from non-thermal electrons streaming from the corona and propagating down towards the chromosphere \citep[see also][]{Cheng2015b, Lee2017}. On the other hand, recent \iris\ observations of a WLF, showing spectral features typical of UV bursts, such as enhanced and broadened \ion{Si}{iv} and \ion{C}{ii} profiles superimposed by chromospheric absorption lines,
 combined with the lack of significant HXR flux in \rhessi, have also provided evidence for a different scenario, in which the WLF is caused by reconnection occurring locally in the lower atmosphere \citep{Song2020}. More observations of WLFs as well as comparison with models will be necessary to investigate these different scenarios. 

As demonstrated by the wealth of results summarized above, \iris\ has allowed significant progress in the understanding of how flare energy is propagated and deposited in the lower atmosphere, as well as providing challenging new constraints for current flare heating models. 
However, many unsolved puzzles still remain and several new questions have also been
raised. For instance, current hydrodynamic models still cannot fully and consistently explain the gradual decay of the evaporation flows (Fig.~\ref{fig:flare_graham15}) and the large symmetric broadenings (Fig.~\ref{fig:flare_polito19}) observed in the \ion{Fe}{xxi} line, as well as the persistent redshifts that are sometimes observed in the transition region lines, and the puzzling complex and broad \ion{Mg}{ii} profiles (Fig.~\ref{fig:flare_panos18}). In addition, important questions remain regarding the details of the energy propagation and dissipation in flares of different classes including WLFs, and the importance of Alfv\'en waves vs electron-beam and thermal conduction heating. 


In the next few years there will be exciting opportunities to investigate some of these puzzles even further, thanks to new advancements in the models, including recent multithreaded and arcade models \citep[e.g.][]{Polito2019, Kerr2020, Reep2020, Mandage2020}, improvements in the modelling and synthesis of the chromospheric diagnostics \citep[e.g.,][]{Zhu2019, Kerr2019a, Graham2020}, as well as new approaches to the analysis of the observations based on statistical studies and machine-learning techniques \citep[e.g.,][]{Panos2018,Panos2020,Sadykov2019}. 
With the Sun going towards a long period of increased activity, many new coordinated observations of flares with existing ground-based (including \sst, \alma, \bbso, \gregor) and space-based (e.g., \sdo, \hinode, \psp, \solo) satellites, as well as the novel \dkist\ instrument, will also be available.

\subsection{Magnetic reconnection in flares}
\label{flare_reco}
Reconnection is widely believed to be the dominant mechanism to convert magnetic energy into the kinetic and thermal energy that triggers flares, although the details of how such processes occur are still not fully understood \citep[see e.g.][for a review]{Fletcher2011}.
The high spatial resolution of \iris\ has provided a
better understanding of the ways in which reconnection can occur in 
three dimensions during solar flares, as discussed in some of the examples below.  

In an earlier study, \citet{LiY2016} used \iris\ 1330 \AA\ slit jaw images
to follow the peculiar motions of an unusual X-shaped flare ribbon, that suggest the presence of 
separatrices in the coronal field, with
reconnection occurring all the way down to the Sun's surface. In the same event, \citet{LiY2017} observed significantly broadened wings extending to 200 km s$^{-1}$ in the spectra of the \ion{Si}{iv} line close to the X-point, which they interpret as bidirectional flows produced by the separator reconnection. Such flows have rarely been observed by previous spectrometers. 
Another example of bi-directional flows was recently reported by \citet{Yang2020}, who observed broadened \iris\ \ion{Si}{iv} profiles with extended wings superimposed by several blueshifted absorption lines at the null
point of a fan-spine magnetic topology during a small B-class flare. These absorption features were interpreted as the signature of cool upward moving material above the reconnection region, and their depth provided an indication of the amount of cooler material superimposed along the line of sight. This result implies that the \iris\ spectral profiles can help study the plasma dynamics in the reconnection site. 


In another study, \citet{Jiang2015} use \iris\ slit jaw 1400 \AA\ images
to trace the evolution of a small active region with a fan-spine
topology as an emerging flux region appears. \iris\ observations of
reconnection and many recurring jets as the new flux emerges near the fan-spine structure  indicate the latter is eventually destroyed. These observations are consistent with 
models that show that fan-spine topologies can be destroyed if a
current sheet forms at the null point of the magnetic field.  

\begin{figure*}[t]
\centering
\includegraphics[width=\textwidth]{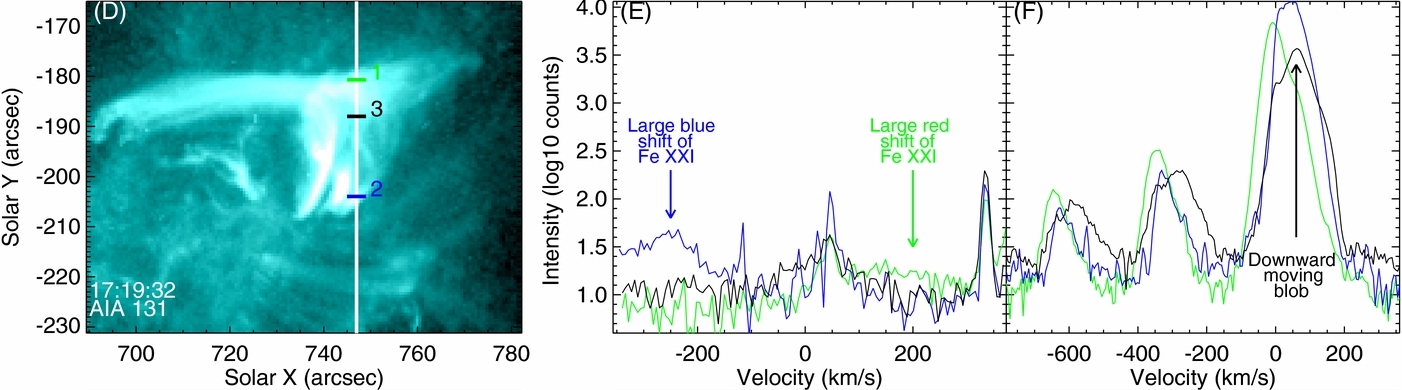}
\caption{Signatures of large flows above the flare loop-top. Left: \aia\ 131\AA~image of the hot flare loops with the position of the \iris\ slit overlaid. Right: \ion{Fe}{xxi} spectra observed in the locations indicated in the left panel. In particular, the green spectra shows the redshifted line above the loop-top. Adapted from \citet{Tian2014d}.} 
\label{fig:flare_tian} 
\end{figure*}

New exciting results on reconnection theory have also been obtained by exploiting
high-resolution observations of the \ion{Fe}{xxi} hot line above the flare
loop-tops.
An observation by \citet{Tian2014d} showed red-shifted \ion{Fe}{xxi} features that coincided with an X-ray source as observed
with \rhessi\ (Fig.~\ref{fig:flare_tian}).  The line of sight speed of this feature as calculated
from the Doppler shift was $\sim$125 km s$^{-1}$,   with $\approx$~100 km~s$^{-1}$ non-thermal width, and it was interpreted as a possible signature of the downward-moving reconnection outflows or the hot retracting loops, as predicted by the flare models. 

\begin{figure*}[t]
\centering
\includegraphics[width=\textwidth]{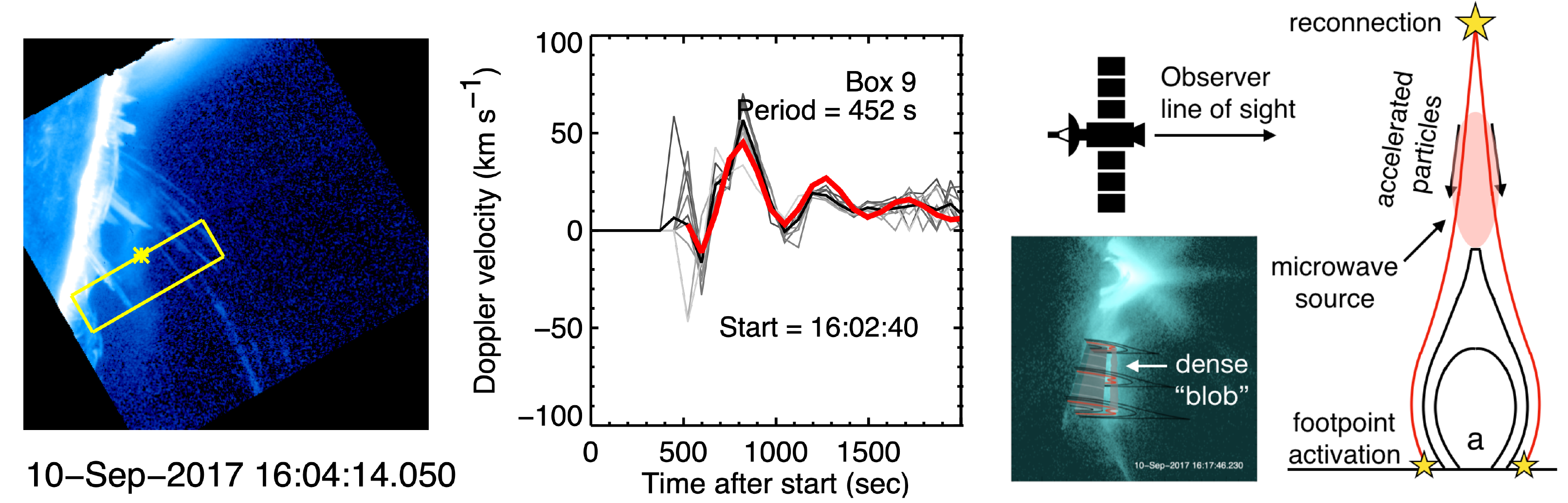}
\caption{Oscillations in the hot plasma above the flare arcade. Left panel: \iris\ SJI 1330\AA~image during the flare showing the hot plasma emission above the loops, with superimposed the field of view of the \iris\ raster. Middle panel: example of damped oscillations observed by \iris\ in the \ion{Fe}{xxi} line. Right panel: schematic representation of the magnetic tuning fork model suggested by the authors as possible interpretation for the observed oscillations. Adapted from \cite{Reeves2020}.} 
\label{fig:flare_kathy} 
\end{figure*}

Large redshifts and simultaneous blueshifts ($\approx$~200--250 km s$^{-1}$) of the \ion{Fe}{xxi} in the loop-top region were also observed by \citet{Polito2018b} during the well-studied 2014 March 29 X-class flare. The locations of the flows were also coincident with HXR loop-top sources observed by \rhessi, and were interpreted by the authors as possible signatures of the deflection flows in the downstream of a flare termination shock, as predicted by \citet{Guo2017}.
While
fast-mode shocks and associated flows are expected by numerical simulations, 
unequivocal evidence has been scarce because of the limited
spatio-temporal resolution of previous spectrographs. Hints of the presence of a
termination shock in supra-arcade fans were also reported by \citet{Cai2019} during the X-class flare on 2017 September 10, the
 second largest flare of this solar cycle.  Using \iris\ SJI images, \citet{Cai2019} found quasi-periodic oscillating features above the flare arcade which
 were suggestive of a bursty and turbulent reconnection process and possibly the
 presence of a termination shock. 
Another recent study reported puzzling oscillations in the hot plasma flows observed in the \ion{Fe}{xxi} line above the loop-top region in the same flare event \citep[][Fig.~\ref{fig:flare_kathy}]{Reeves2020}. The \iris\ resolution made it possible to closely follow the evolution of these flows, which ranged between 20--60 km~s$^{-1}$ in a damped oscillation pattern with periods of ~400 s,  independent of the loop length. These observations seem to be consistent with a scenario in which the dynamics of the loop-top plasma are disturbed by the reconnection outflows impinging on the closed loops below, in agreement with the “magnetic tuning fork” model of \citet{Takasao2016}.

Evidence of transverse MHD waves in flaring loops has also been provided by \iris\  in the \ion{Fe}{xxi}~1354.08~\AA~line. The first fully resolved standing fundamental fast sausage mode was detected by \citet{Tian2016a} during an M1.6 flare, characterised by the $\pi/2$ phase difference between the Doppler velocity and the intensity, with a short period around 25~s. 
Such waves are leading candidates to explain the quasi-periodic pulsation (QPP) phenomena and represent unique seismological tools of flaring loops \citep[see also][]{LiB_2020SSRv..216..136L}. 
Interestingly, an increase of the wave period with time was detected, suggesting a continuous generation of fast sausage modes in the gradually longer flaring loops issued from the reconnection process. Another example was provided by \citet{LiD2017b} with \iris\ observations of a fundamental standing kink mode in a M7.1 flare. These waves are characterised by a periodic Doppler signal (with 3.1~min period) and negligible intensity variation. 

Inspired by the work of \citet{Tian2016a}, \citet{Shi2019} forward modelled the response of the \iris\ \ion{Fe}{xxi} line to fast sausage modes in flare loops under NEI conditions. They found that the synthetic spectral signatures, in particular the phase difference between the intensity and Doppler shift oscillations, are mostly consistent with the \iris\ observations. However, differences between the model predictions and observations still remain, such as 
the presence of asymmetries in the wave period of the Doppler widths (which results from the competition between the broadening due to the superposition of bulk flows and that due to temperature), which are not found in the observations. Additional high cadence spectral observations of the hot flare loops with \iris\ might help addressing some of these discrepancies.

\begin{figure*}[tp]
\centering
\includegraphics[width=0.6\textwidth]{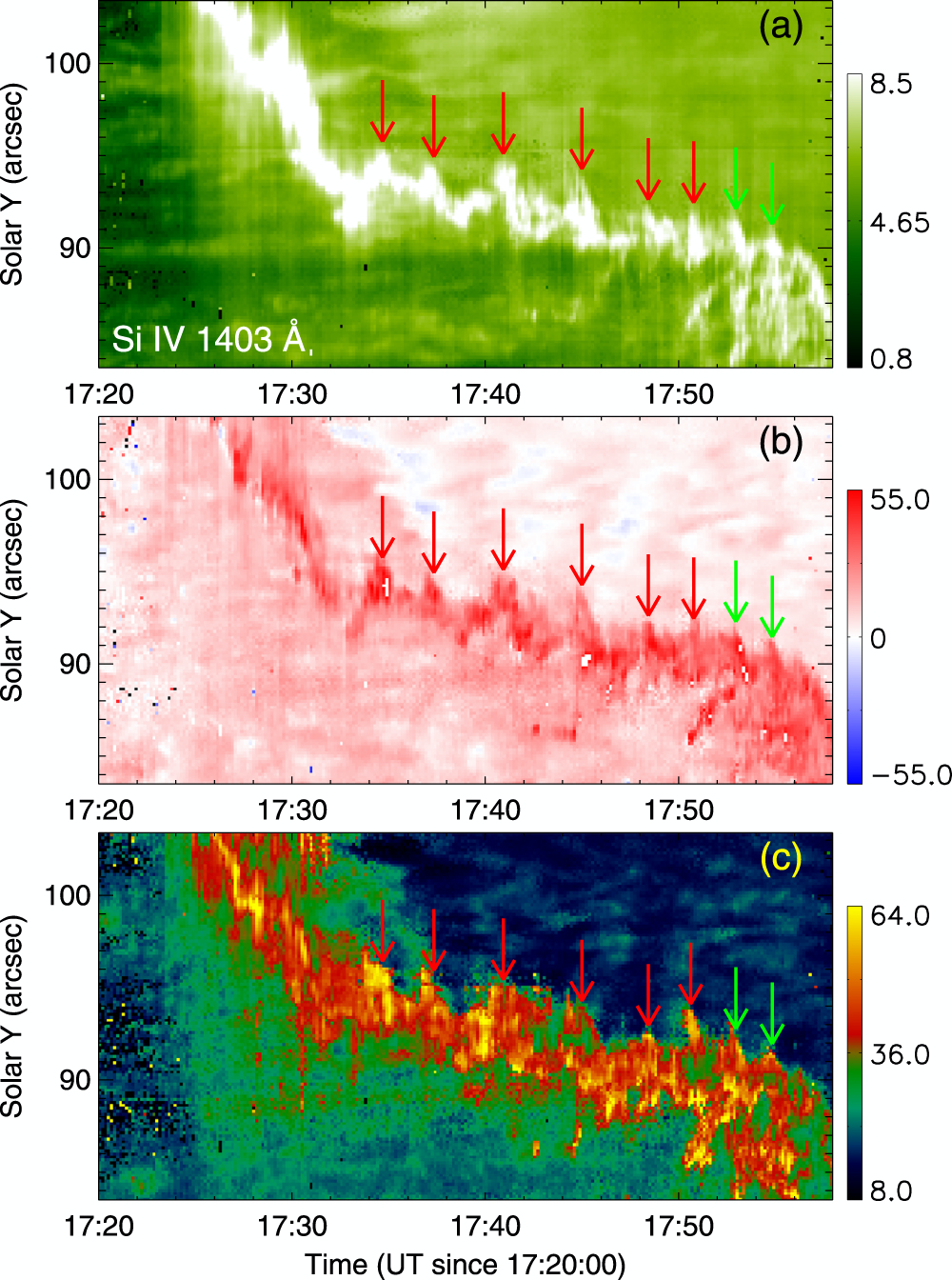}
\caption{Quasi-periodic slipping motions at the ribbons of the 2014 September 10 flare. Temporal evolution of peak intensity (a), Doppler shift (b), and line width (c) of the \iris\ \ion{Si}{iv} line. From \citet{LiT2015a}. } 
\label{fig:flare_slipping} 
\end{figure*}

Other types of oscillations detected with \iris\ have been associated with heating signatures in the lower atmosphere. For example,  \citet{LiD2015a} observed 4-minute QPPs in HXR, radio, EUV and UV light curves, which  corresponded to broadened and redshifted profiles in the \ion{C}{i}, \ion{Si}{iv}, \ion{O}{iv} and \ion{Fe}{xxi} lines observed by \iris\ at the flare ribbons. These observations support the scenario in which the QPPs are produced by non-thermal electron beams accelerated from the corona, where quasi-periodic magnetic reconnection occurs. A possible explanation put forward by the authors is that the reconnection is quasi-periodically modulated by slow p-modes, although other scenarios (for example involving fast waves) cannot be ruled out. 


Signatures of quasi-periodic slipping reconnection has been found by analyzing the motions observed in the \iris\ images and spectra throughout the impulsive phase of the 2014 September 10 X-class flare \citep[e.g.,][]{LiT2015a,Dudik2016}. In this flare event, the footpoints of the slipping flare loops appeared as small-scale bright knots moving along the ribbon following a quasi-periodic pattern, as best observed in the \iris\ SJI 1400\AA~images. While the bright slipping footpoints progressively travelled under the \iris\ slit, the \ion{Si}{iv} spectra also revealed that the line is redshifted and broadened at those locations (Fig.~\ref{fig:flare_slipping}). 
These observations were interpreted by \citet{Dudik2016} in terms of the standard solar flare model in 3D \citep[e.g.,][]{Janvier2015}.
On the other hand, puzzling observations of quasi-periodic slipping motions of ribbon substructures for a non-eruptive flare were reported by \citet{LiT2018}. Despite these observations being reminiscent of the signatures of slipping reconnection, they cannot be simply explained by the 3D standard flare picture, which requires the presence of an eruptive flux rope. Alternative explanations for the observed quasi-periodic property of the flare kernels include MHD waves or the tearing mode instability. For instance, \citet{Parker2017} invoked this instability to explain \iris\ observations of a flare exhibiting sawtooth-like ribbon motion and associated periodicity \citep{Brannon2015, Brosius2015}. Additional high-resolution observations of quasi-periodic motions in both confined and eruptive flares and advanced modelling are required to distinguish between these competing hypotheses and settle this issue.

\subsection{Initiation of coronal mass ejections and flares}
\label{trigger}

Coronal mass ejections occur when energy is built up in the corona in excess of the energy in the potential field, and then suddenly released. Determinations of this ``free energy'' are typically based on non-linear force-free field (NLFFF) extrapolations that use photospheric magnetograms as a lower boundary condition \citep[e.g.][]{Sun2012,Gibb2014}. However, since the photosphere and lower chromosphere are not force-free, the reliability of such extrapolations is questionable.  There are two distinct approaches to improving NLFFF extrapolations, one using coronal constraints, and the other constrained by chromospheric observables \citep[e.g., from \iris, ][]{Aschwanden2015,Aschwanden2016}. However, these approaches lead to very different estimates of the free energy. The resolution of this dilemma may have to await the development of reliable magnetic field mapping in the upper chromosphere or corona.  
 
 Determining the mechanism that suddenly releases the free energy in the corona is fundamental to understanding coronal mass ejection initiation.  Observational precursors of CMEs cataloged prior to the \iris\ mission can be found in the review by \citet{Chen2011}.
 \iris\ has observed evidence for several CME initiation mechanisms, suggesting a variety of processes may be important in triggering eruptions. For example, pre-eruption reconnection (often referred to as ``tether-cutting'') has been identified as a possible trigger in several instances \citep[e.g.,][]{LiT2015b, Dudik2016, Cheng2016}. Many of these examples exploit the sensitivity of \iris\ to reconnection-induced flows and non-thermal motions to provide support to the tether cutting scenario in which flux ropes are formed prior to the eruption \citep[e.g.,][using coordinated observations from \iris, NST and \sdo]{Kumar2015}.  \citet{Zhou2016} interpret oscillations in the \ion{Fe}{xxi} line in a sigmoid minutes before an eruption as evidence for reconnection between the flux rope and the overlying
field. Blue-shifted \ion{Fe}{xxi} line profiles and non-thermal line broadening are observed in another flux rope \citep{Cheng2016} just prior to the eruption, suggesting that reconnection heats threads of the flux rope to 10 MK. In another event, \citet{Cheng2015a} interpret red shifts and enhanced line widths in \iris\ lines at the footpoint of a magnetic flux rope prior to the eruption as evidence for pre-event reconnection along the flux rope. 

\begin{figure*}
    \centering
    \includegraphics[width=\textwidth]{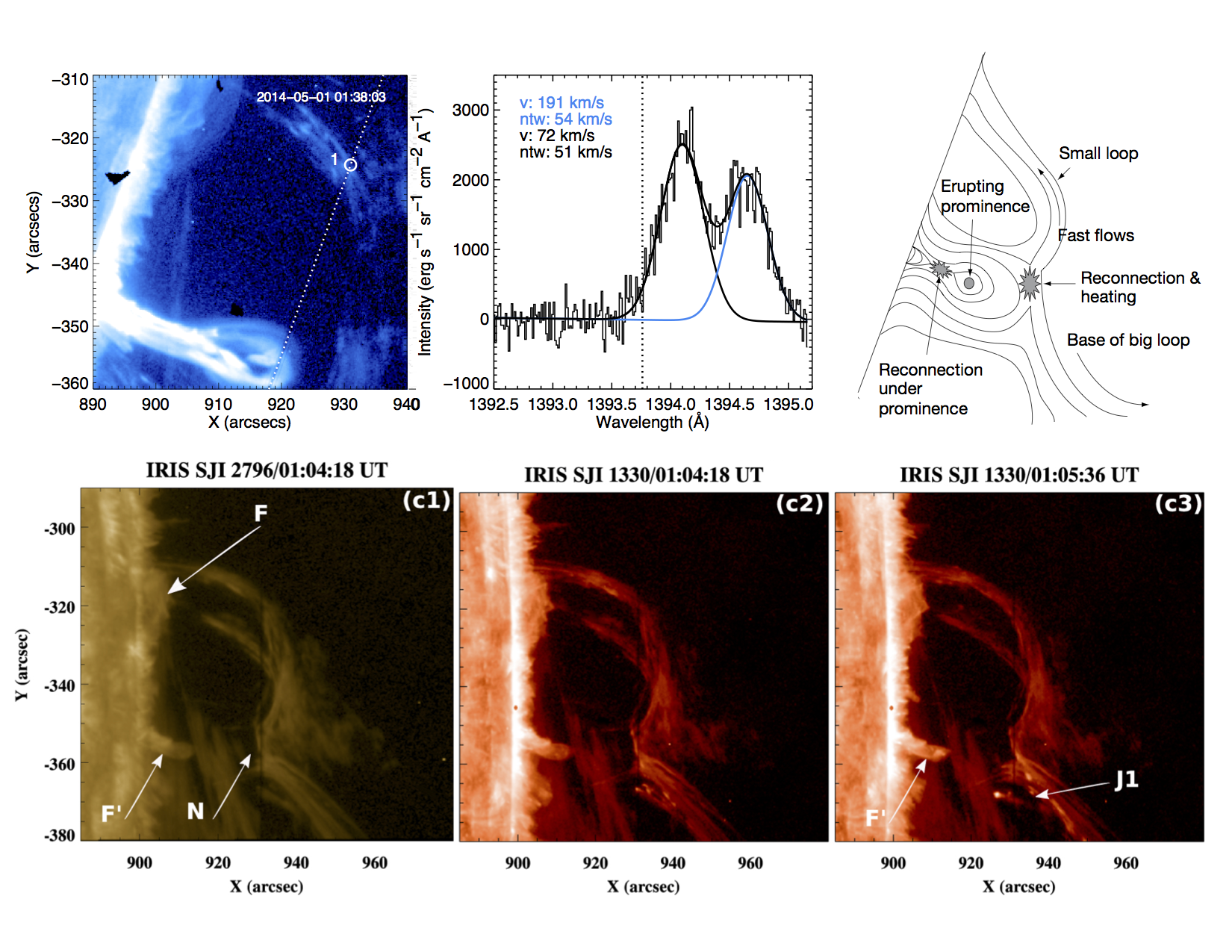}
    \caption{ \iris\ observations of a CME (top left) reveals fast outflows (top middle) as the erupting prominence collided with
the overlying loops containing coronal rain. \citet{Reeves2015} find a brightening under the prominence before the eruption, suggesting that a tether cutting mechanism
 is responsible for initiating
the eruption (top right). An alternate interpretation is given by \citet{Kumar2019}.  The pre-eruption configuration of the filament and coronal rain in the 2796 \AA\ channel (bottom left) shows the ends of the filament (F and F') and a null point (N).  The evolution of the coronal rain in the 1330 \AA\ channel (bottom middle and bottom right) shows brightenings near J1, which are interpreted as plasmoids resulting from reconnection above the prominence, an indication of breakout reconnection.}
    \label{fig_kathy2015}
\end{figure*}
There is also evidence for pre-eruption slipping magnetic reconnection, found by analyzing the motions of flare ribbon kernels in the \iris\ images and spectra \citep[e.g.,][]{LiT2015a,Dudik2016}.  \iris\ observes substructures moving bi-directionally along the flare ribbons in a couple of events, indicating that reconnection in the quasi-separatrix layer between two flux systems has caused slip-running reconnection to occur \citep{LiT2018, LiY2019}.  A fan-spine topology was identified in another event using \iris\ 2796 \AA\ SJI observations, and because of this topology, slip-running reconnection is identified as a possible trigger mechanism \citep{Zuccarello2020}.

Emerging flux can destabilize the magnetic field and cause an eruption. In an event documented by \citet{Hou2019}, an embedded fan-spine structure forms via flux emergence and subsequent reconnection with the overlying flux, triggering a confined flare. In this case, high spatial resolution images from \iris\ confirm predictions from NLFFF modeling that imply the existence of a small fan-spine system embedded in a larger one.  In another event reported by \citet{Kleint2015}, emerging flux destabilizes a filament, leading to a flare.  Increasing Doppler velocities in the \iris\ Si IV line indicate a very high acceleration of the filament (3-5 km~s$^{-2}$) before the impulsive phase of the flare, indicating that the flare was a consequence of the filament eruption and not a cause.  

 Another possible eruption trigger is breakout reconnection, where reconnection high in the corona destabilizes a sheared magnetic field, causing an eruption.  
 Breakout reconnection has been identified as the trigger for a small eruption at the limb based on bi-directional flows and small blobs observed in \iris\ slit-jaw images \citep{Kumar2019}.  However, other authors attribute the triggering of this eruption to tether cutting, based on the identification of brightenings seen in the slit jaw images that occur just as the fast rise phase of the eruption begins \citep[][Fig.~\ref{fig_kathy2015}]{Reeves2015}. In a different event, \citet{Woods2020} observe a brightening in \iris\ located between two flux systems for a series of three flares, the last of which was associated with a CME.  Magnetic modeling indicates that a breakout like mechanism is most likely occurring in this case.


In a few cases, ideal MHD instabilities have been identified as eruption triggers.  An observation of an erupting prominence at the limb finds alternating patterns of red and blue shifts in the \iris\ \ion{Mg}{ii} line, which is interpreted as initiation via the kink instability \citep{Zhang2019b}.  \citet{Woods2018} find that a combination of tether cutting reconnection and the recently proposed double arc instability \citep{Ishiguro2017} is the likely trigger for an X-class flare, based on morphology and brightenings observed by \iris.


A key component in understanding the (in)stability of flux ropes is the long-term evolution of prominences, which often erupt and drive CMEs. Recent combined \iris\ and \hinodes\ observations exploit the high sensitivity and spatial resolution of \iris\ Doppler shift measurements in quiescent prominences to show how these common features can shed helicity through reconnection with the ambient coronal field, thus preventing eruptions
\citep{Okamoto2016}. Coordinated \iris\ observations with ground-based magnetic field measurements have also provided new insights into the injection of helicity at the base of prominences \citep{Levens2016}.  \citet{LiuW2015} examines a prominence eruption at the limb associated with a X1.6 class flare.  The combination of Doppler shifts and cork-screw shaped threads imaged in the \iris\ SJIs lead to the conclusion that the prominence is a left-handed helical structure that unwinds in the counter-clockwise direction during the eruption.  This result is interesting because the source active region is in the southern hemisphere, and left-handedness would be opposite to the prevailing magnetic helicity for that hemisphere \citep[e.g.,][]{Pevtsov1995}. The authors speculate that this anomalous chirality may contribute to the prominence's eruptive potential.

\begin{figure*}
    \centering
    \includegraphics[width=0.8\textwidth]{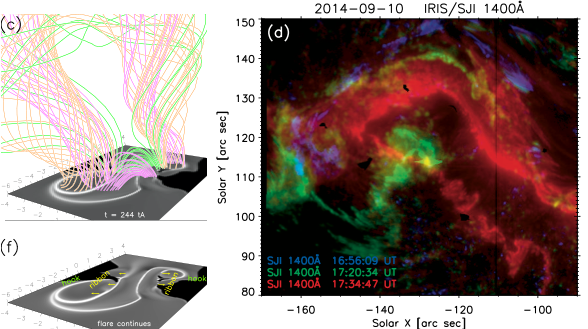}
    \caption{Drifting motions and fine structure (e.g., hook-shapes) of flare
     ribbons observed at the high \iris\ spatial
resolution (right), through comparisons with 3D MHD models (left)
\citep{Aulanier2019}, provide key information about reconnection
occurring in the corona during an eruption and the structure of flux
ropes as they evolve to form ICMEs, critical in the era of \psp\ and \solo.
}
    \label{fig_guillaume}
\end{figure*}

Most attempts at flare forecasting prior to IRIS primarily used the line of sight magnetic field data \citep[see e.g.][]{Barnes2016}, but recent \iris\ results show exciting promise for accurate predictions of the timing of the onset of solar flares using UV spectroscopic data. \citet{Panos2020} applied machine learning
techniques to show how chromospheric spectra
(\ion{Mg}{ii} h/k and, in particular, triplet emission)
in pre-flare regions (40 min before flare) are clearly distinct, 
and could, in principle, be used to predict flare occurrence
with high accuracy.  

\citet{WoodsM2017} studied the pre-flare phase of an X1 flare observed by \iris\ and found strongly blue shifted plasma flows in the \iris\ \ion{Si}{iv} line with velocities up to 200 km~s$^{-1}$ 40 minutes before the eruption.  These flows are along the active region filament, and modeling indicates that the flows are due to internal reconnection in the flux rope or tether cutting along the filament.  \citet{Bamba2017} examine an X1.6 flare and find blue shifts in a variey of \iris\ and \hinodee\ lines at the location of an isolated positive polarity region about 30 minutes prior to the flare.  They find blue shifts of $\sim$100 kms$^{-1}$ in all the \iris\ and \eis\ lines, indicating a jet-like reconnection outflow which may change the topology of the magnetic field enough to trigger the subsequent eruption.

The initiation and propagation of CMEs will continue to be a robust topic of study for \iris\ as solar cycle 25 ramps up, especially considering the recent launches of \psp\ and \solo. 
Future work will combine the powerful spectroscopic diagnostics of \iris\ with 
highly complementary high-resolution transition region and coronal measurements with instruments on \solo. \iris\ observations of eruptions focused on possible source regions during \psp\ perihelia will provide the groundwork for holistic views of eruptions from the chromosphere to the heliosphere.  Combinations of \iris\ observations and MHD modeling have implied that the flare ribbon geometry is an indicator of how the CME evolves as it erupts \citep{Aulanier2019}, and coordinated observations between \iris, \psp\ and \solo\ will be the perfect test bed for this idea.  

\iris\ observations to date have provided evidence for a variety of triggering mechanisms for solar eruptions.  It would be very interesting to conduct 
statistical studies using existing (and future, as the cycle ramps up)
\iris\ observations of eruptions, in coordination with \saia, \hinodex, and \solo\ imaging, in order to understand which, if any, of the triggering mechanisms that have been investigated so far are the most common.  Future efforts could also focus on the formation and evolution of flux ropes, the seed structures of solar eruptions. These studies would benefit from a combination of \iris, HSO, and GBO observations, and 3D numerical models to study the build-up, nature and trigger
mechanisms of pre-eruptive magnetic configurations. GBO measurements of the field in the photosphere and chromosphere, as
well as signatures of cancellation, reconnection, and heating (e.g.,
\iris\ observations of UV bursts, rapid outflows) around the polarity inversion line are key to study flux rope formation, and this research will benefit from
new coordinated observations, especially from \dkist.

\section{Solar cycle variations of far- and near ultraviolet radiation}
\label{irradiance}

Like many solar radiative outputs, the \iris\ far and near ultraviolet passbands are subject to variation over the course of the sunspot cycle. Variations of the solar ultraviolet emission impact the Earth's atmosphere and space environment in a variety of ways, including thermospheric density changes that impact satellite drag and the chemistry of stratospheric ozone. The accessibility of strong ultraviolet \ion{Mg}{ii} emission and the relative ease of measurement from low earth orbits led to the creation of the ``\ion{Mg}{ii} index'', the ratio between the \ion{Mg}{ii} h/k lines and their nearby continua \citep[see, e.g.,][]{1997JGR...102.2597D}. This index has been measured by NOAA on a daily basis since 1978 because it is a strong proxy of solar variability over the entire UV spectral domain, but especially at solar minimum when coronal activity proxies, like the 10.7cm radio flux (``f10.7’’), are insensitive. 

However, since little is understood about the physical underpinnings of ultraviolet variability or the \ion{Mg}{ii} index \-- especially its spatial, temporal and spectral relationship to the magnetized and unmagnetized solar atmosphere \-- the \iris\ synoptic program was initiated shortly after first light. \iris\ regularly obtains a variety of observations to characterize the cyclical variations of spatially resolved NUV and FUV spectra, including quiet Sun measurements, full-disk spectroheliograms, active region disk passages, etc where each will help us to characterize the origins of solar variability.

The time since \iris\ launch has covered the descent of Sunspot Cycle 24, the depths of solar minimum and the very earliest sprouts of Sunspot Cycle 25. Over this time \iris\ has completed more than ninety full disk spectroheliograms \cite[see, e.g.,][]{2015ApJ...811..127S, 2020ApJ...905L..33B}. These observations stitch together a large number of individual \iris\ pointings to build a mosaic of the solar atmosphere at high spectral and spatial resolution over the course of a day, depending on the characteristics of the \iris\ observations employed. Preliminary studies of the full-disk spectroheliogram archive have been focused on the chromospheric footprint of coronal holes \citep[][]{2020ApJ...905L..33B} and the impact of magnetism \citep[][]{2015ApJ...811..127S} on the nature of chromospheric spectra. In addition, \citet{Kayshap2018b} have found  a significant difference, for a given magnetic field strength, between the \ion{Mg}{ii} h/k spectral properties (and thus \ion{Mg}{ii} index) of QS and CH, when observed at the 1\AA\ resolution of space weather measurements with the commonly used ``sun-as-a-star'' \solstice\ instrument \citep{Kayshap2018b}. Since the \ion{Mg}{ii} index is often used as a proxy for plage areas (which vary greatly throughout the cycle), this intensity contrast between CH and QS must be taken into account in future irradiance studies that use this index, in order to properly account for solar activity. 

The comprehensive archive of full disk \iris\ spectroheliograms sees research launching on the impact of center-to-limb effects, active region plage, coronal holes, and sunspots on the \ion{Mg}{ii} index, thereby elucidating what drives the \ion{Mg}{ii} index variability throughout the cycle, and especially in the rising phase of Sunspot Cycle 25.

\begin{figure}
\includegraphics[width=1.0\linewidth]{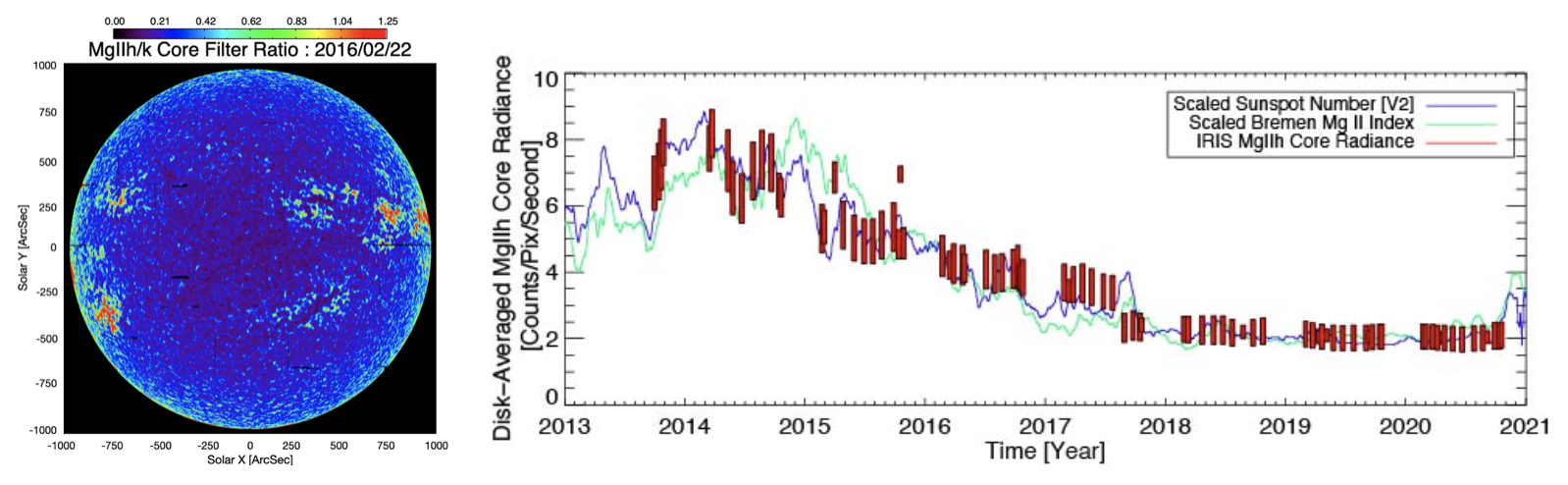}
\caption{\small \iris\ spectroheliograms provide the opportunity to explore the relationships between the Ultraviolet irradiance over the course of the solar cycle. (Left) An image constructed from the ratio of the integrated \ion{Mg}{ii} h and k line cores. (Right) the variance of the disk-averaged (integrated) \ion{Mg}{ii} h line core (red) since launch from the 88 \iris\ spectroheliograms to date in comparison with scaled versions of the Sunspot number (blue) and the University of Bremen \ion{Mg}{ii} index over the same epoch, we see that the \iris\ measurements follow the observed trend in the \ion{Mg}{ii} index, permitting future investigations into the physical origins of \ion{Mg}{ii} variance. \label{fig_scott}}
\end{figure}

\iris\ has obtained a timeseries of full-disk spectroheliograms on a monthly basis since 2013, and will continue to obtain such data. This unique timeseries provides spatially resolved spectra in the NUV and FUV, including \ion{Mg}{ii} h \& k spectra, which can be used to determine the \ion{Mg}{ii} index, a commonly used proxy to gauge impact of variations of solar irradiance on the Earth's upper atmosphere. Comparisons between high-resolution \iris\ spectra and \hmi{} magnetic fields can be used to determine how variability in magnetic structures (plage, ARs) impact the \ion{Mg}{ii} index, and through \solstice\ sun-as-a-star irradiance spectra, develop an improved \ion{Mg}{ii} proxy for irradiance modeling.

\iris\ will continue weekly synoptic observations of the chromospheric counterparts to coronal bright points to help study the onset and strength of the next solar cycle, properties that have been difficult to predict \citep[e.g.,][]{2020SoPh..295..163M}. This will build on recent research suggesting that such bright points appear at mid-latitudes long before onset of the new cycle as part of an ``extended solar cycle’’ \citep[][]{2018FrASS...5...38S}. Weekly observations (part of two IHOPs) are focused on detecting small magnetic features that, according to recent results \citep{McIntosh2014}, arise due to coherent patterns of flux emergence that are related to the dynamo process. These features are hypothesized to relate to underlying toroidal magnetic flux systems. The \iris\ synoptic effort will help to contrast changes in structure, emission and timing of these magnetic systems between the active latitudes and the equator \citep[][]{2019SoPh..294...88M}. The precise location of these magnetic systems may be important for predicting the timing of solar minimum, the strength of the upcoming cycle and, more generally, the long-term variability of the Sun. Since 2017 these IHOPs have been run on a weekly basis, and will continue to be run as the new cycle ramps up activity. Detailed analysis of the timing and location of bright point activity may provide constraints on dynamo models.

\section{Solar-Stellar Connections}
\label{stellar}

The Sun is the one star where we can dig down into the fine-scale nature of the
top-level processes evident in full-disk stellar spectra and broadband photometry \citep[for general reviews see][]{EngvoldBook, Hall2008}. Observations of the solar outer atmosphere offer a close-up view of the physical processes at work in stellar atmospheres, and it allows us to explore the extent and limitation of the solar-stellar analogy ("solar-stellar connection") \citep[e.g.][and references therein]{Testa2015}. Previous work on the solar-stellar connection has focused on spectral irradiance variations over magnetic cycles \citep{Baliunas1995, BV2007, Egeland2015, Ayres2020}, magnetic field dynamo modeling \citep{Charbonneau2013,BrunBrowning2017}, surface magnetic flux transport modeling \citep{Schrijver2001}, magnetic activity diagnostics in young solar analogs \citep{Gudel2007}, asteroseismology \citep{vanSaders2016, Garcia2019}, stellar winds and mass loss \citep{Wood2004, Matt2012, Cranmer2017}, abundance effects \citep{Laming2015}, the link between Sun-as-a-star flare light curves and coronal mass ejection occurrence \citep{Harra2016}, solar flare processes such as the Neupert effect \citep{Hawley1995, Gudel1996, Gudel2004}, and flare scaling relationships \citep{Shibata2002, Namekata2017, Notsu2019, Okamoto2021}.  These comparisons have largely resulted from observations of hotter plasma emitting in X-rays (GOES, Chandra, XMM-Newton), the extreme ultraviolet (e.g., EVE, EUVE), and a narrow spectral region around \ion{Ca}{ii} H and K.  Yet, many types of lower-atmospheric solar phenomena may have stellar analogs that are only revealed and constrained in the FUV and NUV with broad spectral coverage, such as from IUE and the HST. Complementary, spatially resolved solar spectra have long been lacking at these wavelengths, precluding direct comparisons to stellar observations.  
And now, \iris\ provides for the first time the combination of high 
spectral resolution in crucial ultraviolet bands, comparable to that 
delivered by the best modern space instruments for the stars, 
as well as high spatial resolution and fast temporal response.  With \iris, the vagaries that result from mixing radiation from different locations along a flare ribbon, or from basal and active region components, are not an issue.  The  heating properties as a function of atmospheric depth can be uniquely determined on the Sun.

The basic properties of stars, such as radius, temperature, gravity, mass, magnetic field, and chemical composition, are critically important to establishing their
age, formation history, and angular momentum evolution, and consequently the history of the Galaxy as a whole. Most stellar parameter determinations rely on interpreting the spectra of stars, exploiting the record it carries of the photospheric temperature and pressure, and the imprint of its elemental composition, all recorded in the strengths
of absorption features due to individual lines. The basic properties
of stars are then determined from comparisons of observed stellar
spectra with calculations from first principles, i.e., stellar
spectral synthesis codes: temperature from the relative strength of lines with low and high excitation;
gravity from lines of low and high ionization; elemental abundances follow
from predictions of ionization and excitation \citep[see e.g.,][for recent work in the UV and NIR]{Walkowicz2008PhD, Souto2020}.

However, synthesis codes are complex, adopted line parameters vary widely among models, and it is unclear whether the inherent limitations of non-magnetic 1D model calculations can adequately reproduce line formation over the entire photosphere and chromosphere \citep{Uitenbroek2011,Lebzelter2012}. Furthermore, most spectral
synthesis analyses must simultaneously determine temperature, gravity, magnetic field, and filling factors all at once. For the Sun,
however, the temperature and gravity are known accurately through other means so
that deficiencies in the spectral line modeling can be more easily
identified. 
Spatially resolved \iris\ observations of photospheric lines thus provide a unique opportunity to benchmark spectral observations and help determine non-LTE corrections to existing
model opacities in the ultraviolet \citep[e.g.,][]{Bruls1992, Criscuoli2020}, one of the main challenges in stellar atmospheric physics and solar and stellar irradiance modeling.
Furthermore, the lack of direct spatial information for other stars requires unconstrained filling factors of different magnetic field strengths \citep[e.g.,][]{JK2000} and mixed emission components from spots, plage, and basal regions \citep{Walkowicz2008PhD, Ayres2020}, whereas the spatial extent of many solar  features can be directly measured \citep[e.g.,][]{Fontenla2015}.

\iris\ results have also been useful to understand
stellar flux-flux relations.   Correlations between magnetic field and 
\iris\ diagnostics 
show many similarities with those found for stars, but also
intriguing discrepancies for TR lines
\citep{Barczynski2018}. The spatially resolved nature of the \iris\
observations suggests that these differences are caused by
global averaging inherent in stellar observations, e.g.,
center-to-limb variations. Other studies exploit 
\iris\ and SORCE solar observations, and HST and Chandra
observations of $\alpha$ Cen A and B to reveal novel flux-flux power
laws over activity cycles, allowing improvements 
of proxy-based irradiance models, estimates of ionizing
radiation of exoplanet hosts, and quantitative assessment of multi-component magnetic dynamo models of the Sun and solar analogs over stellar evolution \citep{Ayres2020}.

\begin{figure*}[t]
\centering
\includegraphics[width=\textwidth]{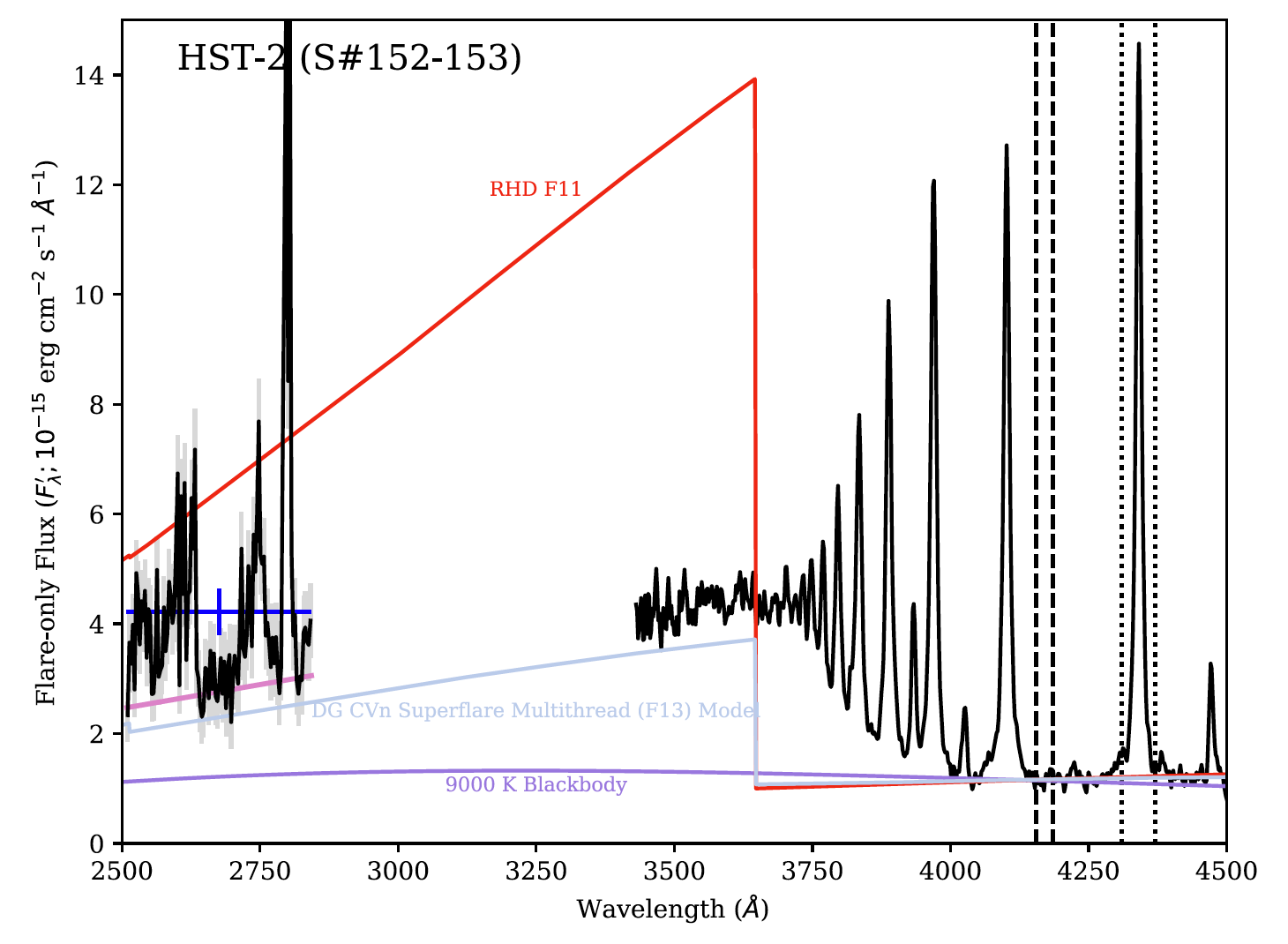}
\caption{Composite NUV flare spectrum of an event from the dM4e star GJ 1243 (from \cite{Kowalski2019HST}), observed by the Hubble Space Telescope/COS ($\lambda < 2840$\AA) and the ARC 3.5m at the Apache Point Observatory ($\lambda>3400$\AA).  This observation shows that the $U$-band consists of  hydrogen Balmer recombination radiation that decreases toward the \iris\ NUV range where  additional \ion{Mg}{ii} and \ion{Fe}{ii} radiation is prominent.  However, the Balmer jump is much smaller than predicted by a model (red, RHDF11) of chromospheric heating by a reasonable flux of non-thermal electrons. This flare has an energy of $E_U \sim 2\times10^{31}$ erg, which is likely comparable to a large solar flare.  } 
\label{fig:stellarflare} 
\end{figure*}

\iris\ NUV spectra during solar flares provide constraints on a major missing observational aspect
of a dominant mode of radiative energy release in both solar and stellar flares \citep{Osten2015}.
Whereas spectra of flares on other stars (primarily
dwarf M emission (dMe) stars) cover a broad wavelength range that includes the optical
and NUV wavelengths that are observable from the ground (i.e.,
wavelengths longer than $\sim350$ nm), \iris\ spectra probe shorter ultraviolet wavelengths where currently 
the Hubble Space Telescope can safely observe the dMe flares that are relatively small in amplitude with a bright Balmer continuum component, as in Figure \ref{fig:stellarflare} (instrumental bright limits prevent frequent monitoring of nearby stars in the NUV due to the possibility of a large flare; Osten 2017, Instrument Science Reports 2017-02)\footnote{\url{https://www.stsci.edu/hst/instrumentation/stis/documentation/instrument-science-reports}}.  From extrapolations of optical
spectra of much more impulsive (and often, larger and more energetic) dMe flares, the peak of the
white-light continuum radiation may occur much further into the ultraviolet, near 300 nm or shorter \citep{HF92, Hawley2003, Fuhrmeister2008, Kowalski2013, Kowalski2016}, which is a critical challenge for models of solar flare heating \citep[see also \S~\ref{Flare_diagnostics} ][]{HF92, Allred2005, Allred2006}.

Detailed modeling of the \iris\ observations of NUV properties of solar flares  serve as ``ground truth'' for  modeling approaches to these white-light properties in stellar flares.  
In particular, \citet{Kowalski2017Mar29} and \citet{Graham2020} studied \iris, \rhessi, and Fermi/GBM 
data of two X-class flares to constrain RADYN simulations of the
chromospheric response to the impact of moderate-to-high fluxes of electron beams and found that the NUV continuum intensity and red-shifted emission line satellite components were rather well-reproduced
by two transient flaring layers at
chromospheric heights:  a stationary heated layer  below a chromospheric condensation (see also \S~\ref{flares_evaporation}).

A similar scenario was recently proposed to explain the 10,000 K blackbody-like white light radiation \citep{HP91} and a peak near 300 nm in flares in dMe stars:  
 RADYN simulations of M dwarf flares with even higher electron beam fluxes show a similar  scenario with two flaring layers, which are deeper and denser in the M dwarf atmosphere \citep{Kowalski2015}. 
 
 Because dMe flares are rarely observed with both high cadence and high resolving power (and never with any direct spatial resolution),  the red-shifted satellite components that are expected from  extreme \citep[and possibly unrealistic;][]{Krucker2011} electron beam fluxes are difficult to validate using only stellar data.  A line-to-continuum ratio in \iris\ NUV spectra was recently investigated to better diagnose the amount of heating in the low chromosphere or upper photosphere in solar flares \citep{Kowalski2019}, demonstrating the relative heating rate of these lower layers to be significant relative to that of the upper chromosphere.
 Other sources of evidence have previously hinted at significant heating below the upper chromosphere \citep{MartinezOliveros2012, Warren2014};  resolving this issue on the Sun would provide important clues for modeling the origin of 10,000 K blackbody-like component\footnote{\cite{Kretzschmar2011} reported that Sun-as-a-star data of many solar flares exhibits a color temperature of 9000 K, but this likely results from an error in the analysis \citep{Hawley1995, Kleint2016, Castellanos2020}.} that clearly dominates the energetics in some dMe flares.

 \iris\ provides robust continuum measurements of flares in the FUV as well \citep{Tian2015}.  Multifarious types of FUV spectral shapes (in units of \AA$^{-1}$) have been reported in stellar flares:  slightly rising toward shorter wavelengths \citep{Froning2019}, nearly flat continua \citep{HP91} and continua falling toward shorter wavelengths \citep{Loyd2018}.  However, in all cases the spectra are bluer than an extrapolation of the NUV optically thin Balmer continuum recombination models, as inferred in \iris\ data of solar flares \citep[][see also \S~\ref{Flare_diagnostics}]{Heinzel2014, Kowalski2017Mar29}.  Also, the FUV continuum radiation is always observed to evolve on shorter timescales than the NUV and optical continuum and emission lines radiation \citep{Hawley2003}, which can be investigated further with \iris\ \citep[e.g.,][]{Tian2015}.  Another challenging observation of the FUV in large stellar flares comes from the rapidly rotating young solar analog EK Dra, observed serendipitously with HST/COS \citep{Ayres2015EK}.  Superimposed on the decay phase were bright bursts that showed up only in the FUV continuum without a response in higher temperature lines, like \ion{Fe}{xxi} and \ion{C}{ii}.  Curiously, the redshifted transition region line emission at tens of km s$^{-1}$ speeds that are often\footnote{Redshifts of several thousand km s$^{-1}$ have been occasionally reported in HST spectra of TR lines in dMe flares \citep{Woodgate1992, Bookbinder1992}.} reported in stellar flares is interpreted as chromospheric condensation from electron beam impact \citep{Hawley2003} or as condensation in cooling ``post-flare'' loops \citep{Ayres2015EK}, in line with the solar flare paradigm.  
 
\iris\ is uniquely suited for comparisons of solar flare FUV and NUV
spectra to stellar flare spectra. 
For example, intriguing observations of temporal
differences in NUV and FUV continuum emission in M dwarf flares are
also seen in \iris\ spectra of solar flares. Future work will help
determine the cause of these effects in M dwarf flares, building on
preliminary \iris\ results
suggesting that these are related to chromospheric condensations that
occur as a result of impulsive flare dynamics.  Investigations into the physics of very broad \ion{Mg}{ii} lines in solar flares \citep[][see also \S~\ref{Flare_diagnostics}]{Rubio-da-Costa2017, Zhu2019, Kerr2019b} may provide insight into the even more extreme broadening that is observed in archival HST echelle spectra of M dwarf flares in the NUV around \ion{Mg}{ii} \citep{Hawley2007}.

Since many M dwarfs are hosts for planets in the habitable zone, the spectral
variation from FUV to NUV is important for accurate
assessments of biosignatures in future exoplanet transit spectroscopy \citep{Tian2014Bio, Harman2015}.  There is now a large database of time-tagged FUV spectra of old and young exoplanet host stars observed with the HST \citep{Osten2005, France2016, France2018, France2020, Loyd2018, Loyd2018B} as part of the MUSCLES Treasury Survey and HAZMAT program.
Spatially resolved \iris\ spectra of  a wide range of targets (from QS
to AR plage) can also
be used to better understand the Wilson-Bappu effect in the
\ion{Mg}{ii} k line, which ties the line width to the
absolute visual magnitude of a star and can be used for determining
the distance of stars.  Further, to our knowledge, the ubiquitous property of (non-flaring) stellar TR lines to be composed of narrow and broad line components  \citep{Wood1997, Pagano2004, Peter2006} during all activity phases has the great potential to be finally resolved with \iris\ data \citep[see discussions in ][]{Ayres2015Cen}.

 \section{Conclusions}
 \label{conclusions}
 
 \iris\ is the highest resolution observatory to provide seamless coverage of spectra and images from the photosphere into the corona. The unique combination of near and far-UV spectra and images at 0.33-0.4\arcsec\ resolution at high cadence allows the tracing of mass and energy through the critical interface between the solar surface and the corona or solar wind. \iris\ has enabled crucial research into the fundamental physics processes thought to play a role in the low solar atmosphere such as ion-neutral interactions, magnetic reconnection (e.g., resulting from braiding or driving flares and smaller scale events), the generation, propagation, and dissipation of waves, the acceleration of non-thermal particles, and various small-scale instabilities. These new findings have helped provide novel insights into a very wide range of phenomena including the discovery of non-thermal particles in coronal nanoflares, the formation and impact of spicules and other jets, resonant absorption and dissipation of Alfv\'enic waves, energy release and jet-like dynamics associated with braiding of magnetic field lines, the importance of thermal instability in the chromosphere-corona mass and energy cycle, the role of turbulence and the tearing mode instability in magnetic reconnection, the contribution
 of waves, turbulence, and non-thermal particles in the energy deposition during flares and smaller-scale events such as Ellerman bombs or UV bursts, and the role of flux ropes and various other mechanisms in triggering and driving CMEs. 
 
 Machine learning techniques are increasingly being utilized to exploit the very rich \iris\ observations, including a revolution in chromospheric thermodynamical diagnostics through inversions and machine learning, and novel flare predictions that provide insight into the trigger mechanisms behind these space weather events. Advanced numerical simulations and radiative transfer calculations have been key for interpreting the \iris\ observations, including exciting insights into the coldest regions and the key role of ambipolar diffusion and ion-neutral interactions in heating the chromosphere and driving jets, the role of braiding-driven ion-cyclotron waves and dissipation in heating transition region loops, and the nature of the physical mechanisms driving white light flares (including on M-dwarf stars, affecting exoplanet habitability).
  
Many of these results are surprising, challenge current models and provide novel avenues to explore, often requiring new observing modes that utilize the flexibility of \iris\ on different targets or events.

\iris\ observes the region that is at the interface between the photospheric driver of solar atmospheric heating and space weather events and the corona, and \iris\ observations are thus best analyzed in coordination with observations from other space-based observatories and ground-based observatories. The next few years promise an exciting new era of solar research: \iris\ will continue to provide seeing-free imaging and spectroscopy at high spectral resolution of the chromosphere, transition region, and flaring corona, revealing velocity, temperature and density diagnostics at high spatio-temporal resolution in thermal domains that are complementary to existing instruments (\sdo, \hinode, \nustar, \alma, \sst, \bbso) and new instrumentation onboard \solo\ and \psp, and from the ground, such as NSF's 4m DKIST telescope. As outlined in the various sections of this paper, unique coordinated datasets of \iris\ with  novel photospheric, chromospheric, transition region, coronal and solar wind diagnostics from new instrumentation, coupled with increasingly sophisticated numerical models and machine learning and inversion techniques, will be able to build on the advances made with \iris\ in the past few years and address many of the unresolved issues. This will lead to key advances in understanding physical mechanisms such as multi-fluid interactions, reconnection and turbulence, energy transfer between the surface and corona or wind, instability of the solar atmosphere, solar-stellar studies, and the drivers of the irradiance that impacts Earth's upper atmosphere.

{\bf Disclosure of Potential Conflicts of Interest} The authors declare they have no conflicts of interest.

\begin{acknowledgements}
The authors would like to thank the \iris\ science and operations team, including all science planners, for enabling the \iris\ science mission. We are grateful to Ed DeLuca for careful reading of the manuscript. We also would like to thank Tom Ayres, Paul Bryans, Georgios Chintzoglou, Milan Gosic, Chad Madsen,  Navdeep Panesar, Ruth Peterson, Alberto Sainz-Dalda, Alan Title, Sanjiv Tiwari, and Magnus Woods for discussing some of the \iris\ results reviewed here.  We gratefully acknowledge Alberto Sainz Dalda for providing Fig.~\ref{fig_alb}a. \iris\ is a NASA small explorer mission developed and operated by LMSAL with mission operations executed at NASA Ames Research center and major contributions to downlink communications funded by ESA and the Norwegian Space Centre. This work is supported by NASA contract NNG09FA40C (\iris). The simulations have been run on clusters from the Notur project, and the Pleiades cluster through the computing project s1061, s2053 and s8305 from the High End Computing (HEC) division of NASA. This research was in part supported by the Research Council of Norway through its Centres of Excellence scheme, project number 262622, and through grants of computing time from the Programme for Supercomputing. PA acknowledges STFC support through Ernest Rutherford Fellowship grant ST/R004285/2. Numerical computations for Fig.~\ref{fig:resonantkhi} were carried out on Cray XC50 at the Center for Computational Astrophysics, NAOJ.
\end{acknowledgements}

  
\bibliographystyle{spr-mp-sola}



\end{article} 

\end{document}